\documentclass[12pt]{article}
\pdfoutput=1

\usepackage{tikz}
\usetikzlibrary{arrows}
\newcommand{\midarrow}{\tikz \draw[-triangle 90] (0,0) -- +(.1,0);}
\usepackage{comment}

\definecolor{lightgray}{rgb}{0.7421875,0.7421875,0.7421875}
\usepackage{float}
\usepackage{jheppub}
\usepackage{amsmath}
\usepackage{amsfonts}
\usepackage{amssymb}
\usepackage{latexsym}
\usepackage{simplewick}
\usepackage{colonequals}
\usepackage{enumitem}
\usepackage{cancel}
\usetikzlibrary{decorations.pathmorphing}

\tikzset{snake it/.style={decorate, decoration=snake}}

\usepackage{amsthm} 
\usepackage{lipsum}
\theoremstyle{definition}
\newtheorem{lemma}{Result}

\RequirePackage{amssymb,amsmath,amsthm,amsfonts}
\RequirePackage{bbm}
\numberwithin{equation}{section}

\usepackage{mathtools}
\usepackage{simpler-wick}

\usepackage{extarrows}

\usepackage{epsf}
\usepackage{color}
\usepackage{graphicx}
\usepackage{subfig}
\usepackage{dsfont}

\usepackage[export]{adjustbox}
\usepackage{hyperref}
\definecolor{darkred}{rgb}{0.8,0.1,0.1}
\hypersetup{colorlinks=true, linkcolor=darkred, citecolor=blue, linktoc=page}

\graphicspath{{figures/}}
\usepackage[utf8]{inputenc} 

\AtBeginDocument{}

\usepackage{comment}

\title{SYK Model with global symmetries in the double scaling limit}

\author[]{Prithvi Narayan,}
\author[]{Swathi T S}
\emailAdd{prithvi.narayan@gmail.com}
\emailAdd{swathisubramanian3@gmail.com}
\affiliation[]{Department of Physics, Indian Institute of Technology, Palakkad 678557, India}

\abstract{
We discuss the double scaling limit of the SYK model with global symmetries. We develop the chord diagram techniques to compute the moments of the Hamiltonian and the two point function in the presence of arbitrary chemical potential. We also derive a transfer matrix acting on an auxilliary hilbert space which can capture the chord diagram contributions. We present explicit results for the case of classical group symmetries namely  orthogonal, unitary and symplectic groups.  We also find the partition functions at fixed charges.

}

\begin{document}

\setcounter{footnote}{0}

\maketitle

\setcounter{equation}{0}
\setcounter{footnote}{0}

\section{Introduction}\label{sec:intro}
Sachdev Ye Kitaev (SYK) model, a variant of Sachdev-Ye model\cite{PhysRevLett.70.3339} introduced by Kitaev \cite{KitaevTalk1}\cite{PhysRevD.94.106002} is one of the few non trivial exactly solvable models  at strong couplings. It is a quantum mechanical model of $N$ fermions with disordered  interactions which is solvable at large $N$ in low energies and strong couplings. It displays several interesting properties, namely conformal two point function, an emergent reparamentrization symmetry\cite{PhysRevD.94.106002}\cite{Polchinski2016TheSI}, maximal chaos (diagnosed by the maximal allowed Lyapunov exponent) \cite{Sachdev:2010uj,PhysRevD.94.106002,Maldacena:2015waa}, low energy dynamics governed by Schwarzian action etc. These  properties are indicative of the system having a dual gravitational description, since black holes in gravity is maximally chaotic and also Schwarzian action is known to be equivalent 2d dilaton gravity\cite{Maldacena:1997re,10.1093/ptep/ptw124}. 

The Hamiltonian for the SYK model is given by \footnote{The model was originally proposed with $p=4$ \cite{PhysRevLett.70.3339} and was generalized to general $p$ in \cite{PhysRevD.94.106002}}\\
\begin{equation} 
    H= {\mathrm i}^{\frac{p}{2}}\sum \limits_{1 \le i_1 < i_2<\dots < i_p \le N} J_{i_1..i_p}\chi_{i_1} \chi_{i_2} ...\chi_{i_p} \label{eq.1}
\end{equation}
where $\chi_i, i=1, \dots N$ are Majorana fermions satisfying the following anticommutation relations,
\begin{equation}
\{\chi_i,\chi_j\}=\delta_{ij}  \label{eq.2}  
\end{equation}
and $J_{i_{1}i_{2}..i_{p}}$ are real couplings picked randomly from a Gaussian distribution The variance of the random couplings is given by,
\begin{equation}\label{majoranavariance}
    \langle\langle J^2_{i_1..i_p}\rangle\rangle =\frac{\mathcal{J}^2}{{ {N \choose p}}}
\end{equation}
for some constant ${\cal J}$. Here $\langle\langle .. \rangle \rangle$ denotes averaging over the random couplings $J_{i_1..i_p}$. The SYK model was originally studied in the large $N$, fixed $p$ limit. The Feynman diagrams contributing to correlation functions simplify in this limit leading to simple Schwinger Dyson equations which is then  solved at low energy limit.  Various interesting observations such as the conformality of the two point function, the low energy effective field theory then follow. Motivated by the remarkable features mentioned above, by now, several works have already explored many variants of SYK. These involve generalizations in various directions, such as those involving higher symmetries \cite{PhysRevX.5.041025,Gross:2016kjj,Yoon:2017nig,Narayan2018SupersymmetricSM,Gu2020NotesOT,Bhattacharya2017SYKMC,Liu:2019niv}, supersymmetry \cite{Wenbo,Yoon2017SupersymmetricSM,Peng:2017spg,Narayan2018SupersymmetricSM} etc. Note that most of the calculations are at low energies  - extracting the physics at higher energies is considerably harder since the Schwinger Dyson equation can then be solved only numerically. We also note here that at large $p$, one can make more progress in a $1/p$ expansion \cite{PhysRevD.99.026010,Choi:2019bmd,PhysRevD.94.106002,Gross:2017hcz,Bulycheva:2017uqj,Bhattacharya:2017vaz,Bhattacharya:2018nrw,Lensky:2020fqf,Streicher:2019wek,Das:2017hrt}. 

The double scaling limit \cite{Erds2014PhaseTI,Cotler:2016fpe} is another interesting limit of SYK which provides calculational control at all energies. In the double scaling limit of SYK (DS-SYK), instead of working with a fixed $p$, large $N$ limit, one works in a fixed $\lambda \equiv \frac{p^2}{N}$, large $N$ limit. The DS-SYK  model is solved using combinatorial techniques which reduces the calculation of various quantities of interest (such as partition function and correlation functions) to determination of contributions from certain diagrammatic representations termed {\it chord diagrams} - we will review this in detail in the following sections.  Various quantities (such as two point function\cite{2018JHEPNarayan}, 4 point functions\cite{Berkooz2019TowardsAF}, correlation functions of multitrace operators \cite{Berkooz:2020fvm}) can then be computed exactly in $\lambda$ at all energies (or equivalently at all temperatures) at large $N$.   In the $\lambda \rightarrow 0$, low energy limit (taken appropriately), the DS-SYK results go over to the usual SYK low energy physics i.e the Schwarzian theory- see \cite{Goel:2023svz} for $\lambda$ corrections. In fact for $\lambda \ne 0$, the model can be thought of as a certain $q$-deformation of this limit where $q\equiv e^{-\lambda}$.  For instance, the $sl(2,{\mathbb R})$ symmetry (of the schwarzian) now becomes a q-deformed $sl(2,{\mathbb R})$ symmetry \cite{Berkooz:2022mfk}. DS-SYK has recently been conjectured to be dual to gravity on de-sitter space - see \cite{Susskind:2021esx,Susskind:2022dfz,Lin2022InfiniteTN,Susskind:2022bia,Rahman:2022jsf}.  For other works on DS-SYK see \cite{Jiang:2019pam,Khramtsov:2020bvs,Berkooz2020ComplexSM,Berkooz:2020xne,Goel:2021wim,Bhattacharjee:2022ave,Berkooz:2022fso,Wu:2022tcg,Lin:2022rbf}.

Several variants of the DS-SYK model have already been studied - see \cite{Berkooz:2020xne} for  supersymmetric version and \cite{Berkooz2020ComplexSM} for version with complex fermions. While these variants  boil down to computing chord diagrams, the rewriting of the model in terms of chord diagrams is done on a case by case basis. For example the matrix representation of fermions is an essential ingredient in deriving the chord diagram rules. Consequently one might imagine that the derivation of the diagrammatic rules depend strongly on the details of the model for instance the global symmetries of the model. In the present work, we show that this is not the case by studying DS-SYK models with general \footnote{As we will see later, our techniques are applicable whenever the tensor product of fermion representation and the conjugate representation contains a singlet.} global symmetry and deriving the chord diagram prescription which is independent of the details of the global symmetry. 

The motivation for our work is two fold. One is ofcourse to work out the  double scaling limit of a larger class of SYK models. Given that one of the main motivations for studying the SYK models is to extract lessons about holography, it is useful to have the model worked out in a large number of cases. Another motivation for our work is to try and  understand better the structure of chord diagrams with the view to recast the calculations such that chord diagrams emerge in a unified way regardless of the details of the model. The reason it is useful to do this is because while initially the chord diagrams looked like a convenient combinatorial tool to solve the SYK model in the double scaling limit, recent works indicate that the chord diagrams could be more directly related to the bulk gravitational theory. In particular, \cite{Lin:2022rbf} showed that the Hilbert space of (sliced open) chord diagram is the same as two sided wormhole Hilbert space and also \cite{Berkooz:2022mfk} showed that the chord diagrams can be related to the boundary particle moving in the dual $AdS_2$.
Given such interpretations, it might be worthwhile to explore the structure of chord diagrams in various individual models so that the connections to the dual gravitational picture might be better explored. 

In this work we study the SYK model with global symmetries in the double scaling limit. We give the chord diagram prescription for the computation of moments of the Hamiltonian in the presence of general chemical potentials. We relate this chord diagram contribution to a transfer matrix which acts on an auxiliary Hilbert space defined by slicing the chord diagram. We also give the chord diagram prescription for the computation of moments relevant to two point function.  Our formalism is general and can handle various symmetry groups and representation in a unified manner.  We also compute the partition function in fixed charge sectors. 

The paper is organized as follows. In section \ref{sec:recall} we quickly review results of the double scaling limit of SYK models and the corresponding chord diagram prescription. We also recall the version of DS-SYK with complex fermions. In section \ref{sec:The Model} we present the SYK model with general global symmetry, discuss the double scaling limit and derive the chord diagram prescription. In section(\ref{sec:Transfer matrix}), we derive the transfer matrix by slicing the chord diagram. In section(\ref{two point function}), we derive the chord diagram prescription for moments of two point function. In section(\ref{sec:Special cases}) we give explicit results for the case of classical groups namely $SO(M),U(M)$ and $USp(M/2)$.  In section \ref{sec:Partion function} we compute the partition at fixed charge sectors.
In section \ref{sec:conclusions} we conclude with some comments on future directions. In appendices, we give details of some technical aspects of the calculation.

\section{Review of chord diagrams in Double scaled SYK model}\label{sec:recall}

As mentioned in the introduction, the SYK model in the double scaling limit can be solved using combinatorial techniques which reduces the calculation of various quantities of interest to chord diagrams. In this section we review these results and explain the origin of chord diagrams following \cite{2018JHEPNarayan} (see also \cite{Berkooz:2020xne}).  We will first discuss all this in the simplest setting which is the case of SYK model with Majorana fermions in section \ref{sec:Majorana SYK} . In section \ref{complex} we will give the chord diagram rules for the SYK model with complex fermions and highlight the differences.

\subsection{Majorana SYK}\label{sec:Majorana SYK}
The Hamiltonian for the SYK model with Majorana fermions was given in eqn(\ref{eq.1}) and the fermions satisfy  eqn(\ref{eq.2}). It is convenient to introduce a concise notation for the ordered index sets, namely $I\equiv \{i_1,i_2,...i_p\}$ such that $1\leq i_1 < i_2 ....<i_p\leq N$. Using this notation the Hamiltonian can be rewritten compactly as,
\begin{equation}
    H={\mathrm i}^{\frac{p}{2}}\sum\limits_I J_I X_{I}
\end{equation}
where $X_I=\chi_{i_1}\chi_{i_2}..\chi_{i_p}$\footnote{in an abuse of notation, we will sometimes refer to $X_I$ as fermion - as against product of fermions}. The variance for the disorder in this new notation is given by,
\begin{equation}
    \langle\langle J_I J_J \rangle\rangle = \frac{\mathcal{J}^2}{{ {N \choose p}}} \delta_{IJ}\label{variancemajorana}
\end{equation}
We will work with $\mathcal{J}=1$ units - one can always restore it by dimensional analysis if needed. The double scaling limit is defined to be
\begin{equation}
    \lambda = \frac{p^2}{N} \textrm{\ fixed \qquad  \hspace{10mm} as }\quad  N \rightarrow \infty , p \rightarrow \infty
\end{equation} 
Note that $p$ is even since the Hamiltonian has to  have even number of fermions. 

\paragraph{Emergence of chord diagrams}
Suppose we are interested in computing the partition function $Z \equiv \langle \langle Tr\  e^{-\beta H} \rangle \rangle$. Here $Tr$ denotes the trace over entire Hilbert space normalized such that $Tr({\mathds 1}) = 1$. The partition function can be thought of as a generating function for moments of Hamiltonian $\small{m_k \equiv \langle \langle Tr(H^k) \rangle \rangle}$. Hence essentially $m_k$ and $Z$ contains the same information and we shift focus to computing the moments $m_k$.  The structure of chord diagrams arise in the computation of $m_k$. Let us consider the $k=4$ first to see how the chord diagram structure arises. Averaging over the couplings using eqn(\ref{variancemajorana}),
 \begin{equation}
 \begin{split}\label{eq:m4}
     m_4 &= \sum\limits_{I_1,I_2,I_3,I_4} \langle\langle J_{I_1}J_{I_2}J_{I_3}J_{I_4}\rangle\rangle Tr(X_{I_1}X_{I_2}X_{I_3}X_{I_4})\\
     =& \frac{1}{{N\choose p}^2}\sum\limits_{I_1 I_2} \left\lbrace Tr(X_{I_1}X_{I_1}X_{I_2}X_{I_2}) + Tr(X_{I_1}X_{I_2}X_{I_1}X_{I_2}) + Tr(X_{I_1}X_{I_2}X_{I_2}X_{I_1})\right\rbrace
 \end{split}
 \end{equation}

\begin{figure}[H]
	\begin{center}    
		\begin{tikzpicture}[scale=1,cap=round,>=latex]
		\draw[thick] (0,0) circle(1.5 cm);
		\draw[thick,color=black] (0,-1.5) to[bend right] (0,1.5);
		\draw[thick,color=black] (-1.5,0) to[bend right] (1.5,0);
		\node[black] at (1.85,0) {$X_{I_1}$};
		\node[black] at (-1.85,0) {$X_{I_1}$};
		\node[black] at (0,1.75) {$X_{I_2}$};
		\node[black] at (0,-1.8) {$X_{I_2}$};
		\end{tikzpicture}
		\end{center}
  \caption{Chord diagram corresponding to Tr$(X_{I_1}X_{I_2}X_{I_1}X_{I_2})$} \label{fig:Majorana chord diagram}
\end{figure}
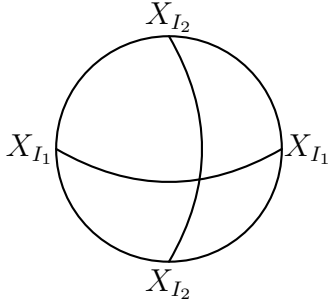

The above relation can be diagrammatically represented using chord diagrams as follows. Noting the cyclic property of trace, we arrange the fermions $X_{I_1},X_{I_2}\dots$ as nodes on a circle. The indices $I_1,I_2\dots$ are now connected via chords representing contraction of indices due to averaging over the coupling constants. Thus a chord diagram with $k$ nodes represents a trace over $k$ fermion products (involving p fermions each) which corresponds to $k$ Hamiltonian insertions. For example, Tr$(X_{I_1}X_{I_2}X_{I_1}X_{I_2})$ which is one of the terms in eq(\ref{eq:m4})  can be represented by the chord diagram given in Figure(\ref{fig:Majorana chord diagram}),
\noindent Thus in terms of chord diagrams $m_4$ is given by,\\

\begin{tikzpicture}
        \node[black] at (-2.75,0) {$m_4 = \frac{1}{ {N \choose p}^2}\sum\limits_{I_1,I_2}$};
        \draw[thick] (0,0) circle(1 cm);
		\draw[thick,color=black] (-1,0) to[bend right] (0,1);
		\draw[thick,color=black] (0,-1) to[bend left] (1,0);
		\node[black] at (1.5,0) {$+$};
		\draw[thick] (3,0) circle(1 cm);
		\draw[thick,color=black] (3,1) to[bend right] (4,0);
		\draw[thick,color=black] (3,-1) to[bend right] (2,0);
		\node[black] at (4.5,0) {$+$};
		\draw[thick] (6,0) circle(1 cm);
		\draw[thick,color=black] (5,0) to[bend right] (7,0);
		\draw[thick,color=black] (6,-1) to[bend right] (6,1);
		  \end{tikzpicture}
\\
The contribution due to each chord diagram needs to be determined. 
 Among the chord diagrams the first two of them have no intersections - the trace structure is just fermion products with same index sets next to each other. Since adjacent fermions with same index gives $X_{I}^2 = (-1)^{\frac{p}{2}}$, each of these chord diagrams contribute just $1$ to the above sum. The analysis of the diagram with intersections corresponding to the trace structure $Tr(X_{I_1}X_{I_2}X_{I_1}X_{I_2})$ is more nontrivial because fermions with same index sets need to be brought next to each other using the anticommutation relations. In other words the chord diagram needs to be `disentangled' via the relations, 
\begin{equation*}
X_IX_J = (-1)^{p^2 - p_{IJ}} X_JX_I 
\end{equation*}
where $p_{IJ}$ is the overlap between the chords I and J. Hence the contribution due to intersection chord diagram is given by $ {N \choose p}^{-2} \sum\limits_{I,J}(-1)^{p_{IJ}}$ which in the double scaling limit evaluates to
\footnote{\begin{equation}
   \textrm{Probability of having J configurations with} |I\cap J|=m =\frac{{N \choose m}{N-m \choose p-m}}{{N \choose p}^2}\xrightarrow{\text{double scaling limit}} \frac{\lambda^{m}}{m!}e^{-\lambda}
\end{equation}},
\begin{equation*}\begin{split}
\frac{1}{ {N \choose p}^2} \sum\limits_{I,J}(-1)^{p_{IJ}} & =\frac{1}{ {N \choose p}^2}\sum\limits_{p_{IJ}=0}^{p} (-1)^{p_{IJ}} \times (\text{number of $J$ configurations with $|I \cap J|  = p_{IJ}$}) \\
& =  \sum\limits_{p_{IJ}=0}^{p}(-1)^{p_{IJ}} \frac{\lambda^{p_{IJ}}}{p_{IJ}!}e^{-\lambda}=e^{-2\lambda}   
\end{split}
\end{equation*}

Thus $m_4$ is given by,
\begin{equation*}
    m_4 = 1 + 1+ e^{-2\lambda}
\end{equation*}

These statements can be generalized appropriately to $k$'th moment also \cite{Erds2014PhaseTI,2018JHEPNarayan} - the double scaling limit is crucial in this. The final prescription to evaluate $m_k$ is as follows,
\begin{itemize}
\item The moments of the Hamiltonian $m_k$ is a sum of the contributions of all possible chord diagrams with $k$ nodes, each denoted by $m_k^{(CD)}$.
\item Foe each chord diagram, $m_k^{(CD)}$ is given by a product of $e^{-2\lambda}$ for every chord intersection\footnote{For instance, the contribution due to every chord diagram without intersection is $1$.}.
\end{itemize}

The $k^{th}$ moment can be now be calculated by determining the numbers corresponding to each type of chord pair and the result is given by\cite{Berkooz2019TowardsAF}, 
\begin{equation}\label{Majorana result}
    m_k = \sum\limits_{CD} e^{-2\kappa_C\lambda} \equiv \sum\limits_{\textrm{CD}}q^{\kappa_C}
\end{equation}
where $\sum\limits_{CD}$ represents sum over chord diagrams, $\kappa_C$ is the number of intersections and we have also defined the parameter $q=e^{-2\lambda}$ which is the weight associated to each intersection. Following \cite{Berkooz2019TowardsAF}, analytic evaluation of the above sum results in,
\begin{equation}\label{Majorana Final}
   m_k = \int\limits_0^\pi \frac{d\theta}{2\pi} (q,e^{\pm 2i\theta};q)_{\infty}\left(\frac{2\cos\theta}{\sqrt{1-q}}\right)^k 
\end{equation}
where $(q,e^{\pm 2i\theta};q)_{\infty}=(q;q)_\infty(e^{ 2i\theta};q)_\infty(e^{-2i\theta};q)_\infty$ is the q-Pochhammer symbol and is defined as, 
\begin{equation}
    (a;q)_{\infty} = \prod\limits_{k=0}^\infty (1-aq^{k}) 
\end{equation}
We will later see that similar statements hold for all the variants of SYK models that we consider in this work. What changes is the structure of chord diagrams and the factors associated with the chord diagram.


\subsection{Complex SYK}\label{complex}
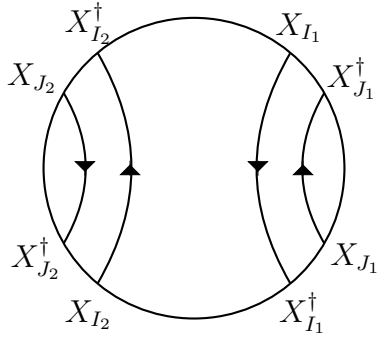
\begin{figure}[H]
\begin{center}
\begin{tikzpicture}
\begin{scope}[thin, every node/.style={sloped,allow upside down}]
\draw[thick] (0,0) circle(2 cm);
\draw[black] (0.85,.1) -- node {\midarrow} (0.85,-.1);
\draw[black] (1.45,-0.1)-- node {\midarrow} (1.45,0.1);
\draw[thick,color=black] (1.28,1.532) to[bend right] (1.28,-1.532);
\draw[thick,color=black] (1.732,1) to[bend right] (1.732,-1);
\draw[thick,color=black] (-1.28,1.532) to[bend left] (-1.28,-1.532);
\draw[thick,color=black] (-1.732,1) to[bend left] (-1.732,-1);
\draw[black](-0.85,-0.1) -- node {\midarrow} (-0.85,0.1);
\draw (-1.45,0.1)-- node {\midarrow} (-1.45,-0.1);
\node[black] at (1.4,1.87) {$X_{I_{1}}$};
\node[black] at (1.45,-1.925) {$X^{\dagger}_{I_1}$};
\node[black] at (2.1,1.2) {$X^{\dagger}_{J_1}$};
\node[black] at (2.15,-1.2) {$X_{J_1}$};
\node[black] at (-1.4,1.932) {$X^{\dagger}_{I_{2}}$};
\node[black] at (-1.4,-1.95) {$X_{I_2}$};
\node[black] at (-2.15,1.2) {$X_{J_2}$};
\node[black] at (-2.1,-1.2) {$X^{\dagger}_{J_2}$};
\end{scope}
\end{tikzpicture}
\end{center}  
\caption{Chord diagram representing $Tr(X^{\dagger}_{I_1}X_{J_1}X^{\dagger}_{J_1}X_{I_1}X^{\dagger}_{I_2}X_{J_2}X^{\dagger}_{J_2}X_{I_2})$ }
\label{Complex chord diagram}
\end{figure}

\begin{figure}[H]
\begin{center}
\begin{tikzpicture}
\begin{scope}[thin, every node/.style={sloped,allow upside down}]
\draw[thick] (0,0) circle(1.5 cm);
\draw[thick,color=black] (1.29,0.75) to[bend right] (1.29,-0.75);
\draw[thick,color=black] (0.96,1.15) to[bend right] (0.96,-1.15);
\draw[thick,color=black] (-1.29,0.75) to[bend left] (-1.29,-0.75);
\draw[thick,color=black] (-0.96,1.15) to[bend left] (-0.96,-1.15);
\draw (-0.62,-0.1)-- node {\midarrow} (-0.61,0.1); 
\draw(-1.07,0.1) -- node {\midarrow} (-1.05,-0.1); 
\draw (0.62,0.1)-- node {\midarrow} (0.61,-0.1); 
\draw (1.07,-0.1)-- node {\midarrow} (1.05,0.1); 
\draw[loosely dotted,thick,color=black] (0.9,1.5) to[bend right] (-0.9,1.5);
\draw[loosely dotted,thick,color=black] (0.9,-1.5) to[bend left] (-0.9,-1.5);
\node[black] at (2,0) {$\equiv$}; 
\draw[thick] (4,0) circle(1.5 cm);
\draw[ultra thick,color=black] (4.96,1.15) to[bend right] (4.96,-1.15);
\draw[ultra thick,color=black] (3.06,1.15) to[bend left] (3.06,-1.15);
\draw[loosely dotted,thick,color=black] (4.96,1.5) to[bend right] (2.96,1.5);
\draw[loosely dotted,thick,color=black] (4.96,-1.5) to[bend left] (2.96,-1.5);
\end{scope}
\end{tikzpicture}
\end{center}
\caption{Chord diagrams with oriented chords represented with H chords}
\label{Oriented to H chord}
\end{figure}
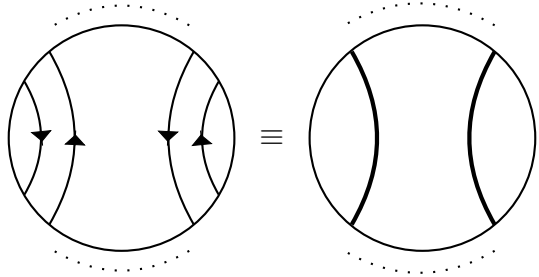

We next present the variant of SYK model with complex fermions. This model was introduced in \cite{PhysRevX.5.041025} and the double scaling limit was discussed in \cite{Berkooz2020ComplexSM}. The model consists of complex fermions $\chi_i$ obeying the following commutation relations,
\begin{equation}
\begin{split}
    \{\chi_i,\chi_j\}&=\{\chi^{\dagger}_i,\chi^{\dagger}_j\}=0\\
    \{\chi_i,\chi^{\dagger}_j\}&=\delta_{ij}
    \end{split}
\end{equation}
Defining $I,J$ as in the previous case, such that $I=\{i_1,i_2,...i_p\}$ with $1\leq i_1<..<i_p\leq N$, the Hamiltonian in terms of index sets  can be written as,
\begin{equation}
    H=\sum\limits_{IJ} J_{IJ} X^{\dagger}_I X_J
\end{equation}
where $J^*_{IJ}=J_{JI}$.
The variance is given by,
\begin{equation}
    \langle \langle J_{IJ}J_{KL}\rangle \rangle = \frac{{\cal J}^2}{ {N \choose p}^2} \delta_{IL}\delta_{JK} \label{variance}
\end{equation}
As in the Majorana case we will work with $\mathcal{J}=1$ units. Since the model has $U(1)$ symmetry, we will also sometimes refer to the model as $U(1)$ symmetric SYK model.  The calculation of moments of Hamiltonian $m_k = \langle \langle Tr H^k \rangle \rangle$ in the double scaling limit i.e $N \rightarrow \infty, \lambda = {p^2 \over N}$ fixed  limit, proceeds as before - $m_k$ has a sum over terms each of which is a specific pairing of $J_{IJ}$. The equivalent chord diagrammatic representation for each term is constructed as follows. Each term now involves a trace with $2k$ fermion products, which can be represented by a circle with $2k$ nodes corresponding to $k$ Hamiltonian insertions (two nodes for each Hamiltonian insertion representing the constituent $X^{\dagger}_I$ and $X_J$). Note that due to disorder averaging only $X$s and $X^\dagger$s can contract unlike the Majorana case. This can be represented by `oriented' chords connecting the corresponding nodes with direction conventionally chosen to be from $X$ to $X^{\dagger}$. For instance the chord diagram representing $Tr(X^{\dagger}_{I_1}X_{J_1}X^{\dagger}_{J_1}X_{I_1}X^{\dagger}_{I_2}X_{J_2}X^{\dagger}_{J_2}X_{I_2})$ is given by Figure (\ref{Complex chord diagram}),\\
Noting that a oriented chords always occur in pairs which can be thought of as a Hamiltonian chord, we can replace a pair of oriented chord by a {\it H chord} denoted by a thickened line - this is shown in Figure (\ref{Oriented to H chord}). The chord diagram corresponding to $m_k$ would now consists of $k/2$ such H chords corresponding to $k$ Hamiltonian insertions.
The prescription to evaluate $m_k$ for the $U(1)$ symmetric SYK model was worked out by \cite{Berkooz2020ComplexSM} which we present below (with some minor rewriting),
\begin{itemize}
    \item The moments of the Hamiltonian $m_k$ is a sum of the contribution of all H chord diagrams with $k$ nodes, each denoted by $m_k^{(CD)}$. 
    \item For a given chord diagram,  $m_k^{(CD)}$ is $e^{-\lambda {k \choose 2}} e^{k\lambda} e^{4\lambda{k/2 \choose 2}}$ times a product of factor corresponding to each possible H chord pair which is computed from the Table below (non intersecting chord pairs give no factor)
   \end{itemize}

    \begin{center}
\begin{tabular}{|l|c|}
    \hline
  {\bf \hspace{20mm} Chord configuration} & {\bf Contribution}\\
  \hline \hline 
 Intersecting chord pair :  \hspace{30mm} \begin{minipage}{3.5cm}
    \begin{center} \begin{tikzpicture}
  \node[black] at (3.25,1) {};
  \draw[ultra thick] (-1,0) arc (0:180:0.75cm);
  \draw[ultra thick] (0.1,0) arc (0:180:0.75cm);
   \node[black] at (3.25,-0.2) {};
  \end{tikzpicture}
 \end{center}
 \end{minipage}     &$e^{-4\lambda}$ \\
  \hline 
\end{tabular}
\end{center}
 
Note that we have phrased the rules here as a factor associated with intersecting chord pairs - this fact will be relevant for the more general version of SYK that we study later. The moments can then be calculated by determining the number corresponding to each chord configuration and is given by,
 \begin{equation}
     m_k = e^{\frac{k}{2}\lambda}\sum\limits_{CD}q_{c}^{\kappa_C}
 \end{equation}
 where $q_c = e^{-4\lambda}$.
Note that the above result is similar to the case with Majorana fermions (eqn(\ref{Majorana result})) except that corresponding to each intersection we now have a factor $q_c$ instead of $q$ and an overall factor of $e^{\frac{k}{2} \lambda}$.
 The model was further studied with a non-zero chemical potential at the double scaling limit and a generalization with U(M) symmetry was also considered in \cite{Berkooz2020ComplexSM}. We will not give the chord diagram rules here, since they will be subsumed under the results given in the next section.

In summary the various variants (Majorana, $U(1)$, $U(M)$) of the DS-SYK described above lead to similar chord diagram prescription. Although we have not made it explicit here, the details of the calculations are quite different - for instance the matrix representation of fermions are used early on in the calculation. SYK models based on more general global symmetries \cite{Yoon:2017nig} are yet to be studied in the double scaling limit. In the present work we study the SYK models with generic global symmetries.  The model and the corresponding the chord diagram prescription are discussed in detail in the following sections.

\section{SYK model with global symmetries and the double scaling limit}\label{sec:The Model}

In this section we present the SYK  model with general global symmetries and discuss it's double scaling limit. 
The model consists of $NM$ fermions given by  $\chi_{i,a}$ where $i \in 1,2,\dots N$ and $a \in 1,2,\dots M$. In what follows, we will be referring the $i$ index as {\it colour} and the $a$ index as {\it flavour}. A (ordered) collection of $p$ colour indices will be denoted by letters $I,J,\dots$ as before and a collection (not necessarily distinct or ordered) of $p$ flavour indices will be denoted by letters $A,B,C,\dots$ (i.e  $A = (a_1,\dots a_p)$). 
The model we will work with has $2p$ fermion interaction terms just like that of $U(1)$ SYK. 
The fermions satisfy the following anticommutation relations for $i \neq j$,
\begin{equation}\label{General Commutations}
    \{\chi^{\dagger}_{i a},\chi_{j b}\}  =0\ \forall i \neq j \hspace{20mm} 
    \{\chi_{i a},\chi_{j b}\} =0 \ \forall i \neq j
\end{equation}
We define fermion products $X_{IA}$ and the corresponding complex conjugate as(with the I and J representing the sets defined above),
\begin{equation}
    X_{IA} \equiv  \chi_{i_1,a_1}\chi_{i_2,a_2}...\chi_{i_p,a_p}, \qquad X^{\dagger}_{IA}=  \chi^{\dagger}_{i_p a_p}...\chi^{\dagger}_{i_2 a_2}\chi^{\dagger}_{i_1 a_1}
\end{equation}
The Hamiltonian for the model is defined to be 
\begin{equation}\label{Generalised SYK Hamiltonian}
    H= \sum_{I,J,A,B} \ J_{IJ} X^{\dagger}_{IA}X_{JB}S^{AB}
\end{equation}
where $S^{AB} \equiv S^{a_1 b_1} \dots S^{a_p b_p}$ and $S^{ab}$ is the invariant tensor\footnote{basically this means that $\sum_{a,b}\chi^\dagger_{ia} \chi_{ib}S^{ab}$ is invariant under the symmetry transformations of the group} for the group under consideration. We will adopt the summation convention for flavour indices for this section - repeated upper and lower flavour indices are assumed to be summed and we drop the explicit sum. For the Hamiltonian to be hermitian, we choose the couplings $J_{IJ}$ such that $J_{IJ}^* = J_{JI}$ and restrict to groups obeying $S^{ab}= \pm {S^{ba}}^*$\footnote{minus signs are allowed since we work with even $p$} . It is clear that any transformation which preserves $\chi^\dagger_{ia} \chi_{jb} S^{ab}$  will be a symmetry of the theory. While this section will be for general symmetries, in the next section we will give explicit results for particular cases of classical group symmetries namely $SO(M),SU(M),USp(M/2)$ for arbitrary $M$. The fermions will then be taken to transform in the fundamental of the group.

The couplings $J_{IJ}$ are random and are picked from an ensemble. We will choose the variance to be,
\begin{equation}\label{Variance of J}
    \langle \langle J_{IJ} J_{KL} \rangle \rangle = \frac{\mathcal{J}^2}{{N \choose p}^2 M^p}\ \delta_{IL}\delta_{JK}
\end{equation}
We will work with $\mathcal{J}=1$ units in the following sections. While other choices of variance are possible, we will choose the simplest one above for now.
We will be working in the double scaled limit. We consider the case with a non zero chemical potential, hence the object of interest is the grand canonical partition function, 
\begin{equation}
    Z = \langle \langle Tr( e^{-\beta H -\sum\limits_{\alpha=1}^r \mu_{\alpha}Q_{\alpha } } )\rangle \rangle \equiv {\cal N}^N \langle \langle Tr_{\Lambda} e^{-\beta H} \rangle \rangle
\end{equation}
where $Tr({\mathds 1}) =1$ and ${\cal N}^N = Tr(e^{-\sum\limits_{\alpha=1}^r \mu_{\alpha}Q_{\alpha }} ) $. $Tr_{\Lambda}$ defined above has the convenient property that $Tr_{\Lambda} ({\mathds 1}) =1$.
$Q_{\alpha}$ above is the conserved cartan charge corresponding to the symmetries of the model, $\mu_{\alpha}$ represents the chemical potentials and $r$ represents the rank.
Hence we will be interested in the quantity\footnote{The moments vanish for odd $k$.}
\begin{equation}
    m_k \equiv {\cal N}^N \langle \langle Tr_{\Lambda}(H^k) \rangle \rangle
\end{equation}

As mentioned in the previous section, various versions of SYK can be seen to be particular cases of the model above.\\
{ $\mathbf{SO(M)}$ \textbf{SYK} : }  Before we proceed to working out the details for the above model, to orient ourselves, we illustrate the model in the particular case of $SO(M)$. The $a$ index of the fermion $\chi_{i,a}$ is a fundamental of $SO(M)$. Since $\sum_a \chi_{ia}\chi_{ja}$ is the invariant of $SO(M)$, the invariant tensor in this case is
    \begin{equation}
        S^{ab} = \delta^{ab}
    \end{equation}
The rank $r$ of $SO(M)$ is  the integer part of $\frac{M}{2}$ and the cartan charges are given by $Q_{\alpha}={ i \over 4} \sum\limits_{i}\left(  \chi_{i 2\alpha-1} \chi_{i 2\alpha} -  \chi_{i2\alpha} \chi_{i 2\alpha-1} \right)$. 
The $SO(M)$ SYK Hamiltonian is then given by,
\begin{equation}    H=\sum\limits_{I,J,A}J_{IJ}X_{IA}X_{JA}
\end{equation}

\subsection{The rise of chord diagrams}
In the rest of this section, we outline the combinatorial technique to evaluate the moments $m_k$. We will closely follow the methods of \cite{Berkooz2020ComplexSM} where it was worked out for the $U(1)$ and $U(M)$ SYK. 

First notice that the moment can be written as a sum over index sets, i,e 
\begin{equation}\label{Moments Initial Formulae}
    \langle \langle Tr_{\Lambda}(H^k) \rangle \rangle = \sum_{\substack{I_1...I_k\\ J_1...J_K}}  \langle \langle J_{I_1 J_1}... J_{I_k J_k} \rangle \rangle \ 
    Tr_{\Lambda}(X^{\dagger}_{I_1 A_1}X_{J_1 B_1}...X^{\dagger}_{I_k A_k}X_{J_k B_k}) S^{A_1 B_1}...S^{A_k B_k}
\end{equation}

\begin{figure}[H]
\begin{center}
\begin{tikzpicture}
\begin{scope}[thin, every node/.style={sloped,allow upside down}]
\draw[thick] (0,0) circle(2 cm);
\filldraw[gray](-1.28557521937,1.53208888624 ) circle(5pt);
\draw[thick,color=black] (0.347,1.969) to[bend right] (0.845,-1.813);
\node[black] at (0.35,2.3) {\small{$J_2G$}};
\node[black] at (0.845,-2.113) {\small{$J_2D$}};
\draw[black] (0.097,1.2)-- node {\midarrow} (0.12,1.3);
\draw[thick,color=black] (-0.347,1.969) to[bend right] (0,-2);
\node[black] at (-0.35,2.25) {\small{$I_2H$}};
\node[black] at (0,-2.3) {\small{$I_2C$}};
\draw[black] (-0.66,1)-- node {\midarrow} (-0.69,0.9);
\draw[thick,color=black] (1.99,0.174) to[bend right] (-1.99,0.174);
\node[black] at (-2.39,0.174) {\small{$I_1A$}};
\node[black] at (2.39,0.174) {\small{$I_1F$}};
\draw[black] (1.22,0.557) -- node {\midarrow} (1.2,0.561);
\draw[thick,color=black] (-1.932,-0.5176) to[bend left] (1.932,-0.5176);
\draw[black] (1.16,-0.145) -- node {\midarrow} (1.18,-0.15);
\node[black] at (-2.432,-0.5176) {\small{$J_1B$}};
\node[black] at (2.432,-0.5176) {\small{$J_1E$}};
\end{scope}
\end{tikzpicture}
\end{center} 
\caption{Chord diagram representing $Tr_{\Lambda}(X^{\dagger}_{I_1A}X_{J_1B}X^{\dagger}_{I_2C}X_{J_2D}X^{\dagger}_{J_1E}X_{I_1F}X^{\dagger}_{J_2G}X_{I_2H})$}
\label{Fig:Chord Diagram Example}
\end{figure}
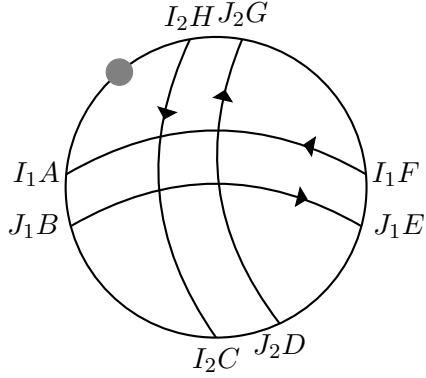

The evaluation of disorder average pairs up the indices I and J. This can be represented by a chord diagram  by arranging the fermions $X^{\dagger}_{I_1 A_1}, X_{J_1 B_1}, \dots$ as nodes on a circle, representing the $\Lambda$ matrix after all nodes (represented by the gray blob in Figure (\ref{Fig:Chord Diagram Example}) ) using the cyclic property of trace and connecting the nodes $I_1,J_1,\dots$ by chords as dictated by the disorder average. This gives rise to the structure of chord diagrams just like the models discussed in the previous sections. Each chord diagram with $2k$ nodes corresponds to $k$ Hamiltonian insertions where each index set appears twice. Also, note that due to the specified averaging, $X$s can only contract with $X^{\dagger}s$(and vice versa), this again results in oriented chord diagrams just like for the $U(1)$ case. Again we choose the convention that the arrows on the chords are  from $X$ to $X^{\dagger}$. As an example we give one of the chord diagram arising in evaluation of $Tr_\Lambda(H^4)$ in Figure (\ref{Fig:Chord Diagram Example}),\\
The evaluation of moments eq(\ref{Moments Initial Formulae}) reduces to the evaluation of the sum over index sets for each possible chord diagrams. More precisely we have,
\begin{equation}
\begin{split}
 m_k & = \sum_{CD} \frac{{\cal N}^N }{M^{\frac{kp}{2}} {N \choose p}^k} \sum_{\substack{I_1..I_{k/2}\\ J_1..J_{k/2}}} S^{A_1 B_1}..S^{A_k B_k}  
 \begin{minipage}{3.5 cm}
 \begin{center}
\begin{tikzpicture}
\begin{scope}[thin, every node/.style={sloped,allow upside down}]
\draw[thick] (0,0) circle(1.75 cm);
\filldraw[gray](-1.34,1.12 ) circle(5pt);
\coordinate (A) at (1.7234,0.303);
\coordinate (A') at (2.1434,0.303);
\coordinate (B) at (0.598,-1.644);
\coordinate (B') at (0.698,-1.944);
\coordinate (C) at (1.5155,0.875);
\coordinate (C') at (1.9655,0.875);
\coordinate (D) at (0.0,-1.75);
\coordinate (D') at (0.0,-2);
\coordinate (E) at (0.303,1.7234); 
\coordinate (E') at (0.394,1.9); 
\coordinate (F) at (1.5155,-0.875); 
\coordinate (F') at (2.0155,-0.875); 
\coordinate (G) at (0.875,1.5155);
\coordinate (G') at (1.085,1.7355);
\coordinate (H) at (1.7234,-0.303);
\coordinate (H') at (2.1634,-0.303);
\coordinate (I) at (-1.5155,0.875); 
\coordinate (I') at (-1.9655,0.875); 
\coordinate (J) at (-1.125,-1.34); 
\coordinate (J') at (-1.175,-1.64); 
\coordinate (K) at (-1.7234,0.303); 
\coordinate (K') at (-2.1234,0.303); 
\coordinate (L) at (-1.5155,-0.875); 
\coordinate (L') at (-1.8955,-0.955); 
\coordinate (I'') at (-1.3788,1.157); 
\coordinate (E'') at (-0.31256,1.7726); 
\draw[thick,color=black] (A) to[bend right] (B);
\node[black] at (A') {\footnotesize{$J_3A_5$}};
\node[black] at (B') {\footnotesize{$J_3B_3$}};
\draw[black](0.76,-0.8) -- node {\midarrow} (0.797,-0.7);
\draw[thick,color=black] (C) to[bend right] (D);
\node[black] at (C') {\footnotesize{$I_3B_5$}};
\node[black] at (D') {\footnotesize{$I_3A_3$}};
\draw[black](0.197,-0.6) -- node {\midarrow} (0.16,-0.7) ;
\draw[thick,color=black] (E) to[bend right] (F);
\node[black] at (E') {\footnotesize{$I_4B_6$}};
\node[black] at (F') {\footnotesize{$I_4A_4$}};
\draw[black](0.36,0.7) -- node {\midarrow} (0.397,0.6);
\draw[thick,color=black] (G) to[bend right] (H);
\node[black] at (G') {\footnotesize{$J_4A_6$}};
\node[black] at (H') {\footnotesize{$J_4B_4$}};
\draw[black](0.96,0.65)-- node {\midarrow} (0.923,0.75) ;
\draw[thick,color=black] (I) to[bend left] (J);
\node[black] at (I') {\footnotesize{$I_1A_1$}};
\node[black] at (J') {\footnotesize{$I_1B_2$}};
\draw[black](-1,-0.2)-- node {\midarrow}  (-1.025,-0.1);
\draw[thick,color=black] (K) to[bend left] (L);
\node[black] at (K') {\footnotesize{$J_1B_1$}};
\node[black] at (L') {\footnotesize{$J_1A_2$}};
\draw[black](-1.46,-0.15)-- node {\midarrow} (-1.4517,-0.25) ;
\draw[loosely dotted,thick,color=black] (-1.2,1.5321) arc (125:90:2);
\end{scope}
\end{tikzpicture}
\end{center}
\end{minipage}
      \\
     & \equiv \sum_{CD} \ m_k^{(CD)}  
\end{split}
\end{equation}

Where $m_k^{(CD)}$ represents the contribution due to each chord diagram. We will now focus on evaluating the contribution to the moment from a given chord diagram $m_k^{(CD)}$. The evaluation of $m_k^{(CD)}$ proceeds in steps. 
\begin{itemize}
    \item[1.] First we re-order the fermions around  within the trace - using their anticommutation relations - such that fermions with the same colour index are next to each other. One can now split the trace as a product of trace for each colour (over the flavour space). These steps are done in section \ref{subsubsec:signs} and the final result is given in eq (\ref{minus signs}). Note that this 
    since this step involves interchanging the ordering of only those fermions with different colour, this result is independent of the symmetries of the model and furthermore is an exact statement independent of double scaling limit.
    \item[2.] In the next step, we perform the colour index sums. We convert the colour index sums into a weighted sum over overlaps of index sets. The double scaling limit is essential here, since it determines the weight with which a particular overlap appears in this sum. 
    These steps are done in section \ref{subsec:index sums}. We note here that the final result is in terms of trace over fermions in the flavour space. 
\end{itemize}
We now proceed to explain the steps outlined above in detail. The reader not interested in details of the calculations can directly look at the results in section (\ref{Final summary}).

\subsubsection{Factorizing into color traces} 
\label{subsubsec:signs}
In the first step, as mentioned before, we bring the fermions of same colour together. To do this we {\it disentangle} the chords in a sense we describe below. In what follows we will find it convenient to work with arbitrary products of $\chi$s and $\chi^\dagger$s given by, 
\begin{equation}
\Phi^S_{IA}\equiv  \chi^{s_1}_{i_1 a_1} \chi^{s_2}_{i_1 a_1} \cdots
\end{equation}
where $S$ denotes the set  $(s_1,s_2,\dots  s_p)$ where $s_i \in (+,-)$ such that $\chi^-_{ia} \equiv \chi_{ia}$ and  $\chi^+_{ia} = \chi^\dagger_{ia}$ - more generally such sets will be denoted by the letters $R,S \dots$. 
For example in this notation $\Phi^{(-,-,\dots,-)}_{IA}$ represents $X_{IA}$. To get some intuition, we consider the case of zero overlap between color index sets. It is easy to see that 
\begin{equation}
    \Phi^{R}_{IA} \Phi^{S}_{JB} = (-1)^{p_I p_J} \Phi_{JB}^{S} \Phi_{IA}^{R}, \hspace{10mm} \mbox{if } |I \cap J | = 0 
\end{equation}
The appropriate generalization of this result to the case when $p_{IJ} \equiv | I \cap J |  \ne 0$ turns out to be (see Appendix(\ref{Disentangling chord diagrams}) for details)
\begin{equation}\label{Unwrap}
    \Phi^{R}_{IA} \Phi^{S}_{JB} = (-1)^{ p_I p_J -p_{IJ}^2 }\    \Phi^{\tilde S}_{J\tilde B} \Phi^{\tilde R}_{I\tilde A} 
\end{equation}
The notation on RHS requires some explaining  : the sets $\tilde A, \tilde B$ are almost $A$ and $B$ respectively, except for the flavor partners of the common color indices, in which case the flavour indices are swapped. Similar statements hold for $R,S$ and $\tilde R,\tilde S$. To illustrate this, consider fermion products with following index sets $I=\{i_1,i_2,i_3,i_4\},  J=\{j_1,j_2,j_3,j_4 \}, A=\{a_1,a_2,a_3,a_4\},B=\{b_1,b_2,b_3,b_4\},R=\{r_1,r_2,r_3,r_4\},S=\{s_1,s_2,s_3,s_4\}$. Let $i_2 = j_1$ and $i_3 = j_3$, swapping the fermion products across each other results in the following configuration,
\begin{equation}
  \Phi^{R}_{IA} \Phi^{S}_{JB} = (-1)^{ p_I p_J - 4} \phi_{j_1 a_2}^{r_2} \phi_{j_2 b_2}^{s_2} \phi_{j_3 a_3}^{r_3} \phi_{j_4 b_4}^{s_4} \quad \phi_{i_1 a_1}^{r_1} \phi_{i_2 b_1}^{s_1} \phi_{i_3 b_3}^{s_3}\phi_{i_4 a_4}^{r_4}
\end{equation}
and hence $\tilde A=\{ a_1,b_1,b_3,a_4\}$, $\tilde B=\{a_2,b_2,a_3,b_4 \}$, $\tilde R=\{ r_1,s_1,s_3,r_4\}$, $\tilde S=\{r_2,s_2,r_3,s_4 \}$.

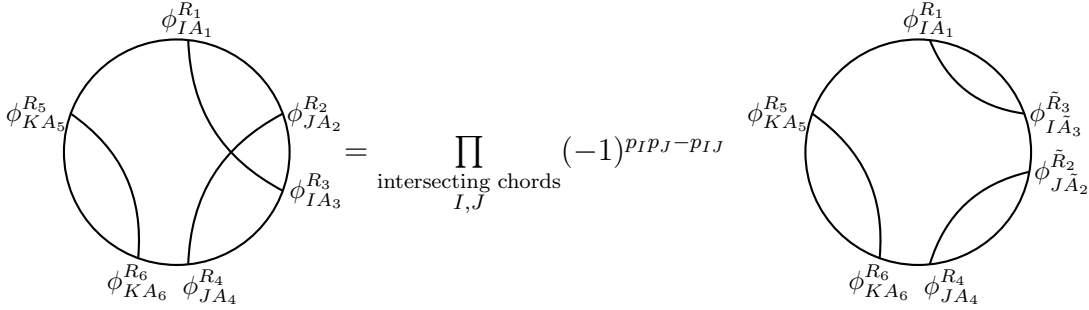
\begin{figure}[H]
 \begin{minipage}[c]{4.5cm}
    \begin{center}
\begin{tikzpicture}
\draw[thick] (0,0) circle(1.5 cm);
\coordinate (A) at (0.1525225498,1.49429204714); 
\coordinate (A') at (0.1525225498,1.76429204714);
\coordinate (B) at (1.40953893118,-0.51303021498); 
\coordinate (B') at (1.85953893118,-0.51303021498);
\coordinate (C) at (1.40953893118,0.51303021498); 
\coordinate (C') at (1.83953893118,0.51303021498); 
\coordinate (D) at (0.1525225498,-1.49429204714 );
\coordinate (D') at (0.4525225498,-1.79429204714 );
\coordinate (E) at (-1.40953893118,0.51303021498); 
\coordinate (E') at (-1.85953893118,0.51303021498); 
\coordinate (F) at (-0.51303021498 ,-1.40953893118 ); 
\coordinate (F') at (-0.51303021498 ,-1.72953893118); 
\draw[thick,color=black] (A) to[bend right] (B);
\node[black] at (A') {\footnotesize{$\phi_{IA_1}^{R_1}$}};
\node[black] at (B') {\footnotesize{$\phi_{IA_3}^{R_3}$}};
\draw[thick,color=black] (C) to[bend right] (D);
\node[black] at (C') {\footnotesize{$\phi_{JA_2}^{R_2}$}};
\node[black] at (D') {\footnotesize{$\phi_{JA_4}^{R_4}$}};
\draw[thick,color=black] (E) to[bend left] (F);
\node[black] at (E') {\footnotesize{$\phi_{KA_5}^{R_5}$}};
\node[black] at (F') {\footnotesize{$\phi_{KA_6}^{R_6}$}};
\end{tikzpicture}
\end{center}
\end{minipage}
$= \prod\limits_{\substack{\text{intersecting chords} \\ I,J} }(-1)^{p_I p_J - p_{IJ}}$
\begin{minipage}{4cm}
    \begin{center}
\begin{tikzpicture}
\draw[thick] (0,0) circle(1.5 cm);
\coordinate (A) at (0.1525225498,1.49429204714); 
\coordinate (A') at (0.1525225498,1.76429204714);
\coordinate (C) at (1.47721162952,-0.2604722665); 
\coordinate (C') at (1.91721162952,-0.2604722665);
\coordinate (B) at (1.40953893118,0.51303021498); 
\coordinate (B') at (1.83953893118,0.51303021498); 
\coordinate (D) at (0.1525225498,-1.49429204714 );
\coordinate (D') at (0.4525225498,-1.79429204714 );
\coordinate (E) at (-1.40953893118,0.51303021498); 
\coordinate (E') at (-1.85953893118,0.51303021498); 
\coordinate (F) at (-0.51303021498 ,-1.40953893118 ); 
\coordinate (F') at (-0.51303021498 ,-1.72953893118); 
\draw[thick,color=black] (A) to[bend right] (B);
\node[black] at (A') {\footnotesize{$\phi_{IA_1}^{R_1}$}};
\node[black] at (B') {\footnotesize{$\phi_{I\tilde{A}_3}^{\tilde{R}_3}$}};
\draw[thick,color=black] (C) to[bend right] (D);
\node[black] at (C') {\footnotesize{$\phi_{J\tilde{A}_2}^{\tilde{R}_2}$}};
\node[black] at (D') {\footnotesize{$\phi_{JA_4}^{R_4}$}};
\draw[thick,color=black] (E) to[bend left] (F);
\node[black] at (E') {\footnotesize{$\phi_{KA_5}^{R_5}$}};
\node[black] at (F') {\footnotesize{$\phi_{KA_6}^{R_6}$}};
\end{tikzpicture}
\end{center}
\end{minipage}
\caption{Disentangling of a Chord diagram}
\label{Disentangling diagram}
\end{figure}

One can think of eq(\ref{Unwrap}) as a way to {\it disentangle} the chord diagram - each chord intersection giving the factor $(-1)^{ p_I p_J -p_{IJ}^2 }$. One can keep doing this until the chord diagram is completely disentangled, for instance see Figure(\ref{Disentangling diagram}), \\
\noindent where we have replaced $(-1)^{p_{IJ}^2}$ by $(-1)^{p_{IJ}}$ since they are the same for integer $p_{IJ}$. Note that the index $\tilde A$ and $\tilde R$ on the colour index will depend on the details of the disentangling procedure. But the final result, as we will see shortly, does not depend on these details. The main point now is that in the disentangled chord diagram, the fermions with the same colour can be brought together with no extra cost of sign. This is true since for each chord $I$, we have 
\begin{equation}
 Tr_{\Lambda}( \cdots \chi^{r_1}_{i_1a_1}\dots \chi^{r_p}_{i_pa_p}\  \chi^{s_1}_{i_1b_1}\dots \chi^{s_p}_{i_pb_p} \cdots   ) =   Tr_{\Lambda}( \cdots \chi^{r_1}_{i_1a_1}\chi^{s_1}_{i_1b_1}\dots \chi^{r_p}_{i_pa_p}\chi^{s_p}_{i_pb_p} \cdots   ) 
\end{equation}
This pairs up the fermions of same colour together - the pair can be moved around so that fermions of same colour are together with no extra cost of sign. At this point, the trace $Tr_{\Lambda}(..)$  factorizes to trace over each colour index $i$  denoted by $tr_{\Lambda_i}(..)$ where $tr_{\Lambda_i}(..) \equiv {\cal N} tr_i(... e^{-\sum\limits_{\alpha}^{r}\mu_\alpha Q_{i\alpha}})$\footnote{Here $tr_i({\mathds 1})=1$. $tr_{\Lambda_i}$ has the convenient property that $tr_{\Lambda_i}({\mathds 1})=1$. An alternative expression for $\cal N$ in terms of $tr_{\Lambda_i}$ is 
\begin{equation}\label{N definition}
    {\cal N} = tr_i(e^{-\sum\limits_{\alpha}^{r}\mu_\alpha Q_{i\alpha}})
\end{equation}} , $Q_{i\alpha}$ represents the i'th component of conserved charge. Note that the ordering of fermions with the same colour (say $i$) inside the trace ($tr_i$) is inherited from the original (entangled) chord diagram. In fact once we read off the signs by using the procedure of disentangling (see for instance Figure \ref{Disentangling diagram}), we can as well revert to the original chord diagram\footnote{An alternate way of thinking about this is as follows : Once the factorization over each color happens, we might as well work with factorized Hilbert space - i.e define new fermions $\psi,\psi^\dagger$ which satisfy $[ \psi_{ia} , \psi_{jb} ] = 0 , \mbox{ if } i \ne j$ and for a given $i$ $\psi_{ia}$ satisfies the same algebra as $\chi_{ia}$ i.e $\{ \psi_{ia} , \psi_{ib}  \} = \{ \chi_{ia}, \chi_{ib} \}$ etc. With these new $\psi$ fermions, we can undo the procedure of disentangling and go back to the original chord diagram. Note that when we re entangle the diagram, we do not get minus signs anymore since the process involves only reordering $\psi$ fermions with different colours, which always anticommute. Hence we would land up on same eq(\ref{minus signs}), with the understanding that the chord diagrams on the RHS there are evaluated with $\psi$ fermions rather than $\chi$ fermions. For simplicity, we take the point of view advocated in the main text}- the only difference is that the chord diagram now represents a product over trace for each colour $tr_i$ rather than $Tr$ over the full Hilbert space. Recalling that the circle of the chord diagram represented $Tr$, we now denote product over trace of individual colours $tr_i$ by a dotted line. In this new notation, we have the following relation for instance,\\

\begin{minipage}[c]{4.5cm}
\begin{center}
\begin{tikzpicture}
\hspace{-10mm}
\begin{scope}[thin, every node/.style={sloped,allow upside down}]
\draw[thick] (0,0) circle(1.5 cm);
\filldraw[gray](-1.06066017178,1.06066017178 ) circle(5pt);
\coordinate (A) at (0.1525225498,1.49429204714); 
\coordinate (A') at (0.1525225498,1.76429204714);
\coordinate (B) at (1.40953893118,-0.51303021498); 
\coordinate (B') at (1.85953893118,-0.51303021498);
\coordinate (C) at (1.40953893118,0.51303021498); 
\coordinate (C') at (1.83953893118,0.51303021498); 
\coordinate (D) at (0.1525225498,-1.49429204714 );
\coordinate (D') at (0.4525225498,-1.79429204714 );
\coordinate (E) at (-1.40953893118,0.51303021498); 
\coordinate (E') at (-1.85953893118,0.51303021498); 
\coordinate (F) at (-0.51303021498 ,-1.40953893118 ); 
\coordinate (F') at (-0.51303021498 ,-1.72953893118); 
\draw[thick,color=black] (A) to[bend right] (B);
\node[black] at (A') {\footnotesize{$X_{KF}$}};
\node[black] at (B') {\footnotesize{$X_{KD}^{\dagger}$}};
\draw[black] (0.35,0.6) -- node {\midarrow}(0.39,0.5);
\draw[thick,color=black] (C) to[bend right] (D);
\node[black] at (C') {\footnotesize{$X_{JE}$}};
\node[black] at (D') {\footnotesize{$X^{\dagger}_{JC}$}};
\draw[black](0.39,-0.5) -- node {\midarrow} (0.35,-0.6);
\draw[thick,color=black] (E) to[bend left] (F);
\node[black] at (E') {\footnotesize{$X^{\dagger}_{IA}$}};
\node[black] at (F') {\footnotesize{$X_{IB}$}};
\draw[black](-0.83,-0.1) -- node {\midarrow} (-0.88,0);
\end{scope}
\end{tikzpicture}
\end{center}
\end{minipage}
 $= \left(  \prod\limits_{\mbox{intersecting chords } I,J }(-1)^{p_I p_J - p_{IJ}}\right)$ 
\begin{minipage}[c]{4.5cm}
\begin{equation}\label{minus signs}
\begin{tikzpicture}
\begin{scope}[thin, every node/.style={sloped,allow upside down}]
\draw[thick,densely dashed] (0,0) circle(1.5 cm);
\filldraw[gray](-1.06066017178,1.06066017178 ) circle(5pt);
\coordinate (A) at (0.1525225498,1.49429204714); 
\coordinate (A') at (0.1525225498,1.76429204714);
\coordinate (B) at (1.40953893118,-0.51303021498); 
\coordinate (B') at (1.85953893118,-0.51303021498);
\coordinate (C) at (1.40953893118,0.51303021498); 
\coordinate (C') at (1.83953893118,0.51303021498); 
\coordinate (D) at (0.1525225498,-1.49429204714 );
\coordinate (D') at (0.4525225498,-1.79429204714 );
\coordinate (E) at (-1.40953893118,0.51303021498); 
\coordinate (E') at (-1.85953893118,0.51303021498); 
\coordinate (F) at (-0.51303021498 ,-1.40953893118 ); 
\coordinate (F') at (-0.51303021498 ,-1.72953893118); 
\draw[thick,color=black] (A) to[bend right] (B);
\node[black] at (A') {\footnotesize{$X_{KF}$}};
\node[black] at (B') {\footnotesize{$X_{KD}^{\dagger}$}};
\draw[black] (0.35,0.6)-- node {\midarrow} (0.39,0.5);
\draw[thick,color=black] (C) to[bend right] (D);
\node[black] at (C') {\footnotesize{$X_{JE}$}};
\node[black] at (D') {\footnotesize{$X^{\dagger}_{JC}$}};
\draw[black] (0.39,-0.5)-- node {\midarrow} (0.35,-0.6);
\draw[thick,color=black] (E) to[bend left] (F);
\node[black] at (E') {\footnotesize{$X_{IA}^{\dagger}$}};
\node[black] at (F') {\footnotesize{$X_{IB}$}};
\draw[black] (-0.83,-0.1)-- node {\midarrow} (-0.88,0);
\end{scope}
\end{tikzpicture}
\end{equation}
\end{minipage} 
Since the trace has now factorised over each color, we can suppress the colour indices when evaluating trace of a given colour (for eg., we will now denote $tr_{\Lambda_i}(\chi_{ia}\chi^{\dagger}_{ib}..)$ by $tr_{\Lambda}(\chi_{a}\chi^{\dagger}_{b}..)$). 

\subsection{Evaluating colour traces}\label{subsec:index sums}

Now that we have factorized the colour index traces and the resulting minus signs, we will proceed to evaluate the colour index sums.  We now have
\begin{equation}\label{eq:Index sum}
    m_k^{(CD)} = \frac{ {\cal N}^N}{{N \choose p}^k M^{kp \over 2}} \sum_{\substack{I_1..I_{k/2}\\ J_1..J_{k/2}}} S^{A_1 B_1}..S^{A_k B_k} \left(\prod_{\substack{\text{intersecting} \\ \text{chords} I,J} }s(p_{IJ})\right)
 \begin{minipage}{4 cm}
  \begin{center}
\begin{tikzpicture}
\begin{scope}[thin, every node/.style={sloped,allow upside down}]
\draw[thick,densely dashed] (0,0) circle(1.75 cm);
\filldraw[gray](-1.34057777546 ,1.12487831695 ) circle(4pt);

\coordinate (A) at (1.7234,0.303); 
\coordinate (A') at (2.1434,0.303);
\coordinate (B) at (0.598,-1.644); 
\coordinate (B') at (0.698,-1.944);
\coordinate (C) at (1.5155,0.875); 
\coordinate (C') at (1.9655,0.875); 
\coordinate (D) at (0.0,-1.75); 
\coordinate (D') at (0.0,-2); 
\coordinate (E) at (0.304,1.723); 
\coordinate (E') at (0.394,1.9); 
\coordinate (F) at (1.5155,-0.875); 
\coordinate (F') at (2.0155,-0.875); 
\coordinate (G) at (0.875,1.5155);
\coordinate (G') at (1.085,1.7355);
\coordinate (H) at (1.7234,-0.303);
\coordinate (H') at (2.1634,-0.303);
\coordinate (I) at (-1.5155,0.875); 
\coordinate (I') at (-1.9655,0.875); 
\coordinate (J) at (-1.125,-1.34); 
\coordinate (J') at (-1.175,-1.64); 
\coordinate (K) at (-1.7234,0.303); 
\coordinate (K') at (-2.1234,0.303); 
\coordinate (L) at (-1.5155,-0.875); 
\coordinate (L') at (-1.8955,-0.955); 
\coordinate (I'') at (-1.3788,1.157); 
\coordinate (E'') at (-0.31256,1.7726); 
\draw[thick,color=black] (A) to[bend right] (B);
\node[black] at (A') {\footnotesize{$J_3A_5$}};
\node[black] at (B') {\footnotesize{$J_3B_3$}};
\draw[black](0.76,-0.8) -- node {\midarrow} (0.797,-0.7);
\draw[thick,color=black] (C) to[bend right] (D);
\node[black] at (C') {\footnotesize{$I_3B_5$}};
\node[black] at (D') {\footnotesize{$I_3A_3$}};
\draw[black](0.197,-0.6) -- node {\midarrow} (0.16,-0.7);
\draw[thick,color=black] (E) to[bend right] (F);
\node[black] at (E') {\footnotesize{$I_4B_6$}};
\node[black] at (F') {\footnotesize{$I_4A_4$}};
\draw[black](0.36,0.7) -- node {\midarrow} (0.397,0.6);
\draw[thick,color=black] (G) to[bend right] (H);
\node[black] at (G') {\footnotesize{$J_4A_6$}};
\node[black] at (H') {\footnotesize{$J_4B_4$}};
\draw[black]  (0.96,0.65)-- node {\midarrow}(0.923,0.75);
\draw[thick,color=black] (I) to[bend left] (J);
\node[black] at (I') {\footnotesize{$I_1A_1$}};
\node[black] at (J') {\footnotesize{$I_1B_2$}};
\draw[black] (-1,-0.2)-- node {\midarrow} (-1.025,-0.1);
\draw[thick,color=black] (K) to[bend left] (L);
\node[black] at (K') {\footnotesize{$J_1B_1$}};
\node[black] at (L') {\footnotesize{$J_1A_2$}};
\draw[black] (-1.46,-0.15)-- node {\midarrow} (-1.4517,-0.25);
\draw[loosely dotted,thick,color=black] (-1.4,1.35) arc (135:90:1.8);
\end{scope}
\end{tikzpicture}
\end{center}
\end{minipage}
\end{equation}
where $s(p_{IJ}) = (-1)^{p_I p_J - p_{IJ}}$.

Recall that the chord diagram appearing above represents the (product over) trace for each colour. As we will see later, in the double scaling limit, the chord diagram depends only on the overlaps between index sets $I_1,\dots J_1, \dots$ (rather than the details of where the overlaps occur, for example the position of the common index etc). Thus the index sum can be effectively written as a sum over overlaps weighed by   the probability of the occurrence of the corresponding overlap (due to factor ${1 \over {N \choose p}^k}$). 

Before proceeding, we introduce some terminology : Note that each H node has chords of opposite orientation emanating from it and the associated fermions contracted by invariant tensor - for instance the node $X^\dagger_{I_1A_1} X_{J_1B_1} S^{A_1B_1}$ in eq (\ref{eq:Index sum}). We term the pair of oriented chords as \textbf{H chords} and the fermions with  contracted flavour indices($\chi_{i_1a_1}$ and $\chi_{j_1b_1}, \chi_{i_2a_2}$ and $\chi_{j_2b_2}\dots$  if the invariant tensors are given by $ S_{a_1 b_1}, S_{a_2 b_2}\dots$) are termed as \textbf{partner fermions}. 

We also assemble below some relevant facts about overlaps of index sets $I,J,\dots$ of length $p_I,p_J,\dots$\footnote{till now all index sets $I,J,\dots$ were of equal length $p_I = p_J=\dots = p$. The case when $p_I \ne p_J$ becomes relevant when we compute the moment of two point function in section \ref{two point function}} in the double scaling limit, $N \rightarrow \infty, p_I \rightarrow  \infty,p_J \rightarrow \infty$ with $ \lambda_{IJ} \equiv {p_I p_J  \over N}$ fixed limit \cite{Erds2014PhaseTI, Cotler:2016fpe,2018JHEPNarayan} will be useful in what follows,
\begin{lemma} \label{lma1}  Any three index sets $I, J,K$ have vanishing overlap, i.e $\mbox{Prob}_{|I \cap J \cap K| \ne 0} = 0$. \end{lemma}

\begin{lemma} \label{lma2}   The probability of overlap between any two index sets are independently distributed, i.e 
    \begin{equation}\label{independent}        
    \mbox{Prob}_{|I\cap J|=p_{IJ}, \forall \mbox{ chord pairs } I,J } = \prod_{\mbox{ chord pairs } I,J}  \mbox{Prob}_{|I \cap J| = p_{IJ}}
    \end{equation}
    Moreover, the probability of having $p_{IJ}$ indices common between two index sets $I$ and $J$ is Poisson distributed, i.e  
    \begin{equation}\label{DoubleScalingProbability}
        \mbox{Prob}_{|I \cap J| = p_{IJ}} = { \lambda_{IJ}^{p_{IJ}} \over p_{IJ} ! } e^{-\lambda_{IJ}}
    \end{equation} 
\end{lemma}

\begin{lemma}\label{lma3} Suppose a fermion with a given colour appears in two index sets in a chord diagram, then the color of it's partner fermion never occurs elsewhere in the chord diagram. 

\end{lemma}
Since Result (\ref{lma3}) is not explicitly mentioned in other works we will briefly explain it here. Let $\chi_{i_m,a_m} \chi^\dagger_{j_m,b_m}$ be partner fermions accompanied by a factor of $S^{a_m b_m}$. Suppose index $i_m$ appears twice in the chord diagram. In the double scaling limit the probability of the partner index $j_m$ occurring in some index set is ${ \binom{N-1}{p-1} \over \binom{N}{p} } $ which vanishes as ${1 \over \sqrt N}$ in the double scaling limit.

\paragraph{The case of no overlaps among index sets :} Before we turn to a more systematic analysis of eq(\ref{eq:Index sum}), let us consider the simplest term in eq(\ref{eq:Index sum}), namely the configuration  in which there are no overlaps /  common index among any of the index sets $I_1,\dots I_{k\over 2}, J_1, \dots J_{k \over 2}$. Since each fermion product comes with flavour indices contracted with another fermion product via the invariant tensor $S^{AB}$ (the pair of fermion products constituting each term in the Hamiltonian, $X^\dagger_{IA} X_{JB}S^{AB}$), oriented chords cannot be considered independently. The chord diagram with no index overlaps between any chord then factors into a trace over a pair of oriented chords i.e., each H-chord. The contribution from a fermion and its partner in an H chord is hence given by $x^2M$ with
\begin{equation}\label{eq:No overlap contribution}
 x^2 M  \equiv  S^{ab} S^{cd} tr_{\Lambda_i}\left( \chi^\dagger_{ia} \chi_{id} \right) tr_{\Lambda_j} \left( \chi_{jb} \chi^\dagger_{jc}  \right)
\end{equation}
Since the contributions now involves a trace over each of the color indices, we may as well drop the color index which we will do from now on. Now, each index set consists of $p$ fermions each, and hence the contribution from each H chord would be $(x^2 M)^p$. Since there are ${k \over 2}$ H chords, the net factor then is $x^{p k} M^{k p \over 2}$. Denoting the probability of such configuration when there is no overlap among any of the chord pairs by $\mbox{Prob}_{|I\cap J|=0, \forall \mbox{ chord pairs } I,J }$, we see that the net contribution to the sum $m_k^{(CD)}$ given in eq(\ref{eq:Index sum}) from such configurations is 
\begin{equation}
    \mbox{Prob}_{|I\cap J|=0, \forall \mbox{ chord pairs } I,J } \ \times \  x^{pk}
\end{equation}
 In the double scaling limit, due to Result \ref{lma1}, the probability above factorizes for each chord pair and which is $e^{-\lambda}$ (see eq(\ref{DoubleScalingProbability})). Hence the net contribution  is given by,
 \begin{equation}\label{eq:Non Overlap Contribution}
    e^{-\lambda {k \choose 2}} x^{pk}
\end{equation}
 
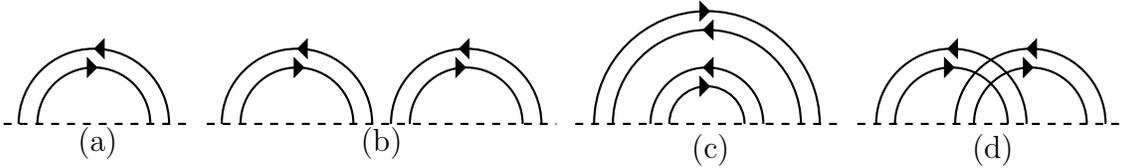
\begin{figure}[H] 
\begin{center}
 \begin{tikzpicture}
\begin{scope}[thin, every node/.style={sloped,allow upside down}]
\draw[thick,black] (-4.2,0) arc (180:0:0.75);
\draw[thick,black] (-4.45,0) arc (180:0:1);
\draw[thick,dashed] (-4.65,0) -- (-2.2,0);
\draw[black] (-3.35,1)-- node {\midarrow} (-3.45,1);
\draw[black] (-3.55,0.75)-- node {\midarrow} (-3.35,0.75);
\node[black] at (-3.4,-0.25) {(a)};
\draw[thick,dashed] (-1.95,0) -- (2.7,0);
\draw[thick,black] (-1.5,0) arc (180:0:0.75);
\draw[thick,black] (-1.75,0) arc (180:0:1);
\draw[thick,black] (0.75,0) arc (180:0:0.75);
\draw[thick,black] (0.5,0) arc (180:0:1);
\draw[black] (-0.65,1)-- node {\midarrow} (-0.75,1);
\draw[black] (-0.75,0.75)-- node {\midarrow} (-0.65,0.75);
\draw[black] (1.5,1)-- node {\midarrow} (1.4,1);
\draw[black] (1.4,0.75)-- node {\midarrow} (1.5,0.75);
\node[black] at (0.375,-0.25) {(b)};
\draw[thick,black,dashed] (2.95,0) -- (6.45,0);
\node[black] at (4.75,-0.35) {(c)};
\draw[thick,black] (3.95,0) arc (180:0:0.75);
\draw[thick,black] (4.2,0) arc (180:0:0.5);
\draw[black] (4.7,0.75)-- node {\midarrow} (4.69,0.75);
\draw[black] (4.69,0.5)-- node {\midarrow} (4.7,0.5);
\draw[thick,black] (3.45,0) arc (180:0:1.25);
\draw[thick,black] (3.2,0) arc (180:0:1.5);
\draw[black] (4.7,1.25)-- node {\midarrow} (4.69,1.25);
\draw[black] (4.69,1.5)-- node {\midarrow} (4.7,1.5);
\draw[thick,black,dashed] (6.7,0) -- (10.25,0);
\node[black] at (8.5,-0.35) {(d)};
\draw[thick,black] (6.95,0) arc (180:0:1);
\draw[thick,black] (8,0) arc (180:0:1);
\draw[thick,black] (7.2,0) arc (180:0:0.75);
\draw[thick,black] (8.25,0) arc (180:0:0.75);
\draw[black] (9,1)-- node {\midarrow} (8.99,1);
\draw[black] (8.99,0.75)-- node {\midarrow} (9,0.75);
\draw[black] (7.95,1)-- node {\midarrow} (7.94,1);
\draw[black] (7.94,0.75)-- node {\midarrow} (7.95,0.75);
\end{scope}
\end{tikzpicture}
\end{center}
 \caption{All pairs}
 \label{all pairs}
\end{figure}

\paragraph{The case of nontrivial overlaps among index sets:}
We now move to the nontrivial case when there are overlaps between various index sets. Since the probabilities of overlaps between any two of the index sets are independently distributed (eq(\ref{independent})), one might naively think that the sum in eq(\ref{eq:Index sum}) factors into a product of all possible pairs of chords. However, the flavour indices spoils the factorization due to the contraction of indices with partner fermions via invariant tensors. For instance, let $\chi_{i_m,a_m} \chi^\dagger_{j_m,b_m}$ be the partner fermions and are thus accompanied by a factor of $S^{a_m b_m}$. Suppose index $i_m$ appears twice in the chord diagram, the factorization is then spoiled if the partner colour index $j_m$ appears elsewhere due to the non-contracted $b_m$ index in $S^{a_m b_m}$. However, in the double scaling limit the partner fermion cannot appear elsewhere (Result \ref{lma3}). As a result, to preserve factorization it is sufficient to include the contribution of the partner fermion (such that no flavour indices remain uncontracted) along with the factors corresponding to overlapping indices. So the contribution from each chord diagram factorizes if we keep track of the $H$ chords in which the pairs of oriented chords appear. Hence the following structure emerges for  eq(\ref{eq:Index sum})
\begin{equation}\label{eq:probability sum}
 \begin{split}
     m_k^{(CD)} &=  {\cal N}^N e^{-\lambda {k \choose 2}}  x^{pk} \ 
          \hspace{30mm}  \\
     & \qquad \times 
\prod_{\mbox{chords } I,J}\left(   \sum_{ p_{IJ} = 0}^\infty 
     \mbox{Prob}_{I \cap J = p_{IJ}}\times  s(p_{IJ})
\times  \begin{minipage}{4 cm}
\begin{tikzpicture}[scale=0.8]
\begin{scope}[thin, every node/.style={sloped,allow upside down}]
\draw[thick,dashed] (0,0) circle(1.75 cm);
\filldraw[gray](-1.34057777546 ,1.12487831695 ) circle(4pt);
 \coordinate (A) at (1.7234,0.303); 
\coordinate (A') at (2.1434,0.303);
\coordinate (B) at (0.598,-1.644); 
\coordinate (B') at (0.698,-1.944);
\coordinate (C) at (1.5155,0.875); 
\coordinate (C') at (0.25,0.2); 
\coordinate (D) at (0.0,-1.75); 
\coordinate (D') at (0.0,-2); 
\coordinate (E) at (0.304,1.723); 
\coordinate (E') at (0.394,1.9); 
\coordinate (F) at (1.5155,-0.875); 
\coordinate (F') at (2.0155,-0.875); 
\coordinate (G) at (0.875,1.5155);
\coordinate (G') at (1.085,1.7355);
\coordinate (H) at (1.7234,-0.303);
\coordinate (H') at (2.1634,-0.303);
\coordinate (I) at (-1.5155,0.875); 
\coordinate (I') at (-0.85,0.25); 
\coordinate (J) at (-1.125,-1.34); 
\coordinate (J') at (-1.175,-1.64); 
\coordinate (K) at (-1.7234,0.303); 
\coordinate (K') at (-2.1234,0.303); 
\coordinate (L) at (-1.5155,-0.875); 
\coordinate (L') at (-1.8955,-0.955); 
\coordinate (I'') at (-1.3788,1.157); 
\coordinate (E'') at (-0.31256,1.7726); 
\draw[thick,color=black] (A) to[bend right] (B);
\draw[black](0.76,-0.8) -- node {\midarrow} (0.797,-0.7);
\draw[thick,color=lightgray] (E) to[bend right] (F);
\draw[lightgray](0.36,0.7) -- node {\midarrow} (0.397,0.6);
\draw[thick,color=lightgray] (G) to[bend right] (H);
\draw[lightgray] (0.96,0.65)-- node {\midarrow} (0.923,0.75);
\draw[thick,color=black] (K) to[bend left] (L);
\draw[black] (-1.46,-0.15) -- node {\midarrow}(-1.4517,-0.25);
\draw[loosely dotted,thick,color=gray] (-1.28557521937,1.53208888624) arc (130:85:2);
\draw[thick,color=black] (C) to[bend right] (D);
\node[black] at (C') {J};
\draw[black] (0.197,-0.6)-- node {\midarrow} (0.16,-0.7);
\draw[thick,color=black] (I) to[bend left] (J);
\node[black] at (I') {I};
\draw[black] (-1,-0.2)-- node {\midarrow} (-1.025,-0.1);
\end{scope}
\end{tikzpicture}
\end{minipage}  
\right)
 \end{split}
\end{equation}
 In the figure of eq(\ref{eq:probability sum}) the diagram highlights the fact that the contribution depends only on the pair of chords $I,J$ and (due to the presence of $S^{ab})$) it's partner chords. Here we have also pulled out the factor corresponding to the non-overlap contribution eq(\ref{eq:Non Overlap Contribution}) and the rest of the chord diagram contribution  will be normalized w.r.t this.  It must be remembered that the actual contribution of the pair of oriented chords to the chord diagram depends on which two H chords the oriented chord $I$ and $J$ sit in - we list out all the possibilities in Figure(\ref{all pairs}). For instance for the choice of chord pairs given in the chord diagram of eq(\ref{eq:probability sum}), the corresponding diagram is Figure(\ref{all pairs}b). 

In the next subsection, we will evaluate the diagram with two oriented chords in detail for each of the cases in Figure (\ref{all pairs}), but here we make some general comments. When we evaluate the chord diagram contribution due to $I,J$ chords as in eq (\ref{eq:probability sum}), note that when a colour index appears both in $I$ and $J$, the corresponding fermions sits in a 4 fermion trace whereas their partner fermions sits in a two fermion trace. Let us denote the resulting product of the two traces by $x'^4 M^2$ - the details of $x'$ depends on which of the chord structures among the diagrams (a)-(d) in Figure(\ref{all pairs}) that the overlapping index appears, and we will evaluate explicitly in upcoming sections. For now, note that this appears in lieu of 4 factors of two fermion traces which evaluated to $x^4 M^2$ (for the overlap case) given in eq(\ref{eq:No overlap contribution}). In effect we have the following factor for every overlapping index 
\begin{equation}
{x'^4 \over  x^4  } 
\end{equation}
If there are $p_{IJ}$ overlaps between two pairs of non-intersecting chords, then the sum over $p_{IJ}$ in eq (\ref{eq:probability sum}) evaluates to 
\begin{equation}\label{overlap}
       \sum_{p_{IJ}=0}^\infty {\lambda^{p_{IJ}} \over p_{IJ}!}              \left(\frac{x'^{4}}{x^4} \right)^{p_{IJ}} = e^{\lambda\left(\frac{x'^{4}}{x^4} \right)}
       \end{equation}
For intersecting chords, there is an extra factor of $s(p_{IJ})=(-1)^{p^2-p_{IJ}}$ which results in $e^{-\lambda\left(\frac{x'^{4}}{x^4} \right)}$. We thus have the moments eq(\ref{eq:probability sum}) to be 
\begin{equation}\label{final contribution}
 \begin{split}
     m_k^{(CD)} &=  {\cal N}^N e^{-\lambda {k \choose 2}}  x^{pk}  \prod_{\mbox{chord pair } I,J} e^{\pm {\lambda x'^4 \over x^4}}
 \end{split}
\end{equation}
where $x'$ is the contribution of the chord pair and $+,-$ sign depends on whether the chords are intersecting or non-intersecting.

What remains is to evaluate is the contribution $x'$ for all pairs of chords given in Figure(\ref{all pairs}). In the following subsections we determine the contributions corresponding to figures (\ref{all pairs}a) - (\ref{all pairs}d).

\subsubsection{Oriented chord pairs in H chord : Figure \ref{all pairs}a}
\label{Hchord}
\begin{figure}[h]
       \begin{center}
   \begin{tikzpicture}
         \begin{scope}[thin, every node/.style={sloped,allow upside down}]
\draw[black] (0.1,2)-- node {\midarrow} (0.08,2);
\draw[black] (0.08,1)-- node {\midarrow} (0.1,1);
\draw (1,0) arc (0:180:1cm);
         \draw (2,0) arc (0:180:2cm);
         \node at (1.1,-0.3) {$X^{\dagger}_{JC}$};
         \node at (-0.8,-0.3) {$X_{JB}$};
         \node at (-1.8,-0.3) {$X^{\dagger}_{IA}$};
         \node at (2.2,-0.3) {$X_{ID}$};
          \node at (0.1,0.7) {J};
         \node at (0.1,1.6) {I};
         \end{scope}
        \end{tikzpicture}
        \end{center}
        \caption{Oriented chord pairs in H chord:1(a)}
        \label{Oriented chord pairs in H chord:1(a)}
   \end{figure}
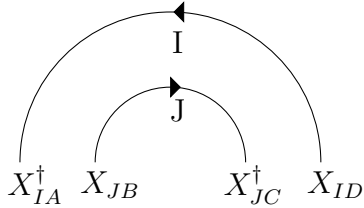

We first look at the contribution due to overlap between  chords constituting an H chord given in Figure(\ref{all pairs}a) which is reproduced in Figure(\ref{Oriented chord pairs in H chord:1(a)}), with fermion indices made explicit,

Let the common indices be $i_k=j_l$ with $i_k \in I, j_l \in J $, and the factor $x'$ corresponding to this case be denoted by $y$. As discussed in the previous section, in the double scaling limit, for an overlap within the H chord, the corresponding colour index from the partner chord $J$ would only appear in a two fermion trace structure, which leads to   
        \begin{equation}\label{General H chord}
        \begin{split}
        y^4 M^2 &\equiv   tr_{\Lambda_{ i_k=j_l}}(\chi^{\dagger}_{i_k a_k}\chi_{j_lb_l}\chi^{\dagger}_{j_lc_l}\chi_{i_kd_k}) tr_{\Lambda_{j_k}}(\chi_{j_k b_k}\chi^{\dagger}_{j_k c_k})tr_{\Lambda_{i_l}}(\chi^{\dagger}_{i_l a_l}\chi_{i_l d_l})S^{a_k b_k}S^{a_l b_l}S^{c_k d_k}S^{c_l d_l}\\
         & = tr_{\Lambda}(\chi^{\dagger}_{a}\chi_{b}\chi^{\dagger}_{c}\chi_{d}) W^{ad} \tilde{W}^{bc}
         \end{split}
        \end{equation}
        
where we have defined $W^{ad}$ and  $\tilde{W}^{ad}$ as follows for concise expressions,
\begin{equation}\begin{split}
     W^{ad} & \equiv S^{ab} tr_\Lambda (\chi_b \chi^\dagger_c) S^{cd} = (W^{da})^* \\
     \tilde W^{ad} & \equiv S^{ba} tr_\Lambda (\chi^\dagger_b \chi_c) S^{dc} = (\tilde W^{da})^*\\
 \end{split}\end{equation}
 
\noindent Following the arguments given near  eq(\ref{overlap}), the contribution of this diagram is given by $e^{\lambda_H}$ where $\lambda_H = {\lambda y^4 \over x^4}$ i.e
\begin{equation}\label{lambda_H}
\begin{split}
    \lambda_H =  \frac{\lambda}{x^4} \  tr_{\Lambda}(\chi^{\dagger}_{a}\chi_{b}\chi^{\dagger}_{c}\chi_{d}) W^{ad} \tilde{W}^{bc}
\end{split}
\end{equation}
where $x$ is given by eq(\ref{eq:No overlap contribution}). Since every chord diagram consists of $\frac{k}{2}$ H chords and this results in an overall factor of $e^{\frac{k}{2}\lambda_H}$.

\subsubsection{Oriented chord pairs in Non nested H chords: Figure \ref{all pairs}b} \label{1(b)}
\begin{figure}[H]
\begin{center}
 \begin{tikzpicture}
\begin{scope}[thin, every node/.style={sloped,allow upside down}]
\draw[thick,black] (-1.25,0) arc (180:0:0.75);
\draw[thick,black] (-2,0) arc (180:0:1.5);
\node[black] at (-1.25,-0.25) {\scriptsize{$X_{JB}$}};
\node[black] at (-2,-0.25) {\scriptsize{$X^{\dagger}_{IA}$}};
\node[black] at (0.25,-0.25) {\scriptsize{$X^{\dagger}_{JC}$}};
\node[black] at (1,-0.25) {\scriptsize{$X_{ID}$}};
\draw[thick,black] (1.7,0) arc (180:0:1.5);
\draw[thick,black] (2.45,0) arc (180:0:0.75);
\node[black] at (1.7,-0.25) {\scriptsize{$X^{\dagger}_{KE}$}};
\node[black] at (2.45,-0.25) {\scriptsize{$X_{LF}$}};
\node[black] at (4.7,-0.25) {\scriptsize{$X_{KH}$}};
\node[black] at (3.95,-0.25) {\scriptsize{$X^{\dagger}_{LG}$}};
\draw[black] (-0.48,1.5)-- node {\midarrow} (-0.5,1.5);
\draw[black] (-0.5,.75)-- node {\midarrow} (-0.48,.75);
\draw[black] (3.22,1.5)-- node {\midarrow} (3.2,1.5);
\draw[black] (3.2,0.75)-- node {\midarrow} (3.22,0.75);
\end{scope}
\end{tikzpicture}
\end{center} 
\caption{Oriented chord pairs in Non nested H chord: 1(b)}
\label{Oriented chord pairs in Non nested H chord: 1(b)}
\end{figure}
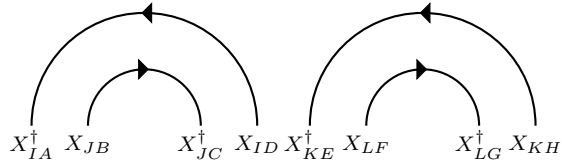

We now start looking at oriented chord pairs sitting in different H chords. There are four\footnote{the chord pairs within the same H-chords (for instance $IJ$ and $KL$ in figure 1(b) have already been accounted for in section(\ref{Hchord}) } possible oriented chord pairs in any two H chords. For instance in the case of non-nested H-chord pair Figure(\ref{Oriented chord pairs in Non nested H chord: 1(b)}), the possible chord pairs are $IK,JK,IL,JL$. 
We now determine the contribution $x'$ corresponding to the case 
of non zero overlap between chord pair $IK$ corresponding to the figure(\ref{Oriented chord pairs in Non nested H chord: 1(b)}). Let the common indices be $i_m = k_n$ with $i_m \in I, k_n \in K $. The contribution is then given by,
       \begin{equation}
       \begin{split}
         x'^4 M^2  &\equiv tr_{\Lambda}(\chi^{\dagger}_{a_m}\chi_{d_m}\chi^{\dagger}_{e_n}\chi_{h_n})tr_{\Lambda}(\chi_{ b_m}\chi^{\dagger}_{ c_m})tr_{\Lambda}(\chi_{ f_n}\chi^{\dagger}_{ g_n})S^{a_m b_m}S^{c_m d_m}S^{e_n f_n} S^{g_n h_n} \\
         &= tr_\Lambda( \chi_a^\dagger \chi_b \chi_c^\dagger \chi_d) W^{ab} W^{cd}
       \end{split}
       \end{equation}

The contributions due to other overlaps in figure (\ref{Oriented chord pairs in Non nested H chord: 1(b)}) $|I\cap L|, |J\cap K|$ and $|J\cap L|$ can also be determined similarly (see Appendix(\ref{General results})) and the net contribution from figure (\ref{Oriented chord pairs in Non nested H chord: 1(b)}) is given by $e^{\lambda_{N\!N}}$, where
\begin{equation}\label{lambda_A}
   \begin{split}
   \lambda_{N\!N} & 
   = \frac{\lambda}{x^4} \left( tr_{\Lambda}(\chi_a^{\dagger}\chi_b \chi_c^\dagger \chi_d)W^{cd}W^{ab} + tr_{\Lambda}(\chi_a\chi_b^{\dagger} \chi_c \chi_d^\dagger)\tilde{W}^{cd}\tilde{W}^{ab} + tr_{\Lambda}(\{\chi_a\chi_b^{\dagger},\chi_c^\dagger \chi_d\})\tilde{W}^{ab}W^{cd} \right)
    \end{split}
\end{equation}

\subsubsection{Oriented chord pairs in Nested H chords : Figure \ref{all pairs}c}\label{subsubsec:nested}
\begin{figure}[H]
\begin{center}
 \begin{tikzpicture}
\begin{scope}[thin, every node/.style={sloped,allow upside down}]
\draw[thick,black] (6,0) arc (180:0:2.5);
\draw[thick,black] (6.6,0) arc (180:0:1.9);
\draw[thick,black] (7.5,0) arc (180:0:1);
\draw[thick,black] (8,0) arc (180:0:0.5);
\node[black] at (6,-0.25) {\scriptsize{$X^{\dagger}_{IA}$}};
\node[black] at (6.6,-0.25) {\scriptsize{$X_{JB}$}};
\node[black] at (7.4,-0.25) {\scriptsize{$X^{\dagger}_{KE}$}};
\node[black] at (8.1,-0.25) {\scriptsize{$X_{LF}$}};

\node[black] at (9,-0.25) {\scriptsize{$X^{\dagger}_{LG}$}};
\node[black] at (9.75,-0.25) {\scriptsize{$X_{KH}$}};
\node[black] at (10.4,-0.25) {\scriptsize{$X^{\dagger}_{JC}$}};
\node[black] at (11,-0.25) {\scriptsize{$X_{ID}$}};
\draw[black] (8.5,2.5)-- node {\midarrow} (8.48,2.5);
\draw[black] (8.48,1.9)-- node {\midarrow} (8.5,1.9);
\draw[black] (8.5,1)-- node {\midarrow} (8.48,1);
\draw[black] (8.48,0.5)-- node {\midarrow} (8.5,0.5);
\end{scope}
\end{tikzpicture}
\end{center}  
\caption{Oriented chord pairs in Nested H chord: 1(c)}
\label{Oriented chord pairs in Nested H chord: 1(c)}
\end{figure}
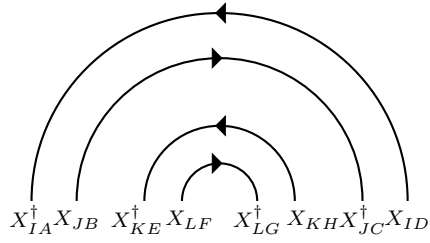
The contributions due to chord pairs in the nested non intersecting chord diagram can also be calculated  similarly (details are in Appendix(\ref{General results})). We denote the  contribution due chord pairs in figure (\ref{Oriented chord pairs in Nested H chord: 1(c)}) by $e^{\lambda_{N}+ \lambda_{NN}}$ where $\lambda_{N} $ is given by ,
\begin{equation}\label{lambda_B}
\begin{split}
    \lambda_{N} =& \frac{\lambda}{x^4 } \left(  tr_{\Lambda}(\chi^\dagger_a\chi^\dagger_c \chi_d \chi_b) W^{ab} W^{cd} +  tr_{\Lambda}(\chi_a\chi_c \chi_d^\dagger \chi_b^\dagger) \tilde{W}^{ab} \tilde{W}^{cd} \right. \\
    &\hspace{2cm} \left. + \left(  tr_{\Lambda}(\chi^\dagger_c\chi_a \chi_b^\dagger \chi_d) + tr_{\Lambda}(\chi_a\chi^\dagger_c \chi_d \chi_b^\dagger)  \right) \tilde{W}^{ab} W^{cd} \right)- \lambda_{NN}\\
    =& \frac{\lambda}{x^4} \left( tr_{\Lambda}(\chi_a^{\dagger}[ \chi_b, \chi_c^\dagger \chi_d] ) W^{ab}W^{cd} + tr_{\Lambda}(\chi_a^{\dagger}[\chi_b, \chi_c \chi_d^\dagger])W^{ab}\tilde W^{cd} 
    \right. \\
    & \qquad    \left.+ tr_{\Lambda}(\chi_a [\chi_b^\dagger, \chi_c^\dagger \chi_d])\tilde{W}^{ab} W^{cd} + tr_{\Lambda}(\chi_a  [\chi_b^\dagger , \chi_c \chi_d^\dagger] )\tilde{W}^{ab}\tilde W^{cd}   \right)
    \end{split}
\end{equation}

Let us note here that if there is no chemical potential, then there is essentially no difference between nested chord pairs and non-nested chord pairs. This is borne out by the expressions since the cyclicity of trace in this case implies that $\tilde{W}^{ab}=W^{ba}$ as a result of which $\lambda_N = 0$. In fact we will later see that even in the case  with a non zero chemical potential, this holds for all the classical groups that we consider.

\subsubsection{Oriented chord pairs in Intersecting H chords diagrams:Figure \ref{all pairs}d}
 \label{1(d)}       
\begin{figure}[H]
 \centering
 \begin{tikzpicture}
 \begin{scope}[thin, every node/.style={sloped,allow upside down}]
         \draw (1.5,0) arc (0:180:1.5cm);
         \draw (2,0) arc (0:180:2cm);
         \draw (3.5,0) arc (0:180:1.5cm);
         \draw (4,0) arc (0:180:2 cm);
         \draw[black] (0,1.5)-- node {\midarrow} (.2,1.5);
        \draw[black] (0.2,2)-- node {\midarrow} (0,2);
        \draw[black] (2,1.5)-- node {\midarrow} (2.2,1.5);
        \draw[black] (2.2,2)-- node {\midarrow} (2,2);
         \node at (-2,-0.4) {\scriptsize{$X^{\dagger}_{IA}$}};
        \node at (-1.4,-0.4) {\scriptsize{$X_{JB}$}};
        \node at (0,-0.4) {\scriptsize{$X^{\dagger}_{KE}$}};
        \node at (0.6,-0.4) {\scriptsize{$X_{LF}$}};
        \node at (1.6,-0.4) {\scriptsize{$X^{\dagger}_{JC}$}};
        \node at (2.2,-0.4) {\scriptsize{$X_{ID}$}};  
        \node at (3.4,-0.4) {\scriptsize{$X^{\dagger}_{LG}$}};
        \node at (4,-0.4) {\scriptsize{$X_{KH}$}}; 
  \end{scope}
  \end{tikzpicture}
  \caption{Oriented chord pairs in Intersecting H chords:1(d)}
  \label{Oriented chord pairs in Intersecting H chords:1(d)}
        \end{figure}
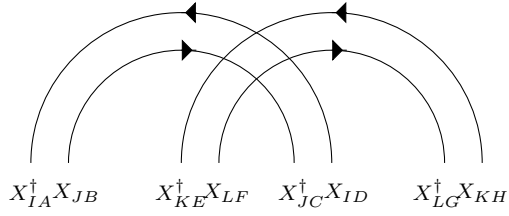

 The next possible chord diagram structure involves chord intersections and is given by Figure(\ref{Oriented chord pairs in Intersecting H chords:1(d)})\
 \vspace{1mm}\\
The calculations proceeds exactly as in the previous cases apart from an additional sign due to the chord intersection of chord pairs as as given by eqn(\ref{final contribution}). 
Thus the net contribution  corresponding to the chord pairs in intersecting H chords is given by $e^{\lambda_C + \lambda_{NN} }$ where
\begin{equation}\label{lambda_C}
\begin{split}
       \lambda_C &= -\frac{\lambda}{x^4 } \left( tr_{\Lambda} (\chi_a^\dagger \chi_c^\dagger \chi_b \chi_d ) W^{ab} W^{cd} + tr_{\Lambda} (\chi_a \chi_c \chi_b^\dagger \chi_d^\dagger ) \tilde{W}^{ab} \tilde{W}^{cd} \right. \\
       & \hspace{1cm} + \left. \left( tr_{\Lambda} (\chi^\dagger_{c}\chi_a\chi_d\chi^\dagger_b) + tr_{\Lambda} (\chi_a\chi^\dagger_{c}\chi^\dagger_b\chi_d)  \right)  \tilde{W}^{ab} W^{cd} \right) - \lambda_{NN}\\
       &= -\frac{\lambda}{x^4} \left( tr_{\Lambda}(\chi_a^{\dagger} \{ \chi_b ,  \chi_c^\dagger \} \chi_d) W^{ab}W^{cd} + tr_{\Lambda}(\chi_a^{\dagger}\{\chi_b ,\chi_c\} \chi_d^\dagger)W^{ab}\tilde W^{cd} 
    \right. \\
& \qquad    \left.+ tr_{\Lambda}(\chi_a \{ \chi_b^\dagger ,\chi_c^\dagger \} \chi_d)\tilde{W}^{ab} W^{cd} + tr_{\Lambda}(\chi_a \{ \chi_b^\dagger , \chi_c \} \chi_d^\dagger )\tilde{W}^{ab}\tilde W^{cd}   \right)
    \end{split}
\end{equation}

\noindent The details of the above calculations is given in the appendix(\ref{General results}).

\subsection{Summary of  the chord diagram prescription for computing $m_k$}\label{Final summary}

We now summarize our results for computing the moments of the Hamiltonian $m_k$. $m_k$ is a sum over contribution from all possible chord diagrams with $k$ nodes. The rules for computing the contribution to a given chord diagram is similar but more intricate that the rules given in section(\ref{complex}) for complex SYK. The contribution depends not only on the number of intersecting chords, but also on the structure of non-intersecting chords, i,e whether they are nested / non-nested\footnote{Since there are $\binom{k/2}{2}$ chord pairs in total, we can trade the dependence on say nested chord pairs with other types of chord pairs.}. We summarize the rules below
\begin{itemize}
\item The moments of the Hamiltonian $m_k$ is a sum of the contribution of all chord diagrams with $k$ nodes denoted by $m_k^{(CD)}$. The chords are termed as H chords. 
\item For a given chord diagram, the contribution $m_k^{(CD)}$ is ${\cal N}^Nx^{pk}e^{-\lambda {k \choose 2}} e^{k\lambda_H \over 2} e^{\lambda_{N\!N}{k/2 \choose 2}}$ (these parameters are defined in equations (\ref{N definition}),(\ref{eq:No overlap contribution}), (\ref{lambda_H}) and (\ref{lambda_A})) times a factor depending on the details of the given chord diagram which is computed from the table below,
\begin{center}
\begin{tabular}{|l|c|}
    \hline
  {\bf \hspace{20mm} Chord configuration} & {\bf Contribution}\\
  \hline \hline 
 Every nested chord pair :  \qquad  \begin{minipage}{1.5cm}
    \begin{center} \begin{tikzpicture}
 \node[black] at (3.25,1.2) {};
 \draw[ultra thick] (4,0) arc (0:180:1cm);
  \draw[ultra thick] (3.5,0) arc (0:180:0.5cm);
  \node[black] at (3.25,-0.2) {};
 \end{tikzpicture}
 \end{center}
 \end{minipage} & $e^{\lambda_{N}}$ \\
 \hline 
 Every intersecting chord pair : \hspace{25mm} \begin{minipage}{3.5cm}
    \begin{center} \begin{tikzpicture}
  \node[black] at (3.25,1.2) {};
  \draw[ultra thick] (-1,0) arc (0:180:0.75cm);
  \draw[ultra thick] (0.1,0) arc (0:180:0.75cm);
   \node[black] at (3.25,-0.2) {};
  \end{tikzpicture}
 \end{center}
 \end{minipage}     & $e^{\lambda_C}$  \\
  \hline 
\end{tabular}
\end{center}
where $\lambda_{N}, \lambda_C$ are given in equations (\ref{lambda_B}) and (\ref{lambda_C}).

\end{itemize}

Note that in the absence of chemical potential and for classical groups with arbitrary chemical potential, the parameter $\lambda_N=0$ as mentioned in section \ref{subsubsec:nested}. In this case the rules become similar to Majorana case in the sense that contribution to $m_k^{CD}$ only depends on intersection number. 
\subsection{Transfer Matrix }\label{sec:Transfer matrix}
 Having described the rules of the chord diagram, in this subsection we show capture it by a {\it Transfer matrix} which acts on an auxiliary Hilbert space on which it acts. We will follow \cite{Berkooz2019TowardsAF}. In this subsection we will term $H$ chords simply as chords.

\begin{figure}[h]
    \centering
    \def\s{1.5}
 \begin{tikzpicture}
\begin{scope}[thin, every node/.style={sloped,allow upside down}]
\draw[thick] (0,0) circle(1.75 cm);
\filldraw[gray](-1.34,1.12 ) circle(5pt);
\coordinate (A) at (1.7234,0.303); 
\coordinate (A') at (2.1434,0.303);
\coordinate (B) at (0.598,-1.644); 
\coordinate (B') at (0.698,-1.944);
\coordinate (C) at (1.5155,0.875); 
\coordinate (C') at (1.9655,0.875); 
\coordinate (D) at (0.303,1.7234);
\coordinate (D') at (0.503,1.95);
\coordinate (E) at (0.304,1.723); 
\coordinate (E') at (0.394,1.9); 
\coordinate (F) at (1.5155,-0.875); 
\coordinate (F') at (2.2155,-0.875); 
\coordinate (G) at (0.875,1.5155);
\coordinate (G') at (1.085,1.7355);
\coordinate (H) at (1.7234,-0.303);
\coordinate (H') at (2.1634,-0.303);
\coordinate (I) at (-1.5155,0.875); 
\coordinate (I') at (-1.9655,0.875); 
\coordinate (J) at (-1.125,-1.34); 
\coordinate (J') at (-1.175,-1.64); 
\coordinate (K) at (-1.7234,0.303); 
\coordinate (K') at (-2.1234,0.303); 
\coordinate (L) at (-1.5155,-0.875); 
\coordinate (L') at (-1.8955,-0.955); 
\coordinate (I'') at (-1.3788,1.157); 
\coordinate (E'') at (-0.31256,1.7726); 
\coordinate (X) at (-0.875,1.5155); 
\coordinate (X') at (-0.9,1.85); 

\draw[ultra thick,color=black] (A) to[bend right] (B);
\draw[ultra thick,color=black] (X) to[bend right] (F);
\draw[ultra thick,color=black] (I) to[bend left] (J);
\draw[ultra thick,color=black] (C) to[bend left] (D);
\draw [ultra thick,-stealth](2.25,0) -- (3.3,0);

 \draw[thick] (3.5,0) -- (9.8,0);

         \filldraw[gray](3.75,0 ) circle(4pt);
        \foreach \x in {4.5,7.5}
        {
        \draw[ultra thick] (\x,0) arc(180:90:0.8);
        }
        \draw[ultra thick] (5.3,0.8)--(5.7,0.8);
        \draw[ultra thick] (8.3,0.8)--(8.7,0.8);

       \foreach \x in {6.5,9.5}
        {
        \draw[ultra thick] (\x,0) arc(0:90:0.8);
        }
        
        \draw[ultra thick] (8.5,0) arc(180:90:0.7);
         \draw[ultra thick] (9.2,0.7)--(9.8,0.7);

         \draw[loosely dotted, ultra thick] (9.8,1.25)--(9.8,-0.75);

         \draw[lightgray,thick] (9.8,0)--(12.75,0);

         \draw[lightgray,ultra thick] (9.8,0.7)--(11.8,0.7); 
        
        \draw[lightgray,ultra thick] (10.5,0) arc(180:90:0.55);
        \draw[lightgray,ultra thick] (11.5,0) arc(0:90:0.55);
        \draw[lightgray,ultra thick] (11.05,0.55)--(10.95,0.55);
        
        \draw[lightgray,ultra thick] (12.5,0) arc(0:90:0.7);

        \draw[black, ultra thick, -stealth] (5.5,0.9)--(5.5,2);
        \draw[black, ultra thick, -stealth] (7.9,0.9)--(7.5,1.9);
        \node[black] at (6.5,2.25) {closed chords};

        \draw[black, ultra thick, -stealth] (9.6,0.8)--(9.6,2);
        \node[black] at (9.25,2.25) {open chord};

        \draw[lightgray, ultra thick, -stealth] (11.35,0.5)--(12,2);
        \node[gray] at (11.65,2.25) {future chord};
\end{scope}
\end{tikzpicture}
\caption{Flattening one of the chord diagrams in $m_8$; the flattened chord diagram is cut at 5th node (vertical dotted lines) illustrating the closed, open, and future chords}
\label{Chord diagram representing one of the terms in $m_8$}
\end{figure}
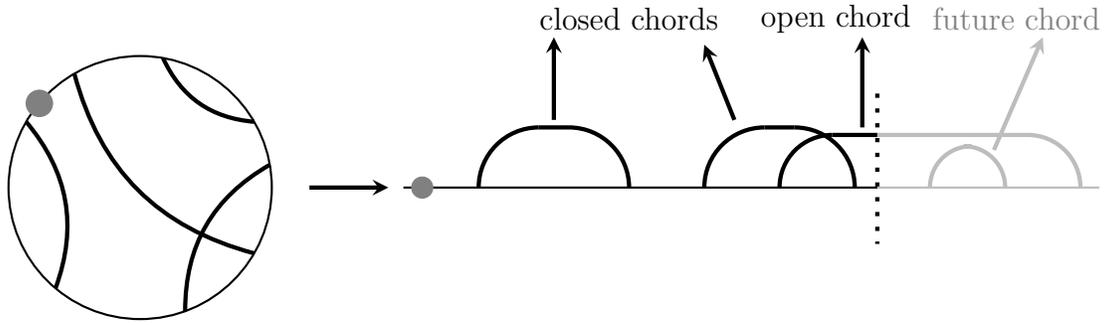

To define the transfer matrix, imagine flattening out the chord diagram by choosing the $H$ node which is right after the chemical potential insertion as the first node. Imagine cutting the flattened chord diagram at $i$'th node. The chords can be classified into three types : chords which already got closed (termed as closed chords), chords which have started but yet to close (termed as open chords) and chords which have not yet started yet (termed as future chords). This is illustrated with the example of one of the chord diagrams in $m_8$ in Figure(\ref{Chord diagram representing one of the terms in $m_8$}) where the chord diagram is cut at the fifth node resulting in closed and open chords - the future chords are also represented in gray.

Recall that to apply the diagrammatic rules for the chord diagram in section \ref{Final summary}, one has to consider all pairs of intersecting and nested chords and weigh the chord diagram appropriately. We can ignore closed-future chord pairs since they are necessarily non-nested. We now define the partial chord diagram contribution at $i$'th node by only applying the chord diagram rules of section \ref{Final summary} for the closed-closed and closed open pairs. Consider all possible chord diagrams such that they have $l$ open chords at $i$'th node. The sum of all partial chord diagram contribution for these diagrams are termed as $v^{(i)}_l$. We will now show that that we can write recursion relations that relate the $v^{(i+1)}$ to $v^{(i)}$. To do this, let us say that we know $v^{(i)}_l$ for all $l$. We can easily compute $v^{(i+1)}_l$ for all $l$ using the following procedure :  Let us a order the open chords by declaring that freshly opened chord is always at the bottom. At $i+1$'th node we can either open a new chord or close one of the existing open chord. Closing the $i$'th chord implies that it crosses all the $i-1$ chords below it and is nested w.r.t all the $l-i$ chords above it. 

\begin{figure}[H]
    \def\s{0.5}
    \centering
    \begin{tikzpicture}    
        \node[black] at (17,1.5) {$l$ open chords};
        \node[black] at (4,3.5) {$l-1$ open chords};
        \node[black] at (4,-0.5) {$l+1$ open chords};

        \foreach \x in {6,...,8}
        {
        \draw[ultra thick] (6.5,\x*\s)--(9,\x*\s);
        \draw[dashed,ultra thick] (6,\x*\s)--(6.5,\x*\s);
        \draw[dashed,ultra thick] (9,\x*\s)--(9.5,\x*\s);
        }

        \draw[ultra thick] (6.5,4.7)--(9,4.7);
        \draw[dashed,ultra thick] (6,4.7)--(6.5,4.7);
        \draw[dashed,ultra thick] (9,4.7)--(9.5,4.7);
        \draw[loosely dotted,ultra thick] (7,4.6)--(7,4.1);
        
        \draw[thick] (6.5,2) -- (9,2);
        \draw[dashed, thick] (6,2)--(6.5,2);
        \draw[dashed, thick] (9,2)--(9.5,2);
        \draw[ultra thick] (7.75,2) arc(180:90:0.5);
        \draw[ultra thick] (8.25,2.5) -- (9,2.5);
        \draw[dashed, ultra thick] (9,2.5)--(9.5,2.5);

        \foreach \x in {-2,-1}
        {
        \draw[ultra thick] (6.5,\x*\s)--(9,\x*\s);
        \draw[dashed,ultra thick] (6,\x*\s)--(6.5,\x*\s);
        \draw[dashed,ultra thick] (9,\x*\s)--(9.5,\x*\s);
        }
        \draw[ultra thick] (6.5,0.75)--(9,0.75);
        \draw[dashed,ultra thick] (6,0.75)--(6.5,0.75);
        \draw[dashed,ultra thick] (9,0.75)--(9.5,0.75);
        \draw[loosely dotted,ultra thick] (7,0.65)--(7,0.1);
        \draw[ultra thick] (6.5,0)--(7.25,0);
        \draw[dashed,ultra thick] (6,0)--(6.5,0);
        \draw[ultra thick] (7.75,-0.5) arc(0:90:0.5);
        \draw[ultra thick] (7.75,-0.5)--(7.75,-1.5);
        \draw[thick] (6.5,-1.5)--(9,-1.5);
        \draw[dashed,thick] (6,-1.5)--(6.5,-1.5);
        \draw[dashed,thick] (9,-1.5)--(9.5,-1.5);

        \node[black] at (13.75,-2.2) {$(i+1)^{th}$ node};
        \node[black] at (7.75,-2.2) {$i^{th}$ node};

        \draw [ultra thick](10.5,2.75) -- (10,2.75);
        \draw [ultra thick]((10.5,0) -- ((10,0);
        \draw[ultra thick] (10.5,2.75)--(10.5,0);
        \draw[ultra thick,-stealth] (10.5,1.5)--(11,1.5);
        
        \foreach \x in {1,...,4}
        {
        \draw[ultra thick] (12.5,\x*\s)--(15,\x*\s);
        \draw[dashed,ultra thick] (12,\x*\s)--(12.5,\x*\s);
        \draw[dashed,ultra thick] (15,\x*\s)--(15.5,\x*\s);
        }
        \draw[ultra thick] (12.5,2.7)--(15,2.7);
        \draw[dashed,ultra thick] (12,2.7)--(12.5,2.7);
        \draw[dashed,ultra thick] (15,2.7)--(15.5,2.7);
        \draw[loosely dotted,ultra thick] (13.5,2.1)--(13.5,2.65);
        \draw[thick] (12.5,0)--(15,0);
        \draw[dashed,thick] (12,0)--(12.5,0);
        \draw[dashed,thick] (15,0)--(15.5,0);
        
    \end{tikzpicture}
    \label{fig:transfer matrix}
    \caption{ A chord diagram with $l$ open chords at $i+1$'th node can arise from two possible chord structures at $i$'th node: opening a new chord in a structure with $l-1$ chords or closing one of the open chords in a structure with $l+1$ chords. }
\end{figure}
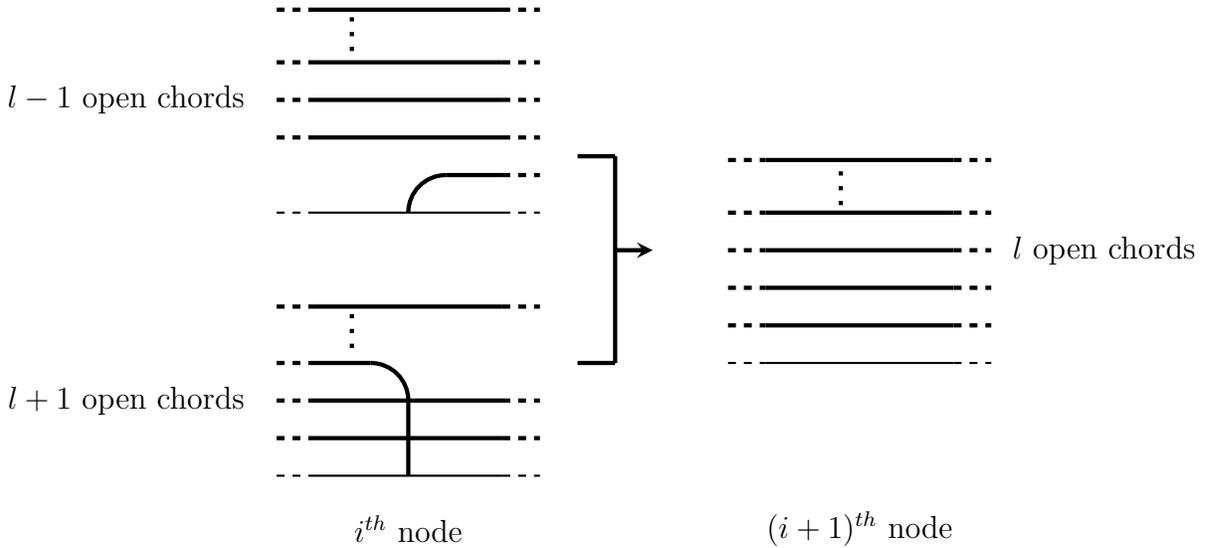

Thus we can write a recursion relation for $v^{(i)}_l$ as 
\begin{equation}
    v^{(i+1)}_{l} = v^{(i)}_{l-1} + \underbrace{ \left( e^{\lambda_C l} + e^{\lambda_C(l-1)+\lambda_N}+ \dots + e^{\lambda_C+\lambda_N(l-1)} + e^{\lambda_N l}\right)}_{\equiv \eta_l} v^{(i)}_{l+1} \\
\end{equation}
We can rewrite $\eta_l$ as follows
\begin{equation}
 \eta_l \equiv { e^{\lambda_N(l+1)}  - e^{\lambda_C(l+1)} \over e^{\lambda_N} - e^{\lambda_C} } =  { (q_N^{l+1}-q^{l+1}) \over q_N-q }
\end{equation}
where $q_N \equiv e^{\lambda_N},q=e^{\lambda_C}$. Recursion relation then becomes 
\begin{equation}
    v^{(i+1)}_{l} = v^{(i)}_{l-1} +  \eta_l\  v^{(i)}_{l+1}
\end{equation}
We now define an infinite dimensional matrix which plays the role of transfer matrix namely
\begin{equation}\label{eq:Transfer Matrix}
T \equiv     \begin{pmatrix}
        0 &  \eta_0 & 0 & \dots \\
        1 & 0 &  \eta_1 & 0 \dots \\
        0 & \ddots & \ddots & \ddots & \ddots \ddots \\
    \end{pmatrix}
\end{equation}
Noting that the flattened chord diagrams ends with a configuration with no chords after $k$ nodes, we have
\begin{equation}
    m_k = \langle 0 | T^k | 0 \rangle
\end{equation}
where $| 0 \rangle$ is the state with no chords 
\begin{equation}
    | 0 \rangle = \begin{pmatrix}
        1 & 0 & 0 & \dots
    \end{pmatrix}^T
\end{equation}

\paragraph{Diagonalizing $T$ matrix ?}
If $v^{\mu}_l$ is the eigenvalue of $T$ matrix with eigenvalue ${2\mu \over \sqrt{q_N-q}}$, then $v^\mu_l$ satisfies the recursion relation
\begin{equation}
    {2\mu \over \sqrt{q_N-q}} v^\mu_l = v^\mu_{l-1} +  \ {(q_N^{l+1}-q^{l+1}) \over q_N-q} v^\mu_{l+1}
\end{equation}
Defining $u^\mu_l$ via
\begin{equation}
    v^\mu_l = {(q_N - q)^{l \over 2} \over (( q_N ;q))_l   } \ u^\mu_l , \qquad \qquad  ((q_N ;q))_l \equiv  \prod_{i=1}^l (q_N^i - q^i) 
\end{equation}
then $u^\mu_l$ satisfies the recursion relation
\begin{equation}
    2\mu u^{\mu}_l = (q_N^l-q^l) u^\mu_{l-1} +   u^\mu_{l+1} ,\qquad u^{\mu}_0  =1, u^\mu_{-1}= 0
\end{equation}
Note that if $q_N=1$, this is the recursion relation obeyed by q-hermite polynomial and this fact was used to diagonalize $T$ matrix in \cite{2018JHEPNarayan}. The recursion relation above with $q_N \ne 1$, was  encountered before in \cite{pqdifferentiation}, but there is not much known about solutions. We leave a detailed study of such recursion relation for future work. However, we mention here that for the case of classical group symmetries see section(\ref{sec:Special cases}), the parameter $q_N$ evaluates to $1$ reducing to the known case mentioned before.

\subsection{Chord diagram relevant for the Two Point function}\label{two point function}

In this subsection we set up the chord diagram prescription for the two-point function of operators. We consider two point function of operator \footnote{This is the simplest operator one could write down. It is possible to consider more general operators than these see for example see \cite{{Berkooz2020ComplexSM}}. We leave the study of such operators for this for future work.}
\begin{equation}
    M_{\tilde A} \equiv \tilde J_{\tilde I} X_{\tilde I \tilde A}
\end{equation}
where $\tilde I$ is a ordered set of $\tilde p$ colour indices and $\tilde A$ is a collection of $\tilde p$ flavour indices. The only difference from the set $I,A$ etc considered before in the definition of Hamiltonian in eq(\ref{Generalised SYK Hamiltonian}) is that now they have $\tilde p$ indices rather than $p$ indices. Here the couplings $\tilde J_{\tilde I}$ are random couplings  picked from an ensemble with variance 
\begin{equation}
    \langle\langle \tilde J_{\tilde I} \  \tilde J_{\tilde K}  \rangle\rangle = { \tilde {\cal J}^2 \over M^{\tilde p \over 2} \binom{N}{\tilde p} }  \delta_{\tilde I \tilde K} 
\end{equation}
The parameter $\tilde p$ is taken to be large with the following scaling
\begin{equation}
 \tilde \lambda \equiv  {p \tilde p \over N}\ \mbox{ fixed }, \hspace{20mm} \mbox{ as } \tilde p \rightarrow \infty , N    \rightarrow \infty
\end{equation}
The computation of two point function invariant under global symmetry i.e\\ $\langle \langle S^{\tilde A \tilde B} Tr_\Lambda \left(  M^\dagger_{\tilde A}(\tau_1) M_{\tilde B}(\tau_2)\right)\rangle \rangle$ involves evaluating the moments 
\begin{equation}
    \tilde m_{k,k_1} = \langle\langle Tr_\Lambda \left( M^\dagger_{\tilde A} H^{k_1} M_{\tilde B} H^{k - k_1}\right)\rangle\rangle S^{\tilde A \tilde B}
\end{equation}
Again, the chord diagram structure arises due to the wick contractions of random couplings. The new ingredient now is that there is a M-chord (which is an oriented chord due to our conventions that arrow goes from $X$ to $X^\dagger$) corresponding to the contraction of $\tilde J$ couplings, which we denote by a {\it wavy chord}. As an example, we give a chord diagram representing
$Tr_\Lambda(M^\dagger H M H^3 )$ below,
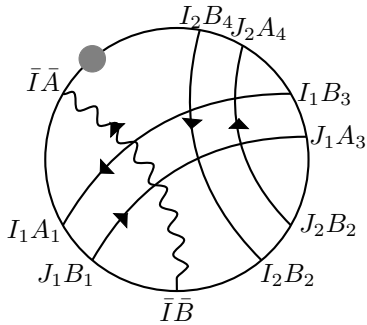
\begin{figure}[H]
\begin{center}
 \begin{tikzpicture}
\begin{scope}[thin, every node/.style={sloped,allow upside down}]
\draw[thick] (0,0) circle(1.75 cm);

\coordinate (A) at (1.7234,0.303); 
\coordinate (A') at (-1.475,-1.54);
\coordinate (B) at (-1.125,1.34); 
\coordinate (B') at (2.144 ,0.298);
\coordinate (C) at (1.5155,0.875); 
\coordinate (C') at (1.9655,0.875); 
\coordinate (D) at (0.0,-1.75); 
\coordinate (D') at (0.0,-2); 
\coordinate (E) at (0.304,1.723); 
\coordinate (E') at (0.394,1.9); 
\coordinate (F) at (1.5155,-0.875); 
\coordinate (F') at (2.0155,-0.875); 
\coordinate (G) at (0.875,1.5155);
\coordinate (G') at (1.085,1.7355);
\coordinate (H) at (1.125,-1.34);
\coordinate (H') at (2.1634,-0.303);
\coordinate (I) at (-1.5155,0.875); 
\coordinate (I') at (1.475,-1.54); 
\coordinate (J) at (-1.125,-1.34); 
\coordinate (J') at (-1.175,-1.64); 
\coordinate (K) at (-1.7234,0.303); 
\coordinate (K') at (-2.1234,0.303); 
\coordinate (L) at (-1.5155,-0.875); 
\coordinate (L') at (-1.8955,-0.955); 
\coordinate (I'') at (-1.7788,1.057); 
\coordinate (E'') at (-0.31256,1.7726); 
\coordinate (X) at (-1.5155,0.875); 0
\coordinate (Y) at (0,-1.75); 

\draw[thick,color=black] (A) to[bend right] (J);
\filldraw[gray](B) circle(5pt);
\draw[thick,color=black] (C) to[bend right] (L);

\draw[black] (-0.80,0.4)-- node {\midarrow} (-0.897,0.50);
\draw[thick,color=black] (E) to[bend right] (H);
\draw[black] (0.2,0.5) -- node {\midarrow} ( 0.19,0.4);
\draw[thick,color=black] (G) to[bend right] (F);
\draw[black] (0.823,0.45)-- node {\midarrow} (0.825,0.55);

\draw[black] (-0.95,-0.115)-- node {\midarrow} (-1.03,-0.19);
\draw[black] (-0.73,-0.79)-- node {\midarrow} (-0.65,-0.715);
\node[black] at (L') {\footnotesize{$I_1A_1$}};
\node[black] at (A') {\footnotesize{$J_1B_1$}};
\node[black] at (I'') {\footnotesize{$\bar{I}\bar{A}$}};
\node[black] at (D') {\footnotesize{$\bar{I}\bar{B}$}};
\node[black] at (I') {\footnotesize{$I_2B_2$}};
\node[black] at (F') {\footnotesize{$J_2B_2$}};
\node[black] at (B') {\footnotesize{$J_1A_3$}};
\node[black] at (C') {\footnotesize{$I_1B_3$}};
\node[black] at (G') {\footnotesize{$J_2A_4$}};
\node[black] at (E') {\footnotesize{$I_2B_4$}};

\draw[thick,color=black, snake it] (X) to[bend left] (Y);

\end{scope}
\end{tikzpicture}
\end{center}  
\caption{Chord diagram representing $Tr_{\Lambda}(M^\dagger H M H^3 )$}
\label{Chord diagram representing two point function moment }
\end{figure}
We have that the moments is,
\begin{equation}
\begin{split}
 \tilde m_{k,k_1} & = \sum_{CD} \frac{{\cal N}^N \tilde {\cal J}^2 }{  {N \choose p}^k {N \choose \tilde p}  M^{{pk\over 2}+ {\tilde{p}\over 2}} } \sum_{\substack{I_1..I_{\frac{k}{2}}\tilde I\\ J_1..J_{\frac{k}{2}}\tilde J}} S^{A_1 B_1}..S^{A_k B_k} S^{\tilde A \tilde B} 
 \begin{minipage}{3.5 cm}
 \begin{center}
\begin{tikzpicture}
\begin{scope}[thin, every node/.style={sloped,allow upside down}]
\draw[thick] (0,0) circle(1.75 cm);
\coordinate (A) at (1.7234,0.303); 
\coordinate (A') at (2.1434,0.303);
\coordinate (B) at (0.598,-1.644); 
\coordinate (B') at (0.698,-1.944);
\coordinate (C) at (1.5155,0.875); 
\coordinate (C') at (1.9655,0.875); 
\coordinate (D) at (0.0,-1.75); 
\coordinate (D') at (0.0,-2); 
\coordinate (E) at (0.304,1.723); 
\coordinate (E') at (0.394,1.9); 
\coordinate (F) at (1.5155,-0.875); 
\coordinate (F') at (2,-0.995); 
\coordinate (G) at (0.875,1.5155);
\coordinate (G') at (1.085,1.7355);
\coordinate (H) at (1.7234,-0.303);
\coordinate (H') at (2.1634,-0.303);
\coordinate (I) at (-1.5155,0.875); 
\coordinate (I') at (-1.9655,0.875); 
\coordinate (J) at (-1.125,-1.34); 
\coordinate (J') at (-1.175,-1.64); 
\coordinate (K) at (-1.7234,0.303); 
\coordinate (K') at (-2.1234,0.303); 
\coordinate (L) at (-1.5155,-0.875); 
\coordinate (L') at (-1.8955,-0.955); 
\coordinate (I'') at (-1.3788,1.157); 
\coordinate (E'') at (-0.31256,1.7726); 
\coordinate (X) at (-1.125,1.34);
\coordinate (X') at (-1.48,1.44);
\coordinate (Y) at (-0.598,1.644);
\draw[thick,color=black] (A) to[bend right] (B);
\node[black] at (A') {\footnotesize{$J_3A_5$}};
\node[black] at (B') {\footnotesize{$J_3B_3$}};
\draw[black](0.76,-0.8) -- node {\midarrow} (0.797,-0.7);
\draw[thick,color=black] (C) to[bend right] (D);
\node[black] at (C') {\footnotesize{$I_3B_5$}};
\node[black] at (D') {\footnotesize{$I_3A_3$}};
\draw[black] (0.197,-0.6)-- node {\midarrow} (0.16,-0.7);
\draw[thick,color=black,snake it] (X) to[bend left] (F);
\node[black] at (X') {\footnotesize{$\tilde{I}\tilde{A}$}};
\node[black] at (F') {\footnotesize{$\tilde{I}\tilde{B}$}};
\draw[black] (0.392,0.7)-- node {\midarrow} (0.33,0.8);
\draw[thick,color=black] (I) to[bend left] (J);
\node[black] at (I') {\footnotesize{$I_1A_1$}};
\node[black] at (J') {\footnotesize{$I_1B_2$}};
\draw[black] (-1,-0.2)-- node {\midarrow} (-1.025,-0.1);
\draw[thick,color=black] (K) to[bend left] (L);
\node[black] at (K') {\footnotesize{$J_1B_1$}};
\node[black] at (L') {\footnotesize{$J_1A_2$}};
\filldraw[gray](Y) circle(5pt);
\draw[black](-1.46,-0.15) -- node {\midarrow} (-1.4517,-0.25);
\draw[loosely dotted,thick,color=black] (-0.34729635533,1.96961550602) arc (100:40:2);
\end{scope}
\end{tikzpicture}
\end{center}
\end{minipage}
      \\
     & \equiv \sum_{CD} \ \tilde m_{k,k_1}^{(CD)}  
\end{split}
\end{equation}
To proceed, we retrace the steps we followed for the case of $m_k$, the moments of the Hamiltonian. The factorization of $Tr_\Lambda$ into product of traces $tr_{\Lambda_i}$ for each colour $i$ proceeds just like before. For an intersecting pair of H chords we continue to have $(-1)^{p_I p_J - p_{IJ}}$ as before, whereas for an intersecting pair of H chord $I$ and M chord $\tilde J$, we have $(-1)^{p_I p_{\tilde J} - p_{I\tilde J}}$. The calculation of index sums are now a bit more involved. 
\paragraph{The case of no overlap between index sets}
In this case, apart from the  factor eq(\ref{eq:Non Overlap Contribution}) corresponding to the H chords, there is the extra contribution of M chords. The probability that the index set $\tilde J$ of the M chord does not overlap with any one of the index set on the oriented chord is  $e^{-\tilde \lambda}$ - see eqn(\ref{independent}). Now the  contribution to the of trace from the M chord is
\begin{equation}\label{x tilde definition}
   S^{\tilde a\tilde b} tr_{\Lambda} (\chi^\dagger_{\tilde a} \chi_{\tilde b}) \equiv \tilde x \sqrt{M}
\end{equation}
from a given colour.  Thus the total contribution to $\tilde m_k$ from the case of no overlap between any of the index sets is given by 
\begin{equation}
     e^{-\lambda \binom{k}{2} - \tilde \lambda k } x^{pk} \tilde x^{\tilde p}
\end{equation}

\paragraph{The case of non-trivial overlaps between index sets}
Suppose there is some overlap between index set $\tilde J$ of the M chord and an index set $I$ of the normal oriented chord (which is part of a $H$ chord). Let us denote the trace contribution from the overlapping case to be $\tilde{x}'^3M^{3 \over 2}$. Note that this appears in lieu of $x^2 \tilde{x} M^{3 \over 2} $. In effect, we have the following factor for every overlapping index
\begin{equation}
   {  {\tilde{x}}^{'3} \over   x^{2} \tilde{x}  } 
\end{equation}
If there are $p_{I\tilde J}$ overlaps between the oriented chord and M chord then the sum over $p_{I \tilde J}$ evaluates to 
\begin{equation}
 \sum_{p_{I \tilde J} = 0}^\infty   {(\pm \tilde \lambda)^{p_{I \tilde J}} \over p_{I\tilde J} ! } \left(\tilde x'^3 \over x^2 \tilde x  \right) =
    e^{\pm \tilde{\lambda} \tilde{x}'^3  \over x^2 \tilde{x} }
\end{equation}
where $\pm$ sign depends on whether the oriented chord and the M chord are intersecting or not as before. What remains is to evaluate $\tilde x'$ for various cases, which is worked out in the Appendix \ref{Two point function appendix}.

The rules for determining the moments of the two point functions, $\tilde{m}_{k,k_1}$ are thus similar to those involved in the determination of moments of the Hamiltonian and can be summarised as follows,
\begin{itemize}
    \item Draw all possible chord diagrams with $k$ nodes and an M chord, with the M chord enclosing $k_1$ nodes.
    \item For a given chord diagram, the contribution $m_{k,k_1}^{(CD)}$ is ${\cal N}^Nx^{pk}e^{-\lambda {k \choose 2}}e^{k\lambda_H \over 2} \tilde{x}^{\tilde p}$ (these parameters are defined in equations \ref{N definition}, \ref{eq:No overlap contribution}, \ref{lambda_H} and \ref{x tilde definition}) times a factor associated with all possible $H-H$ (summarised in section(\ref{Final summary}) ) and $H-M$ chord pair configurations. The contributions due to $H-M$ chord are listed below, 
\begin{center}
\begin{tabular}{|l|c|}
    \hline
  {\bf \hspace{20mm} Chord configuration} & {\bf Contribution}\\
  \hline \hline 
  Non intersecting chord pair; non nested:\quad\begin{minipage}{3.5cm}
    \begin{center} \begin{tikzpicture}
    \begin{scope}[thin, every node/.style={sloped,allow upside down}]
 \node[black] at (3.25,0.9) {};
 \draw[thick,snake it] (-0.5,0) arc (0:180:0.75 cm);
 \draw[ultra thick] (1.5,0) arc (0:180:0.75cm);
 \draw[black] (-1.15,0.77)-- node {\midarrow} (-1.245,0.775);
 \node[black] at (3.25,-0.15) {};
 \end{scope}
 \end{tikzpicture}
 \end{center}
 \end{minipage} & $e^{\gamma_1}$ \\
 \hline 
 
 Non intersecting chord pair;non nested:\quad \begin{minipage}{3.5cm}
    \begin{center} \begin{tikzpicture}
    \begin{scope}[thin, every node/.style={sloped,allow upside down}]
 \node[black] at (3.25,0.9) {};
 \draw[ultra thick] (-0.5,0) arc (0:180:0.75 cm);
 \draw[thick,snake it] (1.5,0) arc (0:180:0.75cm);
 \draw[black] (0.85,0.77)-- node {\midarrow} (0.845,0.77);
 \node[black] at (3.25,-0.15) {};
 \end{scope}
 \end{tikzpicture}
 \end{center}
 \end{minipage} & $e^{\gamma_2}$ \\
 \hline 
 Non intersecting chord pair; nested:\quad \qquad  \begin{minipage}{4 cm}
    \begin{center} \begin{tikzpicture}
    \begin{scope}[thin, every node/.style={sloped,allow upside down}]
  \node[black] at (3.25,1) {};
 \draw[thick,snake it] (1.85,0) arc (0:180:0.85 cm);
 \draw[black] (0.9,0.85)-- node {\midarrow} (0.85,0.85);
 \draw[ultra thick] (1.5,0) arc (0:180:0.5cm);
 \node[black] at (3.25,-0.15) {};
 \end{scope}
  \end{tikzpicture}
 \end{center}
 \end{minipage}
 &  $e^{\gamma_3}$\\
 \hline
 Non intersecting chord pair; nested: \quad \qquad   \begin{minipage}{3.5cm}
    \begin{center} \begin{tikzpicture}
    \begin{scope}[thin, every node/.style={sloped,allow upside down}]
 \node[black] at (3.25,1.15) {};
 \draw[ultra thick] (2,0) arc (0:180:1cm);
 \draw[thick,snake it] (1.5,0) arc (0:180:0.5cm);
 \draw[black] (0.975,0.6)-- node {\midarrow} (0.925,0.6);
  \node[black] at (3.5,-0.15) {};
  \end{scope}
   \end{tikzpicture}
 \end{center}
 \end{minipage}
 &  $e^{\gamma_4}$\\
 \hline
 Intersecting chord pair :\hspace{30mm} \begin{minipage}{4cm}
    \begin{center} \begin{tikzpicture}
    \begin{scope}[thin, every node/.style={sloped,allow upside down}]
    \node[black] at (3.25,0.9) {};
 \draw[thick,snake it] (2,0) arc (0:180:0.75cm);
 \draw[black] (1.35,0.8)-- node {\midarrow} (1.3,0.8);
 \draw[ultra thick] (2.75,0) arc (0:180:0.7 cm);
  \node[black] at (3.5,-0.15) {};
  \end{scope}
  \end{tikzpicture}
 \end{center}
 \end{minipage}     & $e^{\gamma_5}$  \\
  \hline 
  Intersecting chord pair : \hspace{30mm} \begin{minipage}{4cm}
    \begin{center} \begin{tikzpicture}
    \begin{scope}[thin, every node/.style={sloped,allow upside down}]
 \node[black] at (3.25,0.9) {};
 \draw[ultra thick] (2,0) arc (0:180:0.75cm);
 \draw[thick,snake it] (2.75,0) arc (0:180:0.7 cm);
 \draw[black] (2,0.8)-- node {\midarrow} (1.95,0.8);
  \node[black] at (3.5,-0.15) {};
  \end{scope}
  \end{tikzpicture}
 \end{center}
 \end{minipage}     & $e^{\gamma_6}$  \\
  \hline 
\end{tabular}
\end{center}

where $\gamma_1,\dots \gamma_6$ are listed in appendix(\ref{Two point function appendix}).
    
\end{itemize}

\section{Explicit results for SYK models with classical group symmetries}\label{sec:Special cases}
As mentioned earlier, variants of SYK model with specific global symmetries can be seen as special cases of the model we consider in this work. We now consider the special case of classical groups, specifically $SO(M)$, $U(M)$ and $USp(M/2)$ groups with fermions transforming in the fundamental in each case and calculate the parameters relevant for the evaluation of the chord diagram and two point function using the rules given in section (\ref{Final summary}) and (\ref{two point function}). More details of the computation is given in Appendix (\ref{appendix: classical group simplification}). 

\subsection{$SO(M)$ SYK}
The $SO(M)$ symmetric SYK model was discussed briefly in section(\ref{sec:The Model}). The model involves Majorana fermions which are real and obey appropriate commutation relations, namely
\begin{equation}\label{eq: SO(M) commutation relations}
\begin{split}
    \mbox{reality condition}\quad  &: {\chi_{a}}^\dagger= \chi_{a}\\ 
    \mbox{commutation relations} \quad & : \{\chi_{a},\chi_{b}\} =2\delta_{ab}
\end{split}
\end{equation}
 The rank $r$ of $SO(M)$ is given by the integer part of $\frac{M}{2}$ denoted by $r=[\frac{M}{2}]$ and the cartan charges are given by $Q_{\alpha}={ i \over 4} \left(  \chi_{2\alpha-1} \chi_{2\alpha} -  \chi_{2\alpha} \chi_{2\alpha-1} \right)$.  
The parameters arising in the chord diagram evaluation are traces over products of fermions. To compute, we will choose the following representation of the algebra (\ref{eq: SO(M) commutation relations}),\begin{equation}
    \begin{split}
        \chi_{1} & =  \sigma_1 \otimes {\mathbb I}  \otimes {\mathbb I} \otimes \cdots \otimes {\mathbb I}\\
        \chi_{2} & =  \sigma_2 \otimes {\mathbb I}  \otimes {\mathbb I} \otimes \cdots \otimes {\mathbb I}\\
        \chi_{3} & =  \sigma_3 \otimes \sigma_1 \otimes {\mathbb I} \otimes \cdots \otimes {\mathbb I}\\
        \chi_{4} & =  \sigma_3 \otimes \sigma_2 \otimes {\mathbb I} \otimes \cdots \otimes {\mathbb I}\\
        \vdots & \\
        \chi_{2[{M \over 2}]} & = \underbrace{ \sigma_3 \otimes \sigma_3 \otimes \sigma_3 \otimes \cdots \otimes \sigma_2}_{[{M \over 2}] \text{ factors}}\\
    \end{split}
\end{equation}
For odd $M$, we further have that $\chi_{M}  = \underbrace{   \sigma_3 \otimes \sigma_3 \otimes \sigma_3 \otimes \cdots \otimes \sigma_3}_{[{M \over 2}] \text{ factors}}$.  The invariant tensor of $SO(M)$ is  $S^{ab} =\delta^{ab}$.
For the $SO(M)$ case, various parameters relevant for the evaluation of chord diagram listed out in section (\ref{Final summary}), can be evaluated and is given below (see Appendix \ref{appendix: classical group simplification} for details), 
\begin{equation}\label{SO(M) m parameters}
    \begin{split}
        {\cal N} &= \prod\limits_{\alpha=1}^r(2\textrm{cosh}(\frac{\mu_{\alpha}}{2})) \\
        x^2 M &=M-2\sum\limits_{\alpha=1}^r \tanh^2(\mu_{\alpha}/2)\\
        \lambda_H &= \lambda+\frac{4\lambda}{x^4M^2}\sum\limits_{\alpha=1}^r \textrm{sech}^2(\mu_\alpha/2)\tanh^2(\mu_\alpha/2) \\    \lambda_{N\!N} & = 4\lambda_H \\
        \lambda_N &= 0\\
  \lambda_C &= -\frac{8\lambda }{x^2M}
    \end{split}
\end{equation}
The parameters relevant for the computation of two point functions outlined in section (\ref{two point function}) are given below 
\begin{equation}
    \begin{split}
        \tilde{x} &= \sqrt{M}\\
        \gamma_1 & = \gamma_2 = \gamma_4  = 2 \tilde \lambda \\
       \gamma_3 &= \frac{8\tilde{\lambda}}{x^2M}- \frac{8\tilde{\lambda}}{M} + 2\tilde{\lambda} \\
       \gamma_5 &= \gamma_6= - \frac{4\tilde{\lambda}}{M} + 2\tilde{\lambda}
    \end{split}
\end{equation}

 \subsection{$U(M)$ SYK}
 The $U(M)$ symmetric SYK model involves complex fermions $\chi_{a}$ and satisfy the following commutation relations,
\begin{equation}
\{\chi_{a},\chi_{b}^{\dagger}\} =\delta_{ab},\qquad 
\{\chi_{a}, \chi_{b}\}=\{\chi^{\dagger}_{a}, \chi_{b}^{\dagger}\}=0
\end{equation}

The rank $r$ of $U(M)$ is given by $r = M$ and the cartan charges are given by $Q_{\alpha} = { 1 \over 2}\left(  \chi^\dagger_{\alpha} \chi_{\alpha} -  \chi_{\alpha} \chi^\dagger_{\alpha} \right)$. The representation corresponding to the above algebra is given by,

\begin{equation}
    \begin{split}
        \chi_{1} & =  \sigma_+ \otimes {\mathbb I}  \otimes {\mathbb I} \otimes \cdots \otimes {\mathbb I}\\
        \chi_{2} & =  \sigma_3 \otimes \sigma_+ \otimes {\mathbb I} \otimes \cdots \otimes {\mathbb I}\\
        & \vdots\\
        \chi_{M} & = \sigma_3 \otimes \sigma_3 \otimes \sigma_3 \otimes \cdots \otimes \sigma_+\\
    \end{split}
\end{equation}
$\chi_a^\dagger$ has the same structure, except that  $\sigma_+$ replaced by $\sigma_-$. The contributions due to various possible chord structures in the $U(M)$ are as in section (\ref{Final summary}), with the parameters given below (see Appendix \ref{appendix: classical group simplification} for details), 

\begin{equation}\label{U(M) m parameters}
    \begin{split}
     {\cal N} &= \prod\limits_{\alpha=1}^r 2\textrm{cosh}\mu_{\alpha}\\
        x^2M  &= \sum\limits_{\alpha=1}^r \textrm{sech}^2\mu_\alpha\\
       \lambda_H &=\lambda + \frac{\lambda}{x^4M^2}\sum\limits_{\alpha=1}^r e^{-2\mu_{\alpha}} \textrm{sech}^4\mu_\alpha\\
        \lambda_{N\!N} &=4\lambda +  \frac{4\lambda}{x^4M^2}\sum\limits_{\alpha=1}^r\tanh^2\mu_\alpha \textrm{sech}^2\mu_\alpha \\
        \lambda_N &= 0 \\
       \lambda_C &= - \frac{4\lambda}{x^2 M}
    \end{split}
\end{equation}

This result was already obtained in \cite{Berkooz2020ComplexSM}. The parameters relevant for the computation of two point functions outlined in section (\ref{two point function}) are given below
\begin{equation}
    \begin{split}
        \tilde{x}\sqrt{M} &= \sum\limits_{\alpha=1}^r e^{-\mu_{\alpha}} \textrm{sech}\mu_{\alpha}\\
        \gamma_1 &= \gamma_2= \gamma_4=  2\tilde{\lambda}+\frac{2\tilde{\lambda}}{x^2\tilde{x}M^{\frac{3}{2}}}\sum\limits_{\alpha=1}^r \textrm{sech}^2\mu_\alpha \tanh \mu_\alpha\\
        \gamma_3 &= \tilde{2\lambda} -\frac{2 \tilde{\lambda}}{x^2 \tilde{x}M^{\frac{3}{2}}}\sum\limits_{\alpha=1}^r e^{-2\mu_\alpha}\textrm{sech}^2\mu_\alpha \tanh \mu_\alpha \\
         \gamma_5 &= \gamma_6= 2\tilde{\lambda} - \frac{2\tilde{\lambda}}{x^2 \tilde{x}M^{\frac{3}{2}}} \sum\limits_{\alpha=1}^r e^{-\mu_\alpha} \textrm{sech}^3\mu_\alpha 
\end{split}
\end{equation}

\subsection{$USp(M/2)$ SYK}
The case of $USp(M/2)$ is illustrative of the fact that our formalism can apply to fermions transforming not necessarily as fundamental. In the present case, we take two copies of fermions in fundamental representation $USp(M/2)$. Namely, we will split the fermion flavour index $a = (\bar{a},  \underline{a})$,  where $\bar{a}  \in 1,\dots {M \over 2}$ and $  \underline{a}  \in 1,2 $. We will take ${M \over 2}$ to be even in what follows. In this subsection, we will use explicit summations and not use summation convention. The reality condition is given by ${\chi_{i\bar a   \underline{a}}}^\dagger =\sum_{\bar b, \underline{b} } \Omega_{\bar a \bar b} \epsilon_{   \underline{a}   \underline{b}} \chi_{i \bar b  \underline{b}}$, where $\epsilon$ is the Levi-Civita symbol with $\epsilon_{01}=1$ and  
\begin{equation}\label{eq:Symp Invariant}
   \Omega  =  \begin{pmatrix}
    0_{\frac{M}{4} \times \frac{M}{4}} & {\mathbb I}_{\frac{M}{4}\times \frac{M}{4}} \\
    -{\mathbb I}_{\frac{M}{4}\times \frac{M}{4}} & 0_{\frac{M}{4}\times \frac{M}{4}} 
    \end{pmatrix}
\end{equation}

The commutation relations are now given by,
\begin{equation}\label{eq:Algebra of Sp fermions}
\begin{split}
    \{ \chi^\dagger_{\bar a  \underline{a} } , \chi_{\bar b   \underline{b} }  \} & = \delta_{\bar a \bar b} \delta_{ \underline{a}  \underline{b}} \\
    \{ \chi_{\bar a  \underline{a} } , \chi_{\bar b  \underline{b} }  \} & = \Omega_{\bar a \bar b} \epsilon_{ \underline{a}  \underline{b}} 
\end{split}
\end{equation}

The rank is given by $r={M \over 4}$ and the cartans are given by  $Q_{\alpha} = {1 \over 4} \sum\limits_{\bar c, \underline{a}\underline{b}}\epsilon_{\underline{a}  \underline{b}} \Omega_{\alpha \bar c}  \left( \chi_{\bar c  \underline{a}} \chi_{\alpha  \underline{b} } - \chi_{\alpha  \underline{b} } \chi_{\bar c  \underline{a}}  \right)$. The representation corresponding to the above algebra is given by (for $1 \le \alpha \le {M \over 4}$),
\begin{equation}
    \begin{split}
        \chi_{\alpha, 1} & = \overbrace{\underbrace{ \sigma_3 \otimes \cdots \otimes \sigma_3  }_{\alpha-1\  \text{times}}\otimes \sigma_+ \otimes  {\mathbb I} \otimes \cdots}^{\frac{M}{2} \text{factors}} \\
         \chi_{\alpha + {M \over 4} , 2} & = \underbrace{ \sigma_3 \otimes \cdots \otimes \sigma_3  }_{\alpha-1\  \text{times}}\otimes \sigma_- \otimes  {\mathbb I} \otimes \cdots \\
         \chi_{\alpha , 2} & = \underbrace{ \sigma_3 \otimes \cdots\cdots \cdots \cdots \otimes \sigma_3  }_{{M \over 4}  + \alpha-1\  \text{times}}\otimes \sigma_- \otimes  {\mathbb I} \otimes \cdots \\
         \chi_{\alpha + {M \over 4} , 1} & = \underbrace{ \sigma_3 \otimes \cdots\cdots\cdots \cdots  \otimes \sigma_3  }_{{M \over 4} + \alpha-1\  \text{times}}\otimes (-\sigma_+) \otimes  {\mathbb I} \otimes \cdots \\
    \end{split}
\end{equation}

The invariant tensor in this case is $S^{ab} = \delta_{\bar a \bar b} \epsilon_{ \underline{a}  \underline{b}}$, since the object $\sum_{\bar a \underline{a} \underline{b} }\chi^{\dagger}_{\bar a   \underline{a}} \chi_{\bar a   \underline{b}} \epsilon_{ \underline{a}   \underline{b}}$ is invariant under $USp(M/2)$ symmetry. The contributions due to various possible chord structures in the $USp(M/2)$ are as in section (\ref{Final summary}), with the parameters given below (see Appendix \ref{appendix: classical group simplification} for details), 
\begin{equation}\label{USp(M/2) m parameters}
    \begin{split}
     x^2  M &=-4\sum\limits_{\alpha=1}^r \textrm{sech}^2 \frac{\mu_{\alpha}}{2} \\
     {\cal N} &= \prod\limits_{\alpha=1}^r 4\textrm{cosh}^2(\frac{\mu_{\alpha}}{2})\\
     \lambda_H &=  \lambda+\frac{8\lambda}{x^4M^2}\sum\limits_{\alpha=1}^r \textrm{sech}^2\frac{\mu_{\alpha}}{2}\tanh^2\frac{\mu_{\alpha}}{2}\\
     \lambda_{N\!N} &= 4\lambda_H\\
     \lambda_N &= 0 \\
    \lambda_C &= \frac{8\lambda}{x^2M}
    \end{split}
    \end{equation}
The two point function defined in section \ref{two point function} vanishes or $USp(M/2)$ case. This happens because the parameter $\tilde{x}$ defined in eq(\ref{x tilde definition}) (this is the factor associated with the case of zero overlap between index sets of $M$ chord and any of the H-chords) vanishes. 
 
Since in the double scaling limit, there will always be some non-overlapping indices between $H$ chords and $M$ chords, the two point function vanishes identically. This means that for the case of $USp(M/2)$ one needs to consider more general class of operators for instance
\begin{equation}
    M_{\tilde A \tilde B} = \tilde J_{\tilde I \tilde J} X^{\dagger}_{\tilde I \tilde A} X_{\tilde J \tilde B}
\end{equation} 
We leave the study of two point functions of such operators for future work.

\section{The Partition Function}\label{sec:Partion function}

We have now obtained the diagrammatic rules to evaluate a given chord diagram contributing to the partition function or two point function. In this section we use these rules to evaluate the partition function. Since the spectrum of the transfer matrix is not known generically (see section \ref{sec:Transfer matrix}), we restrict ourselves to the case of SYK models based on classical groups. The explicit parameters relevant for the evaluation of the moments for these models were given in section \ref{sec:Special cases}. All these models have the property that the parameter $\lambda_N = 0$ which gives rise to transfer matrices with known spectral properties.

For every chord diagram with $k$ Hamiltonian insertions suppose there are $\kappa_{C}$ pairs of intersecting H chords and $\kappa_N$ pairs of nested chords.  Then from the results in section \ref{Final summary} we get,

\begin{equation}\label{moments}
\begin{split}
m_k =& {\cal N}^N x^{pk} e^{-\lambda {k\choose 2}} e^{k\lambda_H \over 2} e^{\lambda_{N\!N}{k/2 \choose 2}} \sum\limits_{CD} e^{\kappa_C \lambda_C} e^{\kappa_N \lambda_N} \\
= & {\cal N}^N \ x^{pk} \  e^{\frac{k^2}{2}({\frac{\lambda_{N\!N}}{4}-\lambda})} e^{\frac{k}{2}({\lambda +\lambda_H-{ \lambda_{N\!N} \over 2} })} \sum\limits_{CD} e^{\kappa_C \lambda_C}
\end{split}
\end{equation}
where in going to the second line, we have used the fact that $\lambda_N=0$ for all the classical groups. This is the same upto factors as the Majorana SYK model - see eq(\ref{Majorana result}). Defining the parameters,

\begin{equation}
\begin{split}
     q_g & \equiv q_g(\{\mu_\alpha\})=e^{\lambda_C}\\
     h &\equiv h(\{\mu_\alpha\})= e^{\lambda_{NN} -4 \lambda \over 8}\\
    f &\equiv f(\{\mu_\alpha\})=e^{\frac{\lambda}{2}-\frac{\lambda_{N\!N}}{4}+\frac{\lambda_H}{2}}\\
\end{split}
\end{equation}
we get a result analogous to the Majorana SYK, namely 
\begin{equation}\label{eq:final m_k}
    m_k = {\cal N}^N h^{k^2}    \int_{0}^{\pi} \frac{d\theta}{2\pi}(q_g,e^{\pm 2i\theta};q_g)_{\infty}\left(\frac{2 f x^p \textrm{cos}\theta}{\sqrt{1-q_g}}\right)^k
\end{equation}
The partition function is given by summing over the moments i.e, 
\begin{equation}
    Z(\beta,\{\mu_\alpha\})=\sum\limits_{k=0}^{\infty}\frac{(-\beta)^k}{k!}m_k(\{\mu_\alpha\})
\end{equation}
The evaluation of this sum, however has an issue, since due to the $h^{k^2}$ factor, the sum does not converge. This issue was encountered before in the case of complex SYK with chemical potential \cite{Berkooz2020ComplexSM}. We leave a deeper understanding of this issue for future work. For now, we just choose parameters such that the factor $h=1$, i.e $\lambda_{NN} = 4 \lambda$. However, the only way of ensuring $\lambda_{NN} = 4 \lambda$ is by turning off all the chemical potentials to vanish (see eq(\ref{SO(M) m parameters}) for $SO(M)$ case, eq(\ref{U(M) m parameters}) for $U(M)$ case and eq(\ref{USp(M/2) m parameters}) for $USp(M/2)$ case) i.e $\mu_\alpha = 0,\forall \alpha=1,\dots r$. Also note that $x^p=1$ for all classical groups when $\mu_\alpha=0$. The final expression for the partition function is 
\begin{equation}
    Z(\beta) = {\cal N}^N \int_{0}^{\pi} \frac{d\theta}{2\pi}(q_g,e^{\pm 2i\theta};q_g)_{\infty} e^{-\beta E(\theta)},\hspace{20mm} E(\theta) \equiv \frac{2 f  \textrm{cos}\theta}{\sqrt{1-q_g}} 
\end{equation}
For the case of $SO(M)$ and $USp({M \over 2})$, we get $q_g = e^{-8 \lambda \over M},f=1$ and for $U(M)$ we get $q_g = e^{-4 \lambda \over M},f={\lambda \over 2 M}$. 

The only signature of classical groups in the final result (as compared to Majorana SYK) seems to be rescaling of energies (by the factor $f$) and a redefinition of $q$ to $q_g$. The rich parameter set of the SYK model with generic global symmetries, namely the chemical potentials disappears once we demand that $h=1$ in eq(\ref{eq:final m_k}). We leave it for future work, to see if there are other groups or representations where the $h=1$ condition is not so restrictive.

\subsection{Partition function at fixed charges}\label{sec:Partition function at fixed charges}
Although the partition function (almost) trivializes since we turned off the chemical potential, it turns out that the partition function at fixed charges exists and makes sense for all charges - this was also seen for the case of complex SYK case in \cite{Berkooz2020ComplexSM}. In this subsection, we compute the partition function at fixed charges $n_\alpha$ where $\alpha = 1, \dots r$ where $r$ is the rank of the symmetry group. 

The canonical partition function at fixed charges can be evaluated by integrating over (analytical continuation of) chemical potential.  More explicitly, defining $z_{\alpha}=e^{\mu_\alpha}=e^{i\chi_\alpha}$. 
The partition function for fixed charges $n_\alpha$ denoted by $Z(\beta ; \{ n_\alpha  \})$ is then given by,
\begin{equation}
     Z(\beta;\{ n_\alpha  \}) = \frac{1}{2\pi i}\oint_{\text{unit circle}} \prod_{\alpha} dz_{\alpha} \frac{Z(\beta,\{ z_{\alpha} \} )}{z^{n_\alpha +1}} 
\end{equation}
Expanding the partition function in $\beta$ and  plugging in the moments we obtained in eq(\ref{eq:final m_k}) we get,
\begin{equation}
\begin{split}
    Z(\beta;\{ n_\alpha  \})&= \sum\limits_{k=0}^{\infty} \frac{(-\beta)^k}{2\pi k!} \int\limits_{-\pi}^{\pi} d\chi_\alpha C_k h^{k^2 }\int\limits_{0}^{\pi}\frac{d\theta}{2\pi}\left(\frac{2 f \cos\theta}{\sqrt{1-q_g}}\right)^k (q_g,e^{\pm 2i\theta};q_g)_{\infty}
\end{split}
\end{equation}

\noindent where $C_k = \textrm{exp}\left(-i\sum\limits_{\alpha}n_\alpha \chi_\alpha + N \log {\cal N} + pk \log |x| \right)$. Assuming that the charges $n_\alpha$ scale as $N$, we work with ${\cal Q}_\alpha \equiv \frac{n_\alpha}{N}$ which is held fixed in the double scaling limit. The integral above is then evaluated at large $N$ using saddle point approximation. 

The explicit evaluation of the integral is complicated. We outline the calculation for $U(M)$ SYK model below and summarize the results for other groups later.  In terms of $\mu_\alpha$ the function $C_k$ corresponding to the $U(M)$ symmetric SYK model,
\begin{equation}
    C_k=\textrm{exp}\left[-\sum \limits_{\alpha =1}^M N{\cal Q}_{\alpha }\mu _{\alpha }+ N\sum \limits_{\alpha =1}^M \log \left(2\cosh \mu _{\alpha }\right) +\frac{1}{2} k \sqrt{\lambda N}  \log \sum \limits_{\alpha =1}^M \frac{\text{sech}^2\mu _{\alpha }}{M} \right]
\end{equation}
The saddle value of $\mu_\alpha$ is given by,
\begin{equation}
\begin{split}
   n_\alpha \coth \mu _{\alpha } &= N+\frac{2 k^2 \lambda  \left(1- {\cal Q}_{\alpha}^2 \right) \left(M {\cal{Q}}_{\alpha}^2-\left( {\cal{Q}}_{\alpha}^2 +1\right) \sum \limits_{\alpha =1}^M {\cal{Q}}_{\alpha}^2+ \sum \limits_{\alpha =1}^M {\cal{Q}}_{\alpha}^4\right)}{ \left(\sum \limits_{\alpha =1}^M \left(1- {\cal{Q}}_{\alpha}^2\right)\right)^3}  \\
   & \hspace{8cm}-\frac{k \sqrt{\lambda N } \left(1-{\cal{Q}}_{\alpha}^2\right)}{ \sum \limits_{\alpha =1}^M \left(1-{\cal{Q}}_{\alpha}^2\right)}
\end{split}
\end{equation}
The functions $C_k$,$f$, $h$ and $q_g$ are then evaluated at the saddle point. The term with $k^2$ power in $C_k$ exactly cancels with $h$ resulting in the partition function having only terms that are linear in $k$ i.e
\begin{equation}
    \begin{split}
&C_k h^{k^2}|_{\text{saddle}}=\textrm{exp}\left[\frac{\sqrt{\lambda N} k }{2}\log \sum \limits_{\alpha =1}^M \frac{N_{\alpha}}{M} -\frac{N}{2} \sum \limits_{\alpha =1}^M \log (N_\alpha /4)
+N\sum \limits_{\alpha =1}^M  {\cal Q}_\alpha\tanh ^{-1}\left( {\cal{Q}}_{\alpha}\right) \right]\\
\end{split}
\end{equation}
where $N_{\alpha}=1-{\cal{Q}}_{\alpha}^2$. Since now that we have only linear terms in $k$ in the exponent, the partition function at fixed charge converges and takes the following form,
\begin{equation}\label{eq:Form of tyhe partition function at fixed charge}
    Z(\beta,{\cal Q})= e^{S_0}\int\limits_{0}^{2\pi} d\theta (q_g({\cal Q}),e^{\pm 2i\theta};q_g({\cal Q}))_{\infty} e^{-\beta E(\theta,{\cal Q})}
\end{equation}
In the above equation, we have denoted the set $\{ {\cal Q}_\alpha\}$ by ${\cal Q}$. The quantities $S_0$, $E(\theta,{\cal Q})$ and $q_g({\cal Q})$ corresponding to the complex SYK model is given in eqn(\ref{U(M) result}). Note that the above result is similar to that of the usual Majorana SYK with zero chemical potential given by eqn(\ref{Majorana Final}) but has a rescaling in the range of energies which is now given by $E(\theta,{\cal Q})$ and the quantity $S_0$ corresponds to the number of states in a fixed charge sector. We also note that the quantity $q$ which corresponds to contributions from each intersection for Majorana SYK model is now replaced by $q_g({\cal Q})$. 
 
 We can repeat the above calculation for all classical groups. The result is again of the form eq(\ref{eq:Form of tyhe partition function at fixed charge}). The specific values of $S_0$, $q_g({\cal Q})$ and $E(\theta,{\cal Q})$ for each model are specified below,
\begin{itemize}
    \item U(M) SYK:
    \begin{equation}\label{U(M) result}
        \begin{split}
        &S_0 = \log \frac{2^{NM}}{\sqrt{2\pi N N_\alpha}}-\frac{N}{2}\sum\limits_\alpha [\log(N_\alpha ) + 2{\cal Q}_{\alpha} \tanh^{-1}({\cal{Q}}_{\alpha})]\\
        &q_g({\cal Q})=\textrm{exp}\left[\frac{-4\lambda }{\sum\limits_\alpha N_\alpha}\right]\\
        &E(\theta,{\cal Q})=\frac{2\cos\theta}{\sqrt{1-q_g({\cal Q})}}\textrm{exp}\left(\frac{\sqrt{\lambda N}}{2}\log\left(\sum\limits_{\alpha}\frac{N_\alpha}{M}\right)+ \frac{\lambda \sum\limits_{\alpha} N_\alpha(N_\alpha -2{\cal Q}_\alpha)}{2(\sum\limits_\alpha N_\alpha)^2}\right)\\
    \end{split}
    \end{equation}
    where $N_{\alpha}=1-{\cal{Q}}_{\alpha}^2$. While the results for $M>1$ are new, we note here that for $M=1$ this matches with the results in \cite{Berkooz2020ComplexSM}.

\item SO(M) SYK
\begin{equation}\label{SO(M) result}
   \begin{split}
            &S_0 = \log \frac{2^{\frac{N M+1}{2}}}{\sqrt{\pi N N_\alpha}}-\frac{N}{2} \sum\limits_{\alpha} \left(\log N_\alpha + 4{\cal{Q}}_{\alpha} \tanh^{-1}(2{\cal{Q}}_{\alpha})  \right)\\
        &q_g({\cal Q})=\textrm{exp}\left[\frac{-4\lambda}{\sum\limits_{\alpha}N_\alpha}\right]\\
        &E(\theta,{\cal Q})=\frac{2\cos\theta}{\sqrt{1-q_g({\cal Q})}}\textrm{exp}\left(\frac{\sqrt{\lambda N}}{2}\log\left(\frac{2}{M}\sum\limits_{\alpha}N_{\alpha}\right) - \frac{2\lambda}{(\sum\limits_{\alpha}N_\alpha)^2} \sum\limits_{\alpha} {\cal{Q}}_{\alpha} ^2 N_\alpha\right)
        \end{split}
    \end{equation}
    where $N_\alpha=1-4{\cal{Q}}_{\alpha} ^2$
\item USp(M/2)
\begin{equation}\label{USp(M/2) result}
    \begin{split}
        &S_0= \log \frac{2^{MN/2}}{\sqrt{\pi N N_\alpha}}-N\sum\limits_{\alpha} \log N_\alpha -2N\sum\limits_\alpha {\cal{Q}}_{\alpha}\tanh^{-1}({\cal{Q}}_{\alpha})\\
    &q_g({\cal Q})=\textrm{exp}\left[\frac{-2\lambda}{\sum\limits_\alpha N_\alpha}\right]\\
    &E(\theta,{\cal Q})=\frac{2\cos\theta}{\sqrt{1-q_g({\cal Q})}}\textrm{exp}\left(\frac{\sqrt{\lambda N}}{2} \log\sum\limits_{\alpha} \frac{4N_{\alpha}}{M}  - \frac{\lambda}{2} \frac{\sum\limits_{\alpha}{\cal{Q}}^2_{\alpha}N_{\alpha}}{(\sum\limits_{\alpha}N_\alpha)^2}\right)
    \end{split}
\end{equation}
where $N_\alpha = 1- {\cal{Q}}_{\alpha} ^2$
\end{itemize}

\section{Conclusions and Outlook}\label{sec:conclusions}

In this work we have studied the double scaling limit of SYK like models with global symmetries. We evaluate the moments of the partition function and two point function in the presence of chemical potential using the chord diagram techniques. Our treatment is agnostic to the details of the global symmetries in the sense that these details are encoded in a few parameters. We give explicit results for the cases of $USp(M/2), SO(M)$ and also filled in some gaps in the existing literature for the $U(M)$ case. 

\paragraph{Outlook and future directions}

On general grounds, the duals of the models we study here can be expected to be JT gravity models coupled to non-abelian gauge fields. It will be interesting to compute the partition function in the gravity side (in absence of chemical potential and also at fixed charge sector) and compare them to our results. 
In this context, we would also like to mention that (as has been noted before in the context of $U(1)$\cite{Berkooz2020ComplexSM}), the partition  function does not converge in the presence of chemical potential for arbitrary global symmetry. It will be interesting to see if there is an analogous statement in gravity. It is also of interest to see whether the $\lambda \rightarrow 0$ limit of our model corresponds to the large $q$ SYK\cite{Goel:2023svz}. If so, one should probably think of the double scaling limit always as a deformation of the SYK case. It would be interesting to see whether the techniques developed here can be applied to other models - for example as mentioned in the introduction, the double scaling limit of various the supersymmetric SYK models have been studied in \cite{Wenbo,Yoon2017SupersymmetricSM,Peng:2017spg,Narayan2018SupersymmetricSM}. It will be interesting to see if  one can organize their calculations in a single framework. If so, then one can also study the supersymmetric SYK models with global symmetries as in \cite{Narayan2018SupersymmetricSM}. An intriguing object we find in this work is the transfer matrix eq(\ref{eq:Transfer Matrix}). It will be interesting to  find the spectrum of the transfer matrix and explore it's consequences for the dual bulk theory. 

\section*{Acknowledgements} We would like to thank M. Berkooz, R. Loganayagam, P Pawar  and J. Yoon for interesting
discussions related to the project. PN
would like to acknowledge SERB grant MTR/2021/000145.

\appendix
\section{Disentangling chord diagrams}\label{Disentangling chord diagrams}

  In this Appenix we determine the negative signs resulting from the disentangling of chords. The case of non zero overlap can be determined by swapping all the constituent fermions across each other. Consider two fermions products $\Phi_{IA}^{R}$ and $\Phi_{JB}^{S}$ with $|\Phi_{IA}^{R}|=p_I$ and $|\Phi_{JB}^{S}|=p_J$. For  $|I \cap J|=0$, flipping these fermions across each other would give,
    \begin{equation}\label{zero overlap}
        \Phi_{IA}^{R} \Phi_{JB}^{S} = (-1)^{p_I p_J} \Phi_{JB}^{S} \Phi_{IA}^{R}
    \end{equation}
However, in the case of non zero overlap, the fermions with same color indices cannot be simply swapped across each other at cost of signs (since they have non trivial commutation relations, the result will not be just a sign). The most we can hope for is for all the fermions with no colour overlap swap across each other at cost of some sign. We will find that this is indeed the case and also determine the sign below. 

To determine the negative signs in this case, we go back to eq(\ref{zero overlap}) and break down factor of $(-1)^{p_I p_J}$ into following steps. We choose two fermions $\alpha_1$, $\alpha_2$ from the fermion product $\Phi_{IA}^{R}$ and two fermions each from each of the fermion products $\beta_1$ and $\beta_2$ from the fermion product $\Phi_{JB}^{S}$ and mark them by putting boxes around them. Note that although we are marking them, there is nothing special about them for now.
\begin{equation*}
        \Phi_{IA}^{R} \Phi_{JB}^{S} = \phi_{i_1 a_1}^{r_1}\dots\boxed{\phi_{i_{\alpha_1} a_{\alpha_1}}^{r_{\alpha_1}}}\dots\boxed{\phi_{i_{\alpha_2} a_{\alpha_2}}^{r_{\alpha_2}}}\dots \phi_{i_{p_I} a_{p_I}}^{r_{p_I}} \quad \phi_{j_1 b_1}^{s_1}\dots \boxed{\phi_{j_{\beta_1} b_{\beta_1}}^{s_{\beta_1}}} \dots \boxed{\phi_{j_{\beta_2} b_{\beta_2}}^{s_{\beta_2}}} \dots \phi_{j_{p_J} b_{p_J}}^{s_{p_J}}
    \end{equation*}
Now we break down the swapping of fermions to obtain eq(\ref{zero overlap}) into the following steps,
\begin{itemize}
    \item[1]  We first move all the fermions constituting the fermion products $\Phi_{IA}^{R}$, and $\Phi_{JB}^{S}$ except those at positions $\alpha_1$, $\alpha_2$, $\beta_1$ and $\beta_2$ (fermion strings of length $p_I-2$ and $p_J-2$) to the leftmost and rightmost edges (the empty boxes below represent the positions from where the marked fermions were removed), 
    \begin{equation*}
        \phi_{i_1 a_1}^{r_1}\dots\boxed{}\dots\boxed{}\dots \phi_{i_{p_I} a_{p_I}}^{r_{p_I}} \quad \phi_{i_{\alpha_1} a_{\alpha_1}}^{r_{\alpha_1}} \phi_{i_{\alpha_2} a_{\alpha_2}}^{r_{\alpha_2}}\phi_{j_{\beta_1} b_{\beta_1}}^{s_{\beta_1}}\phi_{j_{\beta_2} b_{\beta_2}}^{s_{\beta_2}}\quad \phi_{j_1 b_1}^{s_1}\dots \boxed{ } \dots \boxed{} \dots \phi_{j_{p_J} b_{p_J}}^{s_{p_J}}
    \end{equation*}
    \item[2.] The strings of fermions (except the fermions at $\alpha_1$, $\beta_1$, $\alpha_2$ and $\beta_2$) at both edges are now swapped to the opposite edges,
     \begin{equation*}
       \phi_{j_1 b_1}^{s_1}\dots \boxed{} \dots \boxed{} \dots \phi_{j_{p_J} b_{p_J}}^{s_{p_J}}\quad\phi_{i_{\alpha_1} a_{\alpha_1}}^{r_{\alpha_1}}\phi_{i_{\alpha_2} a_{\alpha_2}}^{r_{\alpha_2}}\phi_{j_{\beta_1} b_{\beta_1}}^{s_{\beta_1}}\phi_{j_{\beta_2} b_{\beta_2}}^{s_{\beta_2}}\quad \phi_{i_1 \alpha_1}^{r_1}\dots\boxed{}\dots\boxed{}\dots \phi_{i_{p_I} a_{p_I}}^{r_{p_I}}
    \end{equation*}
   \item[3.] Now we swap the marked fermions from each fermion product, $\phi_{i_{\alpha_1} a_{\alpha_1}}^{r_{\alpha_1}}$, $\phi_{i_{\alpha_2} a_{\alpha_2}}^{r_{\alpha_2}}$ and   $\phi_{j_{\beta_1} b_{\beta_1}}^{s_{\beta_1}}$,  $\phi_{j_{\beta_2} b_{\beta_2}}^{s_{\beta_2}}$ across each other such that,
   \begin{equation*}
        \phi_{j_1 b_1}^{s_1}\dots \boxed{} \dots \boxed{} \dots \phi_{j_{p_J} b_{p_J}}^{s_{p_J}}\quad\phi_{j_{\beta_1} b_{\beta_1}}^{s_{\beta_1}}\phi_{j_{\beta_2} b_{\beta_2}}^{s_{\beta_2}}\phi_{i_{\alpha_1} a_{\alpha_1}}^{r_{\alpha_1}}\phi_{i_{\alpha_2} a_{\alpha_2}}^{r_{\alpha_2}}\quad\phi_{i_1 \alpha_1}^{r_1}\dots\boxed{}\dots\boxed{}\dots \phi_{i_{p_I} a_{p_I}}^{r_{p_I}}
    \end{equation*}
    Note that the rearrangement of marked fermions amounts to additional negative signs, $(-1)^{2^2}$ 
    \item[4.] We now move $\phi_{i_{\alpha_1} a_{\alpha_1}}^{r_{\alpha_1}}$,   $\phi_{j_{\beta_1} b_{\beta_1}}^{s_{\beta_1}}$, $\phi_{i_{\alpha_2} a_{\alpha_2}}^{r_{\alpha_2}}$ and   $\phi_{j_{\beta_2} b_{\beta_2}}^{s_{\beta_2}}$ to their original positions such that we get, 
    \begin{equation*}
    \phi_{j_1 b_1}^{s_1}\dots\boxed{\phi_{j_{\beta_1} b_{\beta_1}}^{s_{\beta_1}}}\dots\boxed{\phi_{j_{\beta_2} b_{\beta_2}}^{s_{\beta_2}}}\dots \phi_{j_{p_J} b_{p_J}}^{s_{p_J}} \quad \phi_{i_1 a_1}^{r_1}\dots \boxed{\phi_{i_{\alpha_1} a_{\alpha_1}}^{r_{\alpha_1}}} \dots \boxed{\phi_{i_{\alpha_2} a_{\alpha_2}}^{r_{\alpha_2}}} \dots \phi_{i_{p_I} a_{p_I}}^{r_{p_I}} \equiv \Phi_{JB}^{S} \Phi_{IA}^{R}
    \end{equation*}
\end{itemize}
The product of minus signs generated in all these steps account for the explicit factor in eq(\ref{zero overlap}). We now modify the above procedure to determine the net negative signs resulting from disentangling of chords with overlap of two indices, $|I \cap J|=2$. Here, we consider the marked fermions to be the fermions with common indices, let $i_{\alpha_1}=j_{\beta_1}$ and $i_{\alpha_2}=j_{b\\beta_2}$.  Disentangling the two chords follows the same procedure as in the previous case except that we do not swap the fermions with same color indices across each other.  Thus to disentangle chords with 2 index overlaps, we follow the same steps as before except step 3 where the marked fermions are swapped across each other. The resulting configuration after doing the steps (except step 3) is given by,
\begin{equation*}
    \phi_{j_1 b_1}^{s_1}\dots\boxed{\phi_{i_{\alpha_1} a_{\alpha_1}}^{r_{\alpha_1}}} \dots \boxed{\phi_{i_{\alpha_2} a_{\alpha_2}}^{r_{\alpha_2}}}\dots \phi_{j_{p_J} b_{p_J}}^{s_{p_J}} \quad \phi_{i_1 a_1}^{r_1}\dots \boxed{\phi_{j_{\beta_1} b_{\beta_1}}^{s_{\beta_1}}}\dots \boxed{\phi_{j_{\beta_2} b_{\beta_2}}^{s_{\beta_2}}} \dots \phi_{i_{p_I} a_{p_I}}^{r_{p_I}}
\end{equation*}
Since step 3 accounted for a factor $(-1)^{2^2}$, it is easy to obtain the net negative signs, i.e 
\begin{equation*}
        \begin{split}
            \Phi_{IA}^{R} \Phi_{JB}^{S} =&  (-1)^{p_I p_J -2^2} \phi_{j_1 b_1}^{s_1}\dots\boxed{\phi_{i_{\alpha_1} a_{\alpha_1}}^{r_{\alpha_1}}} \dots \boxed{\phi_{i_{\alpha_2} a_{\alpha_2}}^{r_{\alpha_2}}}\dots \phi_{j_{p_J} b_{p_J}}^{s_{p_J}} \quad \phi_{i_1 a_1}^{r_1}\dots \boxed{\phi_{j_{\beta_1} b_{\beta_1}}^{s_{\beta_1}}}\dots \times\\
            & \hspace{9.5cm}\dots\boxed{\phi_{j_{\beta_2} b_{\beta_2}}^{s_{\beta_2}}} \dots \phi_{i_{p_I} a_{p_I}}^{r_{p_I}} \\
            =&  (-1)^{p_I p_J -2^2} \phi_{j_1 b_1}^{s_1}\dots\boxed{\phi_{j_{\beta_1} a_{\alpha_1}}^{r_{a_1}}} \dots \boxed{\phi_{j_{\beta_2} a_{\alpha_2}}^{r_{a_2}}}\dots \phi_{j_{p_J} b_{p_J}}^{s_{p_J}} \quad \phi_{i_1 a_1}^{r_1}\dots \boxed{\phi_{i_{\alpha_1} b_{\beta_1}}^{s_{\beta_1}}}\dots\times\\
        &\hspace{9.5cm}\dots\boxed{\phi_{i_{\alpha_2} b_{\beta_2}}^{s_{b_2}}} \dots \phi_{i_{p_I} a_{p_I}}^{r_{p_I}}
        \end{split}
\end{equation*}
where in going to the second line, we have used $i_{\alpha_1}=j_{\beta_1}$ and $i_{\alpha_2}=j_{\beta_2}$. The above fermions products are almost the same as $\Phi_{JB}^{S}$ and $\Phi_{IA}^{R}$ respectively except for flavour indices on fermions with overlapping colour indices which get swapped (the same goes for $R$ and $S$ labels).  The equation can be concisely captured by,
\begin{equation}
\begin{split}
        \Phi_{IA}^{R} \Phi_{JB}^{S} = (-1)^{p_I p_J -2^2} \Phi^{\tilde{S}}_{J\tilde{B}}\Phi^{\tilde{R}}_{I\tilde{A}}
\end{split}
\end{equation}
where the notation $\tilde S, \tilde R$ has been explained near eq(\ref{Unwrap}). Note that the net negative  signs remains the same even if $i_{\alpha_1}=j_{\beta_2}$ and $i_{\alpha_2}=j_{\beta_1}$. This happens since, this case involves an extra step of internal rearrangement of marked fermions which does not result in any  negative signs.  
The same procedure follows for higher number of common indices, $|I \cap J|=m$. In this case, eq(\ref{zero overlap}) generalises to,  
\begin{equation}
\Phi_{IA}^{R} \Phi_{JB}^{S}  = (-1)^{ p_I p_J - m^2 }\   \Phi^{\tilde{S}}_{J\tilde{B}}\Phi^{\tilde{R}}_{I\tilde{A}}
\end{equation}

 \section{Details of chord diagram relevant for partition function}\label{General results}
In this appendix, we evaluate explicitly the contribution to the chord diagram from various chord pairs. As mentioned in section (\ref{subsec:index sums}) - see eqn(\ref{final contribution}) - the relevant object is $x'$  for each of the chord pair configurations listed in Figure (\ref{all pairs}). For the oriented chord pairs in H chord, this was already done in section (\ref{Hchord}) and for the other configurations we give the details below 
\begin{itemize}
     \item \textbf{Oriented chord pairs in non nested chord diagram}: In the non nested chord diagram (Figure(\ref{Oriented chord pairs in Non nested H chord: 1(b)})) there are six possible oriented chord pairs. Since the contribution from pairs within the H chords has already been accounted for, the remaining contribution is due to chord pairs $IK$, $IL$, $JK$ and $JL$ which we evaluate below, 
\begin{equation}\label{non nested general result}
    \begin{split}
        |I\cap K| &\rightarrow \sum_{\substack{a,b,c,d\\ e,f,g,h}}S^{a b} S^{c d} S^{e f} S^{g h}tr_{\Lambda_i}(\chi^{\dagger}_{a}\chi_{d}\chi^{\dagger}_{e}\chi_{h} ) tr_{\Lambda_j}(\chi_b \chi_c^{\dagger} ) tr_{\Lambda_l}(\chi_f \chi_g^{\dagger} )  \\
        & = tr_\Lambda( \chi_a^\dagger \chi_b \chi_c^\dagger \chi_d) W^{ab} W^{cd} \\
    |I\cap L| &\rightarrow \sum_{\substack{a,b,c,d\\e,f,g,h}} S^{ab} S^{cd} S^{ef} S^{gh} tr_{\Lambda_i}(\chi^{\dagger}_{a} \chi_{d}\chi_f \chi^{\dagger}_{g}) tr_{\Lambda_j}(\chi_b \chi_c^{\dagger} ) tr_{\Lambda_k}(\chi_e^{\dagger} \chi_h )  \\
            & = tr_\Lambda( \chi_a^\dagger \chi_b \chi_c \chi_d^\dagger ) W^{ab} \tilde W^{cd} \\
    |J \cap K| &\rightarrow \sum_{\substack{a,b,c,d\\e,f,g,d}} S^{ab} S^{cd} S^{ef} S^{gh} tr_{\Lambda_j}(\chi_b \chi^{\dagger}_c \chi^{\dagger}_{e}\chi_h ) tr_{\Lambda_i}(\chi_a^{\dagger} \chi_d  ) tr_{\Lambda_l}(\chi_f \chi_g^{\dagger} ) \\
     & = tr_\Lambda( \chi_a \chi_b^\dagger \chi_c^\dagger \chi_d ) \tilde W^{ab}  W^{cd} \\
    |J \cap L| &\rightarrow  \sum_{\substack{abcd\\efgd}} S^{ab} S^{cd} S^{ef} S^{gh} tr_{\Lambda_j}(\chi_b \chi^{\dagger}_c \chi_f \chi^{\dagger}_{g}) tr_{\Lambda_i}(\chi_a^{\dagger} \chi_d  ) tr_{\Lambda_k}(\chi_e^{\dagger} \chi_h ) \\
     & = tr_\Lambda( \chi_a \chi_b^\dagger \chi_c \chi_d^\dagger ) \tilde W^{ab} \tilde W^{cd}
    \end{split}    
\end{equation}
Thus the net contribution is given by the sum of all these terms which is given in eq(\ref{lambda_A}).  

 \item \textbf{Oriented chord pairs in nested chord diagram}: As in the previous case the net contribution due to chord overlaps between the chords constituting nested chord diagram (Figure(\ref{Oriented chord pairs in Nested H chord: 1(c)})) are the following,
 \begin{equation}
     \begin{split}
    |I\cap K| &\rightarrow  \sum_{\substack{a,b,c,d\\ e,f,g,h}}S^{a b}S^{c d}S^{e f} S^{g h} tr_i (\chi^{\dagger}_{a}\chi^{\dagger}_{e}\chi_{h}\chi_{d}\lambda_{i}) tr_j(\chi_b \chi_c^{\dagger} \lambda_{j}) tr_l(\chi_f \chi_g^{\dagger} \lambda_{l})  \\
          & = tr_\Lambda( \chi_a^\dagger  \chi_c^\dagger \chi_d \chi_b ) W^{ab} W^{cd} \\
    |I\cap L| &\rightarrow \sum_{\substack{a,b,c,d\\ e,f,g,h}}S^{a b}S^{c d}S^{e f} S^{g h} tr_{i}(\chi^{\dagger}_{a}\chi_{f}\chi^{\dagger}_{g}\chi_{d}\lambda_{i}) tr_j(\chi_b \chi_c^{\dagger} \lambda_{j}) tr_k(\chi_e^{\dagger} \chi_h  \lambda_{k})   \\
     & = tr_\Lambda( \chi_a^\dagger \chi_c \chi_d^\dagger \chi_b) W^{ab} \tilde W^{cd} \\
    |J\cap K| &\rightarrow \sum_{\substack{a,b,c,d\\ e,f,g,h}}S^{a b}S^{c d}S^{e f} S^{g h} tr_j(\chi_{b}\chi^{\dagger}_{e}\chi_{h}\chi^{\dagger}_{c}\lambda_{j}) tr_i(\chi_a^{\dagger} \chi_d  \lambda_{i}) tr_l(\chi_f \chi_g^{\dagger}  \lambda_{l}) \\
     & = tr_\Lambda( \chi_a \chi_c^\dagger \chi_d \chi_b^\dagger) \tilde W^{ab} W^{cd} \\
    |J \cap L| &\rightarrow \sum_{\substack{a,b,c,d\\ e,f,g,h}}S^{a b}S^{c d}S^{e f} S^{g h} tr_j(\chi_{b}\chi_{f}\chi^{\dagger}_{g}\chi^{\dagger}_{c}\lambda_{j}) tr_i(\chi_a^\dagger \chi_d  \lambda_{i}) tr_k(\chi_e^{\dagger} \chi_h   \lambda_{k}) \\
     & = tr_\Lambda( \chi_a \chi_c\chi_d^\dagger \chi_b^\dagger) \tilde W^{ab} \tilde W^{cd} 
     \end{split}
 \end{equation}
 The net contribution is then given by the sum of all these terms and is given in eq(\ref{lambda_B}).
 
\item \textbf{Oriented chord pairs in intersecting chord diagram}: The net contribution due to chord overlaps between the chords constituting intersecting chord diagram (Figure(\ref{Oriented chord pairs in Intersecting H chords:1(d)})) is given by, 
\begin{equation}
    \begin{split}
        |I\cap K| &\rightarrow  \sum_{\substack{a,b,c,d\\ e,f,g,h}}S^{a b}S^{c d}S^{e f} S^{g h} tr_{\Lambda_i}(\chi^{\dagger}_{a}\chi^{\dagger}_{e}\chi_{d}\chi_{h}) tr_{\Lambda_j}( \chi_b \chi_c^{\dagger}  )tr_{\Lambda_l}( \chi_f  \chi_g^{\dagger}) \\
        &= tr_{\Lambda}(\chi^{\dagger}_{a}\chi^{\dagger}_{c}\chi_{b}\chi_{d})W^{ab}W^{cd}\\
        |I\cap L| &\rightarrow  \sum_{\substack{a,b,c,d\\ e,f,g,h}} S^{a b}S^{c d}S^{e f} S^{g h} tr_{\Lambda_j}(\chi^{\dagger}_{a}\chi_{f}\chi_{d} \chi^{\dagger}_{g}) tr_{\Lambda_i}( \chi_b \chi_c^{\dagger})tr_{\Lambda_k}( \chi_e^{\dagger} \chi_h ) \\
        &= tr_{\Lambda}(\chi^{\dagger}_{a}\chi_{c}\chi_{b} \chi^{\dagger}_{d}) W^{ab}\tilde{W}^{cd} \\
        |J\cap K| &\rightarrow  \sum_{\substack{a,b,c,d\\ e,f,g,h}} S^{a b}S^{c d}S^{e f} S^{g h} tr_{\Lambda_j}(\chi_{b}\chi^{\dagger}_{e}\chi^{\dagger}_{c} \chi_{h}) tr_{\Lambda_i}( \chi^{\dagger}_a \chi_d ) tr_{\Lambda_l}( \chi_f \chi_g^{\dagger} ) \\
        &= tr_{\Lambda} (\chi_{a}\chi^{\dagger}_{c}\chi^{\dagger}_{b} \chi_{d}) \tilde{W}^{ab} W^{cd} \\
        |J\cap L| &\rightarrow  \sum_{\substack{a,b,c,d\\ e,f,g,h}} S^{a b}S^{c d}S^{e f} S^{g h} tr_{\Lambda_j}(\chi_{b}\chi_{f}\chi^{\dagger}_{c}\chi^{\dagger}_{g}) tr_{\Lambda_i}(  \chi_a^{\dagger} \chi_d ) tr_{\Lambda_k}( \chi_e^{\dagger} \chi_h )\\
         &= tr_{\Lambda} (\chi_{a}\chi_{c}\chi^{\dagger}_{b}\chi^{\dagger}_{d}) \tilde{W}^{ab}\tilde{W}^{cd}
    \end{split}
\end{equation}
 The net contribution is then given by the sum of all these terms, along with a sign since they are intersecting, and is given in eq(\ref{lambda_C}).

\end{itemize}

\section{Details of chord diagram relevant for two point functions}\label{Two point function appendix}

As seen in section(\ref{two point function}), the chord diagram prescription for the two point function depends on contributions for M-H chord pairs and all possible configurations of M-H chord pairs were listed. In this Appendix, we compute the contribution for each of these configurations. 
\begin{itemize}
    \item[1.] \textbf{Non intersecting Non-Nested chord diagram}: 
For the case of non intersecting non nested chord diagrams, we have two possible chord diagram structures. Consider the first possible chord configuration,
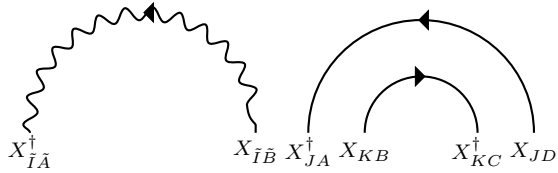
\begin{figure}[H]
\begin{center}
 \begin{tikzpicture}
\begin{scope}[thin, every node/.style={sloped,allow upside down}]
\draw[thick,black,snake it] (-2,0) arc (180:0:1.5);
\node[black] at (-2,-0.25) {\scriptsize{$X^{\dagger}_{\tilde{I}\tilde{A}}$}};
\node[black] at (1,-0.25) {\scriptsize{$X_{\tilde{I}\tilde{B}}$}};
\draw[thick,black] (1.7,0) arc (180:0:1.5);
\draw[thick,black] (2.45,0) arc (180:0:0.75);
\node[black] at (1.7,-0.25) {\scriptsize{$X^{\dagger}_{JA}$}};
\node[black] at (2.45,-0.25) {\scriptsize{$X_{KB}$}};
\node[black] at (4.7,-0.25) {\scriptsize{$X_{JD}$}};
\node[black] at (3.95,-0.25) {\scriptsize{$X^{\dagger}_{KC}$}};
\draw[black] (-0.42,1.6)-- node {\midarrow} (-0.47,1.6);
\draw[black] (3.22,1.5)-- node {\midarrow} (3.2,1.5);
\draw[black] (3.2,0.75)-- node {\midarrow} (3.22,0.75);
\end{scope}
\end{tikzpicture}
\end{center} 
\label{Non nested H chord two point function1}
\end{figure}
Consider the possible overlaps in the first chord diagram structure,
    \begin{equation}
    \begin{split}
        |\tilde{I}\cap J| &\rightarrow tr_{\Lambda_{\tilde{i}}}(\chi^{\dagger}_{\tilde{a}}\chi_{\tilde{b}}\chi_b\chi^{\dagger}_c)tr_{\Lambda_j}(\chi^\dagger_a \chi_d)S^{\tilde{a}\tilde{b}}S^{ab}S^{cd}\\
        &= tr_{\Lambda}(\chi^{\dagger}_{\tilde{a}}\chi_{\tilde{b}}\chi_a\chi^{\dagger}_b)S^{\tilde{a}\tilde{b}}\tilde{W}^{ab}\\
         |\tilde{I}\cap K| &\rightarrow tr_{\Lambda}(\chi^{\dagger}_{\tilde{a}}\chi_{\tilde{b}}\chi^{\dagger}_a\chi_d)tr_{\Lambda_k}(\chi_b \chi^\dagger_c)S^{\tilde{a}\tilde{b}}S^{ab}S^{cd}\\
        & = tr_{\Lambda}(\chi^{\dagger}_{\tilde{a}}\chi_{\tilde{b}}\chi^{\dagger}_a\chi_b)S^{\tilde{a}\tilde{b}}W^{ab}
       \end{split}
     \end{equation}

Hence the net contribution in this case is given by $e^{\gamma_1}$,where,
\begin{equation}
    \gamma_1 = \frac{\tilde{\lambda}}{x^2\tilde{x}}\left( tr_{\Lambda}(\chi^{\dagger}_{\tilde{a}}\chi_{\tilde{b}}\chi^{\dagger}_a\chi_b)S^{\tilde{a}\tilde{b}}W^{ab} + tr_{\Lambda}(\chi^{\dagger}_{\tilde{a}}\chi_{\tilde{b}}\chi_a\chi^{\dagger}_b)S^{\tilde{a}\tilde{b}}\tilde{W}^{ab} \right)
\end{equation}

The second possibility is given by,
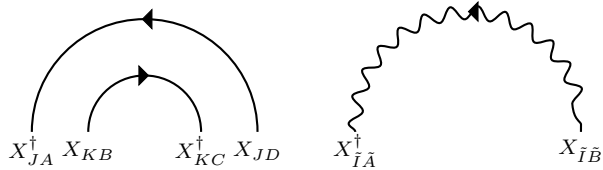
\begin{figure}[H]
\begin{center}
 \begin{tikzpicture}
\begin{scope}[thin, every node/.style={sloped,allow upside down}]
\draw[thick,black,snake it] (6,0) arc (180:0:1.5);
\node[black] at (6,-0.25) {\scriptsize{$X^{\dagger}_{\tilde{I}\tilde{A}}$}};
\node[black] at (9,-0.25) {\scriptsize{$X_{\tilde{I}\tilde{B}}$}};
\draw[thick,black] (1.7,0) arc (180:0:1.5);
\draw[thick,black] (2.45,0) arc (180:0:0.75);
\node[black] at (1.7,-0.25) {\scriptsize{$X^{\dagger}_{JA}$}};
\node[black] at (2.45,-0.25) {\scriptsize{$X_{KB}$}};
\node[black] at (4.7,-0.25) {\scriptsize{$X_{JD}$}};
\node[black] at (3.95,-0.25) {\scriptsize{$X^{\dagger}_{KC}$}};
\draw[black] (7.57,1.6)-- node {\midarrow} (7.52,1.6);
\draw[black] (3.22,1.5)-- node {\midarrow} (3.2,1.5);
\draw[black] (3.2,0.75)-- node {\midarrow} (3.22,0.75);
\end{scope}
\end{tikzpicture}
\end{center} 
\label{Non nested H chord two point function 2}
\end{figure}
The contributions due to the possible overlaps in this case is given by,
    \begin{equation}
    \begin{split}
   |\tilde{I}\cap J| &\rightarrow tr_{\Lambda_{\tilde{i}}}(\chi^{\dagger}_a\chi_d\chi^{\dagger}_{\tilde{a}}\chi_{\tilde{b}})tr_{\Lambda_k}(\chi_b \chi^\dagger_c)S^{\tilde{a}\tilde{b}}S^{ab}S^{cd}\\
    & =  tr_{\Lambda}(\chi^{\dagger}_a\chi_b\chi^{\dagger}_{\tilde{a}}\chi_{\tilde{b}})S^{\tilde{a}\tilde{b}}W^{ab} \\
    |\tilde{I}\cap K| &\rightarrow tr_{\Lambda_{\tilde{i}}}(\chi_b\chi^{\dagger}_c\chi^{\dagger}_{\tilde{a}}\chi_{\tilde{b}})tr_{\Lambda_j}(\chi^\dagger_a \chi_d)S^{\tilde{a}\tilde{b}}S^{ab}S^{cd}\\
        &= tr_{\Lambda}(\chi_a\chi^{\dagger}_b\chi^{\dagger}_{\tilde{a}}\chi_{\tilde{b}})S^{\tilde{a}\tilde{b}}\tilde{W}^{ab}
    \end{split}
    \end{equation}

Hence the net contribution in this case is given by $e^{\gamma_2}$,where,
\begin{equation}
    \gamma_2 = \frac{\tilde{\lambda}}{x^2\tilde{x}}\left( tr_{\Lambda}(\chi^{\dagger}_a\chi_b\chi^{\dagger}_{\tilde{a}}\chi_{\tilde{b}})S^{\tilde{a}\tilde{b}}W^{ab} + tr_{\Lambda}(\chi_a\chi^{\dagger}_b\chi^{\dagger}_{\tilde{a}}\chi_{\tilde{b}})S^{\tilde{a}\tilde{b}}\tilde{W}^{ab} \right)
\end{equation}

\item[2.] \textbf{Non intersecting Nested chord diagram}: The first possible chord structure in this case is given by,
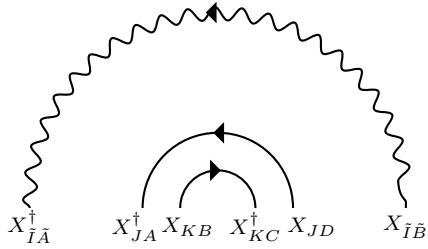
\begin{figure}[H]
\begin{center}
 \begin{tikzpicture}
\begin{scope}[thin, every node/.style={sloped,allow upside down}]
\draw[thick,black,snake it] (6,0) arc (180:0:2.5);
\draw[thick,black] (7.5,0) arc (180:0:1);
\draw[thick,black] (8,0) arc (180:0:0.5);
\node[black] at (6,-0.25) {\scriptsize{$X^{\dagger}_{\tilde{I}\tilde{A}}$}};
\node[black] at (7.4,-0.25) {\scriptsize{$X^{\dagger}_{JA}$}};
\node[black] at (8.1,-0.25) {\scriptsize{$X_{KB}$}};

\node[black] at (9,-0.25) {\scriptsize{$X^{\dagger}_{KC}$}};
\node[black] at (9.75,-0.25) {\scriptsize{$X_{JD}$}};
\node[black] at (11,-0.25) {\scriptsize{$X_{\tilde{I}\tilde{B}}$}};
\draw[black] (8.4,2.6)-- node {\midarrow} (8.38,2.6);
\draw[black] (8.5,1)-- node {\midarrow} (8.48,1);
\draw[black] (8.48,0.5)-- node {\midarrow} (8.5,0.5);
\end{scope}
\end{tikzpicture}
\end{center}  
\label{Nested H chord nested two point 1}
\end{figure}
The contributions due to the $|\tilde{I}\cap J|$ and $|\tilde{I}\cap K|$ overlaps are given by,
 \begin{equation}
    \begin{split}
      |\tilde{I}\cap J| &\rightarrow  tr_{\Lambda_{\tilde{i}}}(\chi^{\dagger}_{\tilde{a}}\chi^{\dagger}_a\chi_d\chi_{\tilde{b}})tr_{\Lambda_k}(\chi_b \chi^\dagger_c)S^{\tilde{a}\tilde{b}}S^{ab}S^{cd}\\
       &=  tr_{\Lambda}(\chi^{\dagger}_{\tilde{a}}\chi_a^{\dagger}\chi_b\chi_{\tilde{b}})S^{\tilde{a}\tilde{b}}W^{ab} \\
    |\tilde{I}\cap K| &\rightarrow tr_{\Lambda_{\tilde{i}}}(\chi^{\dagger}_{\tilde{a}}\chi_b\chi^{\dagger}_c\chi_{\tilde{b}})tr_{\Lambda_j}(\chi^\dagger_a \chi_d)S^{\tilde{a}\tilde{b}}S^{ab}S^{cd}\\
    & =  tr_{\Lambda}(\chi^{\dagger}_{\tilde{a}}\chi_a\chi_b^{\dagger}\chi_{\tilde{b}})S^{\tilde{a}\tilde{b}}\tilde{W}^{ab}
    \end{split}
    \end{equation}

Hence the net contribution in this case is given by $e^{\gamma_3}$,where,
\begin{equation}
    \gamma_3 = \frac{\tilde{\lambda}}{x^2\tilde{x}}\left( tr_{\Lambda}(\chi^{\dagger}_{\tilde{a}}\chi_a^{\dagger}\chi_b\chi_{\tilde{b}})S^{\tilde{a}\tilde{b}}W^{ab} + tr_{\Lambda}(\chi^{\dagger}_{\tilde{a}}\chi_a\chi_b^{\dagger}\chi_{\tilde{b}})S^{\tilde{a}\tilde{b}}\tilde{W}^{ab} \right)
\end{equation}
The second possible chord structure is given by,
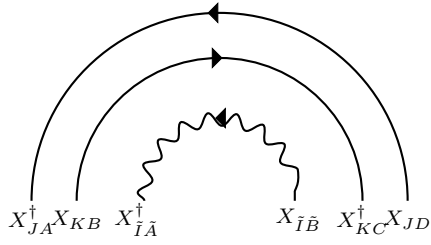
\begin{figure}[H]
\begin{center}
 \begin{tikzpicture}
\begin{scope}[thin, every node/.style={sloped,allow upside down}]
\draw[thick,black] (6,0) arc (180:0:2.5);
\draw[thick,black] (6.6,0) arc (180:0:1.9);
\draw[thick,black,snake it] (7.5,0) arc (180:0:1);
\node[black] at (6,-0.25) {\scriptsize{$X^{\dagger}_{JA}$}};
\node[black] at (6.6,-0.25) {\scriptsize{$X_{KB}$}};
\node[black] at (7.4,-0.25) {\scriptsize{$X^{\dagger}_{\tilde{I}\tilde{A}}$}};

\node[black] at (9.55,-0.25) {\scriptsize{$X_{\tilde{I}\tilde{B}}$}};
\node[black] at (10.4,-0.25) {\scriptsize{$X^{\dagger}_{KC}$}};
\node[black] at (11,-0.25) {\scriptsize{$X_{JD}$}};

\draw[black] (8.4,2.5)-- node {\midarrow} (8.38,2.5);
\draw[black] (8.48,1.9)-- node {\midarrow} (8.5,1.9);
\draw[black] (8.5,1.1)-- node {\midarrow} (8.48,1.1);
\end{scope}
\end{tikzpicture}
\end{center}  
\label{Nested H chord nested two point 2}
\end{figure}
The contributions are given by,
 $|\tilde{I}\cap J|\neq 0$,
    \begin{equation}
    \begin{split}
       |\tilde{I}\cap J| &\rightarrow  tr_{\Lambda_{\tilde{i}}}(\chi^{\dagger}_a\chi^{\dagger}_{\tilde{a}}\chi_{\tilde{b}}\chi_d)tr_{\Lambda_k}(\chi_b \chi^\dagger_c)S^{\tilde{a}\tilde{b}}S^{ab}S^{cd} \\
    & = tr_{\Lambda}(\chi_a^{\dagger}\chi^{\dagger}_{\tilde{a}}\chi_{\tilde{b}}\chi_b)S^{\tilde{a}\tilde{b}}W^{ab} \\
    |\tilde{I}\cap K| &\rightarrow tr_{\Lambda_{\tilde{i}}}(\chi_b\chi^{\dagger}_{\tilde{a}}\chi_{\tilde{b}}\chi^{\dagger}_c)tr_{\Lambda_j}(\chi^\dagger_a \chi_d)S^{\tilde{a}\tilde{b}}S^{ab}S^{cd}   \\
      & = tr_{\Lambda}(\chi_a\chi^{\dagger}_{\tilde{a}}\chi_{\tilde{b}}\chi^{\dagger}_b)S^{\tilde{a}\tilde{b}}\tilde{W}^{ab} 
    \end{split}
    \end{equation}

Hence the net contribution in this case is given by $e^{\gamma_4}$,where,
\begin{equation}
    \gamma_4 = \frac{\tilde{\lambda}}{x^2\tilde{x}}\left( tr_{\Lambda}(\chi_a^{\dagger}\chi^{\dagger}_{\tilde{a}}\chi_{\tilde{b}}\chi_b)S^{\tilde{a}\tilde{b}}W^{ab}  + tr_{\Lambda}(\chi_a\chi^{\dagger}_{\tilde{a}}\chi_{\tilde{b}}\chi^{\dagger}_b)S^{\tilde{a}\tilde{b}}\tilde{W}^{ab}  \right)
\end{equation}

\item[3.] \textbf{Intersecting chord diagram} : 
The first possible chord structure in this case is given by,
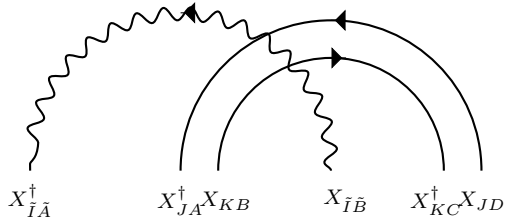
\begin{figure}[H]
 \centering
 \begin{tikzpicture}
 \begin{scope}[thin, every node/.style={sloped,allow upside down}]
         \draw[thick,snake it] (2,0) arc (0:180:2cm);
         \draw[thick] (3.5,0) arc (0:180:1.5cm);
         \draw[thick] (4,0) arc (0:180:2 cm);
        \draw[black] (0.2,2.1)-- node {\midarrow} (0,2.1);
        \draw[black] (2,1.5)-- node {\midarrow} (2.2,1.5);
        \draw[black] (2.2,2)-- node {\midarrow} (2,2);
         \node at (-2,-0.4) {\scriptsize{$X^{\dagger}_{\tilde{I}\tilde{A}}$}};
        \node at (0,-0.4) {\scriptsize{$X^{\dagger}_{JA}$}};
        \node at (0.6,-0.4) {\scriptsize{$X_{KB}$}};
        \node at (2.2,-0.4) {\scriptsize{$X_{\tilde{I}\tilde{B}}$}};  
        \node at (3.4,-0.4) {\scriptsize{$X^{\dagger}_{KC}$}};
        \node at (4,-0.4) {\scriptsize{$X_{JD}$}}; 
  \end{scope}
  \end{tikzpicture}
  \label{Intersecting H chords two point function1}
        \end{figure}
The contributions for the possible overlaps in the above chord diagram is given by,
\begin{equation}
    \begin{split}
        |\tilde{I}\cap J| &\rightarrow tr_{\Lambda_{\tilde{i}}}(\chi^{\dagger}_{\tilde{a}}\chi^{\dagger}_a\chi_{\tilde{b}}\chi_d)tr_{\Lambda_k}(\chi_b \chi^\dagger_c)S^{\tilde{a}\tilde{b}}S^{ab}S^{cd}\\
        &=  tr_{\Lambda}(\chi^{\dagger}_{\tilde{a}}\chi^{\dagger}_a\chi_{\tilde{b}}\chi_b)S^{\tilde{a}\tilde{b}}W^{ab} \\
        |\tilde{I}\cap K| &\rightarrow  tr_{\Lambda_{\tilde{i}}}(\chi^{\dagger}_{\tilde{a}}\chi_b\chi_{\tilde{b}}\chi^{\dagger}_c)tr_{\Lambda_j}(\chi^\dagger_a \chi_d)S^{\tilde{a}\tilde{b}}S^{ab}S^{cd} \\
       &= tr_{\Lambda}(\chi^{\dagger}_{\tilde{a}}\chi_a\chi_{\tilde{b}}\chi^{\dagger}_b)S^{\tilde{a}\tilde{b}}\tilde{W}^{ab} 
    \end{split}
    \end{equation}

Hence the net contribution in this case is given by $e^{\gamma_5}$,where,
\begin{equation}
    \gamma_5 = -\frac{\tilde{\lambda}}{x^2\tilde{x}}\left( tr_{\Lambda}(\chi^{\dagger}_{\tilde{a}}\chi^{\dagger}_a\chi_{\tilde{b}}\chi_b)S^{\tilde{a}\tilde{b}}W^{ab}   + tr_{\Lambda}(\chi^{\dagger}_{\tilde{a}}\chi_a\chi_{\tilde{b}}\chi^{\dagger}_b)S^{\tilde{a}\tilde{b}}\tilde{W}^{ab}  \right)
\end{equation}
The second intersecting chord diagram structure is given by, 
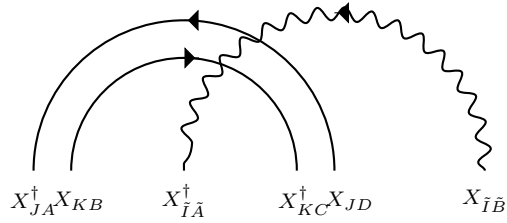
\begin{figure}[H]
 \centering
 \begin{tikzpicture}
 \begin{scope}[thin, every node/.style={sloped,allow upside down}]
         \draw[thick] (1.5,0) arc (0:180:1.5cm);
         \draw[thick] (2,0) arc (0:180:2cm);
         \draw[thick,snake it] (4,0) arc (0:180:2 cm);
         \draw[black] (0,1.5)-- node {\midarrow} (.2,1.5);
        \draw[black] (0.2,2)-- node {\midarrow} (0,2);
        \draw[black] (2.2,2.1)-- node {\midarrow} (2,2.1);
         \node at (-2,-0.4) {\scriptsize{$X^{\dagger}_{JA}$}};
        \node at (-1.4,-0.4) {\scriptsize{$X_{KB}$}};
        \node at (0,-0.4) {\scriptsize{$X^{\dagger}_{\tilde{I}\tilde{A}}$}};
        \node at (1.6,-0.4) {\scriptsize{$X^{\dagger}_{KC}$}};
        \node at (2.2,-0.4) {\scriptsize{$X_{JD}$}};  
        \node at (4,-0.4) {\scriptsize{$X_{\tilde{I}\tilde{B}}$}}; 
  \end{scope}
  \end{tikzpicture}
  \label{Intersecting H chords two point function2}
        \end{figure}
The contributions due to possible overlaps in the above diagram is given by,

    \begin{equation}
    \begin{split}
         |\tilde{I}\cap J| &\rightarrow  tr_{\Lambda_{\tilde{i}}}(\chi^{\dagger}_a\chi^{\dagger}_{\tilde{a}}\chi_d\chi_{\tilde{b}})tr_{\Lambda_k}(\chi_b \chi^\dagger_c)S^{\tilde{a}\tilde{b}}S^{ab}S^{cd}\\
         & = tr_{\Lambda}(\chi^{\dagger}_a\chi^{\dagger}_{\tilde{a}}\chi_b\chi_{\tilde{b}})S^{\tilde{a}\tilde{b}}W^{ab} \\
        |\tilde{I}\cap K| &\rightarrow  tr_{\Lambda_{\tilde{i}}}(\chi_b\chi^{\dagger}_{\tilde{a}}\chi^{\dagger}_c\chi_{\tilde{b}})tr_{\Lambda_j}(\chi^\dagger_a \chi_d)S^{\tilde{a}\tilde{b}}S^{ab}S^{cd}\\
        & =tr_{\Lambda}(\chi_a\chi^{\dagger}_{\tilde{a}}\chi^{\dagger}_b\chi_{\tilde{b}})S^{\tilde{a}\tilde{b}}\tilde{W}^{ab} 
    \end{split}
    \end{equation}

Hence the net contribution in this case is given by $e^{\gamma_6}$,where,
\begin{equation}
    \gamma_6 = -\frac{\tilde{\lambda}}{x^2\tilde{x}}\left( tr_{\Lambda}(\chi^{\dagger}_a\chi^{\dagger}_{\tilde{a}}\chi_b\chi_{\tilde{b}})S^{\tilde{a}\tilde{b}}W^{ab}   + tr_{\Lambda}(\chi_a\chi^{\dagger}_{\tilde{a}}\chi^{\dagger}_b\chi_{\tilde{b}})S^{\tilde{a}\tilde{b}}\tilde{W}^{ab}  \right)
\end{equation}
\end{itemize}

\section{Classical group results}\label{appendix: classical group simplification}
In this section we discuss some simplifications which hold for the case of classical groups, which will be useful in explicitly evaluating various parameters relevant for chord diagram rules. The following two results hold for all the groups we consider,
\begin{itemize}
\item[1.] Four fermion traces simplify to products of two fermion traces as follows,(suppressing color indices)
\begin{equation}\label{trace result}
\begin{split}
    tr_{\Lambda}(\chi^{r_1}_{a} \chi^{r_2}_{b} \chi^{r_3}_{c} \chi^{r_4}_{d})&= tr_{\Lambda}(\chi^{r_1}_{a} \chi^{r_2}_{b})tr_{\Lambda}(\chi^{r_3}_{c}\chi^{r_4}_{d})-tr_{\Lambda}(\chi^{r_1}_{a} \chi^{r_3}_{c})tr_{\Lambda}(\chi^{r_2}_{b}\chi^{r_4}_{d})\\
    & \hspace{4.75 cm} +tr_{\Lambda}(\chi^{r_1}_{a} \chi^{r_4}_{d})tr_{\Lambda}(\chi^{r_2}_{b}\chi^{r_3}_{c})
\end{split}
\end{equation}
where recall $r_i \in +,-$ such that $\chi_a^+ = \chi_a^{\dagger}$ and $\chi_a^- = \chi_a$\\
\item[2.] The two fermions traces satisfy the following property,
\begin{equation}\label{S_AB Contraction}
\begin{split}
    S^{ab} tr_{\Lambda}(\chi_{b} \chi^\dagger_{c}) S^{cd} &= \kappa tr_{\Lambda}(\chi_{a} \chi^\dagger_{d}) \\
    S^{ab} tr_\Lambda(\chi^\dagger_b \chi_c) S^{cd} &= \kappa tr_\Lambda(\chi^\dagger_a \chi_d)
    \end{split}
\end{equation} 
where $\kappa$=1 for $U(M),SO(M)$ and  $\kappa=-1$ for $USp(M/2)$. 
\end{itemize}
As a consequence of the above two results, the parameters corresponding to the partition function (see section(\ref{Final summary})) simplify further and are given by,
\begin{equation}
    \begin{split}
     x^2  M  &= \kappa tr_{\Lambda}(\chi^\dagger_{a}\chi_{d}) tr_{\Lambda}(\chi_{a} \chi^\dagger_{d})\\
     \lambda_H &= \lambda + \frac{\lambda}{x^4 M^2} tr_{\Lambda}(\chi_a \chi^{\dagger}_{d})tr_{\Lambda}(\chi^{\dagger}_b \chi_c)\left(tr_{\Lambda}(\chi^{\dagger}_a \chi_b)tr_{\Lambda}(\chi^{\dagger}_c \chi_d)-tr_{\Lambda}(\chi^{\dagger}_a \chi^{\dagger}_c)tr_{\Lambda}(\chi_b \chi_d)\right)\\
      \lambda_{N\!N} &= 4\lambda+\frac{\lambda}{x^4 M^2} tr_{\Lambda}[\chi_a,\chi^{\dagger}_b]tr_{\Lambda}[\chi_c,\chi^{\dagger}_d]\left( tr_{\Lambda}(\chi_b\chi^{\dagger}_c)tr_{\Lambda}(\chi^{\dagger}_a\chi_d)-tr_{\Lambda}(\chi_b\chi_d)tr_{\Lambda}(\chi^{\dagger}_a\chi^{\dagger}_c)\right)\\
      \lambda_{N} &= 4\lambda+ \frac{\lambda}{x^4 M^2} tr_{\Lambda}[\chi_c,\chi^{\dagger}_d](-tr_{\Lambda}(\chi^{\dagger}_a\chi_d)tr_{\Lambda}(\chi^{\dagger}_c\chi_b)tr_{\Lambda}(\chi_a\chi^{\dagger}_b)+tr_{\Lambda}(\chi_a\chi^{\dagger}_c)tr_{\Lambda}(\chi_d\chi^{\dagger}_b)tr_{\Lambda}(\chi^{\dagger}_a\chi_b)\\
    & \hspace{1cm}+tr_{\Lambda}(\chi^{\dagger}_a\chi^{\dagger}_c)tr_{\Lambda}(\chi_d\chi_b)tr_{\Lambda}(\chi_a\chi^{\dagger}_b)-tr_{\Lambda}(\chi_a\chi_d)tr_{\Lambda}(\chi^{\dagger}_c\chi^{\dagger}_b)tr_{\Lambda}(\chi^{\dagger}_a\chi_b) ) \\
     \lambda_C &= 4\lambda-\frac{\lambda}{x^4 M^2} tr_\Lambda(\chi^\dagger_a \chi^\dagger_b) tr_\Lambda (\chi_c \chi_d)  \left( 
        tr_\Lambda (\chi_a  \chi_c^\dagger) tr_\Lambda( \chi_b \chi_d^\dagger) + tr_\Lambda( \chi_c^\dagger \chi_a) tr_\Lambda (\chi_d^\dagger \chi_b) \right. \\
 & \hspace{30mm}  \left.          + 2 \ tr_\Lambda(\chi_a \chi_d^\dagger) tr_\Lambda(\chi_c^\dagger \chi_b) 
        \right)\\
        & \hspace{20mm}+2\  tr_\Lambda(\chi_a^\dagger \chi_b ) tr_\Lambda( \chi_a \chi^\dagger_c) \left( tr_\Lambda(\chi_c \chi_d^\dagger) tr_\Lambda(\chi_b^\dagger \chi_d ) + tr_\Lambda(\chi_d^\dagger \chi_c) tr_\Lambda(\chi_d \chi_b^\dagger) \right)
\end{split}
\end{equation}
For brevity, we use the convention that repeated indices are summed (regardless of whether they are upper or lower).  Similar simplifications can also be done for the parameters relevant for the two point function but we do not display it since they are not insightful.

\bibliography{DRAFT_BIB}
\bibliographystyle{JHEP}

\end{document}